\documentclass[12pt]{article}
\usepackage{amsmath,amssymb}
\usepackage[english]{babel}
\usepackage[cp866]{inputenc}
\numberwithin{equation}{section}
\newcommand{\bq}{\begin{eqnarray}}
\newcommand{\eq}{\end{eqnarray}}
\newcommand{\be}{\begin{equation}}
\newcommand{\ee}{\end{equation}}
\newcommand{\ra}{\rightarrow}
\newcommand{\la}{\Lambda_Q}
\newcommand{\mph}{\mu_{\Phi}}
\newcommand{\ov}{\overline}
\newcommand{\lym}{\Lambda_{YM}}
\newcommand{\mq}{ m^{\rm pole}_Q}
\newcommand{\bo}{{\rm b_o}}
\newcommand{\nd}{{\ov N}_c}
\newcommand{\mc}{\mu_C}

\newcommand{\mcs}{\mu^{(S)}_C}
\newcommand{\sq}{\textsf{Q}}

\newcommand{\oq}{\ov{\textsf{Q}}}
\newcommand{\dl}{{\ov \mu}^{L}_C}
\newcommand{\ds}{{\ov \mu}^{S}_C}
\newcommand{\bd}{{\rm\ov b}_{\rm o}}
\newcommand{\omp}{\mu^{\rm pole}_{q,1}}

\newcommand{\qq}{{\ov Q}Q}

\newcommand{\mo}{\mu_{\Phi,\rm o}}

\begin{document}
\begin{center}{\bf \large Mass spectra in $\mathbf{{\cal N}=1}$ SQCD with additional fields. I} \end{center}
\vspace{1cm}
\begin{center}\bf Victor L. Chernyak \end{center}
\begin{center}(e-mail: v.l.chernyak@inp.nsk.su) \end{center}
\begin{center} Budker Institut of Nuclear Physics SB RAS and Novosibirsk State University,\\ 630090 Novosibirsk, Russia
\end{center}
\vspace{1cm}
\begin{center}{\bf Abstract} \end{center}
\vspace{1cm}

Considered is the ${\cal N}=1$ SQCD-like theory with $SU(N_c)$ colors and $0< N_F<2N_c$ flavors of light quarks $\,Q_i,{\ov Q}_j$ and with the additional $N^2_F$ colorless flavored fields $\Phi_{ij}$ with the large mass parameter $\mph\gg\la$. The mass spectra of this $\Phi$ - theory (and its dual variant, the $d\Phi$ - theory) are calculated at different values of $\mph/\la$ within the dynamical scenario which implies the (quasi)spontaneous breaking of chiral symmetry. It is shown that, under appropriate conditions, the seemingly heavy and dynamically irrelevant fields $\Phi$ `return back' and there appear two additional generations of light $\Phi$ - particles with small masses $\mu(\Phi)\ll\la$.

Also considered is the $X$ - theory which is the ${\cal N}=2$ SQCD with $SU(N_c)$ colors and $0< N_F<2N_c$ flavors of light quarks, broken down to ${\cal N}=1$ by the large mass parameter of the adjoint scalar field $X$, \, $\mu_X\gg\Lambda_2$. The tight interrelations between these $X$ and $\Phi$ - theories are described, in particular, the conditions under which they are equivalent.

\newpage
{\Large\bf Contents}\\

{\bf 1 \,\,\,  Definitions}\\
1.1 \,\,\, Direct $\mathbf{\Phi}$ - theory \hspace*{12.7cm}\,\,3\\
1.2 \,\,\, Dual $\,\mathbf{d\Phi}$ - theory \hspace*{12.6cm}\,\,4
\vspace{1mm}

{\bf 2 \,\,\,   Mass spectra at $\mathbf{N_F<N_c}$ }\\
2.1 \,\,\,  Unbroken flavor symmetry \hspace*{11.05cm}\,\,5\\
2.2 \,\,\,  Broken flavor symmetry\,: $\mathbf{U(N_F)\ra U(n_1)\times U(n_2)}$ \hspace*{6.4cm}\,\,7
\vspace{1mm}

{\bf 3 \,\,\,  Quark condensates and multiplicity of vacua at~$\mathbf{N_c<N_F<2N_c}$}\hspace*{2.5cm}\,\,11
\vspace{1mm}

{\bf 4 \,\,\, Fions\,:\, one or three generations}\hspace*{8.8cm}\,\,14
\vspace{1mm}

{\bf 5 \,\,\,  Direct theory. Unbroken flavor symmetry}\\
5.1 \,\,\,  $\mathbf L$ - vacua \hspace*{14.0cm}\,\,17\\
5.2 \,\,\,  $\mathbf S$ - vacua \hspace*{14.065cm}\,\,19
\vspace{1mm}

{\bf 6\,\,\,\,\,  Dual theory. Unbroken flavor symmetry}\hspace*{0.7cm}\\
6.1 \,\,\,  $\mathbf L$ - vacua \hspace*{14.05cm}\,\,20\\
6.2 \,\,\,  $\mathbf S$ - vacua \hspace*{14.07cm}\,\,23
\vspace{1mm}

{\bf 7 \,\,\, Direct theory. Broken flavor symmetry. The region $\mathbf{\la\ll\mph\ll\mo}$}\\
7.1 \,\,\, $\mathbf L$ - type vacua \hspace*{13.15cm}\,\,25\\
7.2 \,\,\, ${\rm\bf br2}$ vacua \hspace*{13.95cm}\,\,25\\
7.3 \,\,\, {\bf special} vacua, $\mathbf{n_1=\nd,\,n_2=N_c}$ \hspace*{9.65cm}\,\,29
\vspace{1mm}

{\bf 8 \,\,\,  Direct theory. Broken flavor symmetry. The region $\mathbf{\mo\ll\mph\ll\la^2/m_Q}$}\\
8.1 \,\,\,   $\rm\bf br1$ vacua, $DC_1-DC_2$ phase \hspace*{10.5cm}\,\,30\\
8.2 \,\,\,   $\rm\bf br1$ vacua, $Higgs_1-DC_2$ phase \hspace*{10.0cm}\,\,32\\
8.3 \,\,\,   $\rm\bf br1$ vacua, $Higgs_1-HQ_2$ phase \hspace*{10.0cm}\,\,33\\
8.4 \,\,\,  $\rm \bf br2$ and {\bf special} vacua \hspace*{11.75cm}\,\,34
\vspace{1mm}

{\bf 9 \,\,\, Dual theory. Broken flavor symmetry. The region $\mathbf{\la\ll\mph\ll\mo}$}\\
9.1 \,\,\, $\mathbf L$ - type vacua  \hspace*{13.15cm}\,\,\,34\\
9.2 \,\, ${\rm\bf br2}$ vacua   \hspace*{13.95cm}\,\, 35\\
9.3 \,\, {\bf special} vacua, $\mathbf{n_1=\nd,\, n_2=N_c}$ \hspace*{9.75cm} 38
\vspace{1mm}

{\bf 10 \,\,\, Dual theory. Broken flavor symmetry. The region $\mathbf{\mo\ll\mph\ll\la^2/m_Q}$}\\
10.1 \,\,\,   $\rm\bf br1$ vacua, $DC_1-DC_2$ phase \hspace*{10.3cm}\,\,40\\
10.2 \,\,\,   $\rm\bf br1$ vacua, $HQ_1-DC_2$ phase \hspace*{10.25cm}\,\,41\\
10.3 \,\,  $\rm \bf br2$ and {\bf special} vacua \hspace*{11.6cm}\,\,43
\vspace{1mm}

{\bf 11 \,\,\, Broken $\mathbf{{\cal N}=2}$\, SQCD} \hspace*{10.9cm}\,\,43
\vspace{1mm}

{\bf 12 \,\,\, Conclusions}\hspace*{12.85cm}\,\,  46\\

{\bf Appendix A \,\,\, The RG flow in the $\mathbf \Phi$ - theory at $\mathbf{\mu>\la}$ }\hspace*{4.55cm}\,\,  47

{\bf Appendix B \,\,\, There is no vacua with $\mathbf{\langle S\rangle=0}$ at $\mathbf{m_Q\neq 0}$}
\hspace*{4.76cm}\,\, 49

{\bf References} \hspace*{13.95cm}\,\, 52

\newpage

\section{Definitions and some generalities}

\hspace{1cm} {\bf 1.1.\,\, Direct $\mathbf \Phi$ - theory}\\

The field content of this direct ${\cal N}=1\,\,\,\Phi$ - theory includes $SU(N_c)$ gluons and $0< N_F<2N_c$ flavors of quarks ${\ov Q}_j, Q_i$. Besides, there is $N^2_F$ colorless but flavored fields $\Phi_{ji}$ (fions).

The Lagrangian at scales $\mu\gg\la$ (or at $\mu\gg\mu_H$ if $\mu_H\gg\la$, where $\mu_H$ is the next largest physical mass below $\mu^{\rm pole}_1(\Phi)\gg\la$, see the Appendix A\,;\, $\nd=N_F-N_c$\,, the exponents with gluons in the Kahler term K are implied here and everywhere below) looks as
\bq
K=\frac{1}{f^2}{\rm Tr}\,\Bigl (\Phi^\dagger \Phi\Bigr )+z(\la,\mu){\rm Tr}\Bigl (\,Q^\dagger Q+(Q\ra {\ov Q})\,\Bigr )\,,
\quad W=-\frac{2\pi}{\alpha(\mu,\la)} S+W_{\Phi}+W_Q\,,
\eq
\bq
W_{\Phi}=\frac{\mph}{2}\Biggl [{\rm Tr}\,(\Phi^2)-\frac{1}{\nd}\Bigl ({\rm Tr}\,\Phi\Bigr )^2\Biggr ],\quad
W_Q={\rm Tr}\,{\ov Q}(m_Q-\Phi) Q,\quad
z_Q(\la,\mu)\sim \Bigl (\ln\frac{\mu}{\la}\Bigr )^{N_c/\bo}\gg 1\,.\nonumber
\eq
Here\,: $\mph$ and $m_Q$  are the mass parameters, $S=-W^{a}_{\beta}W^{a,\,\beta}/32\pi^2$ where $W^a_{\beta}$ is the gauge field strength, $a=1...N_c^2-1,\, \beta=1,2$,\, $\alpha(\mu,\la)=g^2(\mu,\la)/4\pi$ is the gauge coupling with its scale factor $\la$, $f$ is the Yukawa coupling, $a_f=N_cf^2/8\pi^2< 1$. {\it This normalization of fields is used everywhere below in the main text}. Besides, the perturbative NSVZ $\beta$-function for massless SUSY theories \cite{NSVZ1,NSVZ2} is used in this paper.

Therefore, finally, the $\Phi$-theory we deal with has the parameters\,: $N_c\,,\,0<N_F<2N_c\,,\,\mph$,\,
$\la,\, m_Q,\, f$, with the {\it strong hierarchies} $\mph\gg\la\gg m_Q$. Everywhere below in the text the mass parameter $\mph$ will be varied while $m_Q$ and $\la$ will stay intact.\\

The Konishi anomalies \cite{Konishi} for the $i$-th flavor look as (${\it i}=1\, ...\, N_F$)
\bq
\langle\Phi_{i}\rangle\langle\frac{\partial W_{\Phi}}{\partial \Phi_{i}}\rangle=0\,,\quad
\langle m_{Q,i}^{\rm tot}\rangle\langle {\ov Q}_i Q_i\rangle=\langle S\rangle\,,\quad \langle m_{Q,\,i}^{\rm tot}\rangle=m_Q-\langle\Phi_{i}\rangle\,, \nonumber
\eq
\bq
\langle\Phi_{ij}\rangle=\frac{1}{\mph}\Biggl ( \langle{\ov Q}_j Q_i \rangle-\delta_{ji}\frac{1}{N_c}{\rm Tr}\,\langle\qq\rangle\Biggr )\,,\quad \langle{\ov Q}_j Q_i \rangle=\delta_{ji}\langle{\ov Q}_i Q_i \rangle\,,
\eq
and, in cases with $\mu_H<\la$,\, $\langle m_{Q,l}^{\rm tot}\rangle$ is the value of the quark running mass at $\mu=\la$.

At all scales until the field $\Phi$ remains too heavy and non-dynamical, i.e. until its perturbative running mass $\mu_{\Phi}^{\rm pert}(\mu)>\mu$, it can be integrated out and the Lagrangian takes the form
\bq
K=z_Q(\la,\mu){\rm Tr}\Bigl (Q^\dagger Q+Q\ra {\ov Q}\Bigr ),\quad W=-\frac{2\pi}{\alpha(\mu,\la)}S+W_Q\,, \nonumber
\eq
\bq
W_Q=m_Q{\rm Tr}({\ov Q} Q)-\frac{1}{2\mph}\Biggl ({\rm Tr}\,({\ov Q}Q)^2-\frac{1}{N_c}\Bigl({\rm Tr}\,{\ov Q} Q \Bigr)^2 \Biggr ).
\eq

The Konishi anomalies for the Lagrangian (1.3) look as
\bq
\langle S\rangle=\langle {\ov Q}_i\frac{\partial W_Q}{\partial {\ov Q}_i}\rangle=m_Q\langle {\ov Q}_i Q_i \rangle-
\frac{1}{\mph}\Biggl (\sum_j\langle{\ov Q}_i Q_j\rangle\langle{\ov Q}_j Q_i\rangle-\frac{1}{N_c}\langle{\ov Q}_i Q_i\rangle\langle {\rm Tr}\,{\ov Q} Q \rangle\Biggr )=\nonumber
\eq
\bq
=\langle {\ov Q}_i Q_i \rangle\Biggl [\,m_Q-\frac{1}{\mph}\Biggl ( \langle {\ov Q}_i Q_i \rangle-\frac{1}{N_c}
\langle {\rm Tr}\,{\ov Q} Q \rangle\Biggr )  \Biggr ]\,,\quad i=1\,...\, N_F\,,\quad
\langle S\rangle=\langle\frac{\lambda\lambda}{32\pi^2}\rangle\,,
\eq
\bq
0=\langle {\ov Q}_i\frac{\partial W_Q}{\partial {\ov Q}_i}-{\ov Q}_j\frac{\partial W_Q}{\partial {\ov Q}_j}\rangle=\langle
{\ov Q}_i Q_i -{\ov Q}_j Q_j\rangle \Biggl [\,m_Q-\frac{1}{\mph}\Biggl ( \langle {\ov Q}_i Q_i+{\ov Q}_j Q_j\rangle
-\frac{1}{N_c}\langle {\rm Tr}\,{\ov Q} Q \rangle\Biggr )  \Biggr ]\,.\nonumber
\eq
It is most easily seen from (1.4) that there are only two types of vacua\,: a) the vacua with the unbroken flavor symmetry,\, b) the vacua with the spontaneously broken flavor symmetry, and the breaking is of the type $U(N_F)\ra U(n_1)\times U(n_2)$ only. In these vacua one obtains from (1.4)
\bq
\langle {\ov Q}_1 Q_1+{\ov Q}_2 Q_2-\frac{1}{N_c}{\rm Tr}\, {\ov Q} Q\rangle_{\rm br}=m_Q\mph,\quad
\langle S\rangle_{\rm br}=\frac{1}{\mph}\langle {\ov Q}_1 Q_1 \rangle_{\rm br}\langle {\ov Q}_2 Q_2 \rangle_{\rm br},\quad
\langle {\ov Q}_1 Q_1 \rangle_{\rm br}\neq\langle {\ov Q}_2 Q_2 \rangle_{\rm br}\,,\nonumber
\eq
\bq
\langle m^{\rm tot}_{Q,1}\rangle_{\rm br}=m_Q-\langle\Phi_1\rangle_{\rm br}=\frac{\langle{\ov Q}_2 Q_2\rangle_{\rm br}}{\mph},\quad \langle m^{\rm tot}_{Q,2}\rangle_{\rm br}=m_Q-\langle\Phi_2\rangle_{\rm br}=\frac{\langle{\ov Q}_1 Q_1\rangle_{\rm br}}{\mph}\,.
\eq

\hspace{1cm} {\bf 1.2.\,\, Dual $\mathbf {d\Phi}$ - theory}\\

In parallel with the direct $\Phi$ - theory with $N_c<N_F<2N_c$\,,\, we consider also the Seiberg dual variant \cite{S1, S2} (the $d\Phi$ - theory), with the dual Lagrangian at $\mu=\la$
\bq
K=\frac{1}{f^2}{\rm Tr}\,\Phi^\dagger \Phi+ {\rm Tr}\Bigl ( q^\dagger q + (q\ra\ov q)\, \Bigr )+{\rm Tr}\, \frac{M^{\dagger}M}{\mu_2^2}\,,\quad
W=\, -\,\frac{2\pi}{\ov \alpha(\mu=\la)}\, {\ov s}+{\ov W}_M+W_q\,,\nonumber
\eq
\bq
{\ov W}_M=\frac{\mph}{2}\Biggl [{\rm Tr}\,(\Phi^2)-\frac{1}{\nd}\Bigl ({\rm Tr}\,\Phi\Bigr )^2\Biggr ]+ {\rm Tr}\,
M(m_Q-\Phi),\quad W_q= -\,\frac{1}{\mu_1}\,\rm {Tr} \Bigl ({\ov q}\,M\, q \Bigr )\,.
\eq
Here\,:\, the number of dual colors is ${\ov N}_c=(N_F-N_c)$ and $M_{ij}$ are the $N_F^2$  elementary mion fields, ${\ov a}(\mu)=\nd{\ov \alpha}(\mu)/2\pi=\nd{\ov g}^2(\mu)/8\pi^2$ is the dual running gauge coupling (with its scale parameter $\Lambda_q$),\,\,${\ov s}=-{\rm \ov w}^{b}_{\beta}{\rm \ov w}^{b,\,\beta}/32\pi^2$,\,\, ${\rm \ov w}^b_{\beta}$ is the dual gluon field strength. The gluino condensates of the direct and dual theories are matched, $\langle{-\,\ov s}\rangle=\langle S\rangle=\lym^3$, as well as $\langle M_{ji}(\mu=\la)\rangle=\langle{\ov Q}_j Q_i (\mu=\la)\rangle$, and the scale parameter $\Lambda_q$ of the dual gauge coupling is taken as $|\Lambda_q|\sim\la$, see appendix in \cite{ch2}. At $3/2<N_F/N_c<2$ this dual theory can be taken as UV free at $\mu\gg\la$, and this requires that its Yukawa coupling at
$\mu=\la,\, f(\mu=\la)=\mu_2/\mu_1$, cannot be larger than its gauge coupling ${\ov g}(\mu=\la)$, i.e. $\mu_2/\mu_1=O(1)$. The same requirement to the value of the Yukawa coupling follows from the conformal behavior of this theory at $3/2<N_F/N_c<2$ and $\mu\ll\la$, i.e. $f(\mu=\la)\simeq f_*=O(1)$.  We consider below this dual theory at $\mu\leq\la$ only, where it claims to be equivalent to the direct $\Phi$ - theory. As was explained in \cite{ch2}, one has to take $\mu_1\sim\la$ at $\bd/\nd=O(1)$ in (1.6) to match the gluino condensates in the direct and dual theories. Therefore, $\mu_2\sim\mu_1\sim\la$ also.

Really, the fields $\Phi$ remain always too heavy and dynamically irrelevant in this $d\Phi$ - theory, so that they can be integrated out once and forever and, finally, we write the Lagrangian of the dual theory at $\mu=\la$ in the form
\bq
K= {\rm Tr}\Bigl ( q^\dagger q +(q\ra\ov q) \Bigr )+{\rm Tr}\,\frac{M^{\dagger}M}{\la^2}\,,\quad
W=\, -\,\frac{2\pi}{\ov \alpha(\mu=\la)}\, {\ov s}+W_M+W_q\,,\nonumber
\eq
\bq
W_M=m_Q{\rm Tr}\,M -\frac{1}{2\mph}\Biggl [{\rm Tr}\, (M^2)- \frac{1}{N_c}({\rm Tr}\, M)^2 \Biggr ]\,,\quad
W_q= -\,\frac{1}{\la}\,\rm {Tr} \Bigl ({\ov q}\,M\, q \Bigr )\,.
\eq

The Konishi anomalies for the $i$-th flavor look here as (${\it i}=1\, ...\, N_F$)
\bq
\langle M_i\rangle\Bigl (\langle N_i\rangle\equiv\langle{\ov q}_i q_i(\mu=\la)\rangle \Bigr )=\la\langle S\rangle\,,
\quad \frac{\langle N_i\rangle}{\la}=m_Q-\frac{1}{\mph}\Bigl (\langle M_i-\frac{1}{N_c}{\rm Tr}\,M \rangle\Bigr )=
\langle m_{Q,i}^{\rm tot}\rangle\,.
\eq

In vacua with the broken flavor symmetry these can be rewritten as
\bq
\langle M_1+M_2-\frac{1}{N_c}{\rm Tr}\, M\rangle_{\rm br}=m_Q\mph,\quad
\langle S\rangle_{\rm br}=\frac{1}{\mph}\langle M_1\rangle_{\rm br}\langle M_2\rangle_{\rm br},\quad
\langle M_1\rangle_{\rm br}\neq\langle M_2\rangle_{\rm br}\,,\nonumber
\eq
\bq
\frac{\langle N_1\rangle_{\rm br}}{\la}=\frac{\langle S\rangle_{\rm br}}{\langle M_{1}\rangle_{\rm br}}=\frac{\langle M_{2}\rangle_{\rm br}}{\mph}=m_Q-\frac{1}{\mph}\Bigl (\langle M_{1}-\frac{1}{N_c}{\rm Tr}\, M\rangle_{\rm br} \Bigr )=\langle m^{\rm tot}_{Q,1}\rangle_{\rm br}\,,
\eq
\bq
\frac{\langle N_2\rangle_{\rm br}}{\la}=\frac{\langle S\rangle_{\rm br}}{\langle M_{2}\rangle_{\rm br}}=\frac{\langle M_{1}\rangle_{\rm br}}{\mph}=m_Q-\frac{1}{\mph}\Bigl (\langle M_{2}-\frac{1}{N_c}{\rm Tr}\, M\rangle_{\rm br} \Bigr )=\langle m^{\rm tot}_{Q,1}\rangle_{\rm br}\,.\nonumber
\eq
\vspace{3mm}

Our purpose in this paper is to calculate the mass spectra in the two above theories, $\mathbf\Phi$ and $\mathbf{d\Phi}$. At present, to calculate the mass spectra in ${\cal N}=1$ SQCD-like theories, one has to rely on a definite dynamical scenario. Two different scenarios have been considered in \cite{ch1, ch2, ch3} and the mass spectra were calculated in the standard direct ${\cal N}=1$ SQCD with the superpotential $W={\rm Tr}\,(\,{\ov Q}m_Q Q)$ and in its dual variant \cite{S1, S2}.  It was shown that the direct theory and its Seiberg dual variant are not equivalent in both scenarios. In this paper we calculate the mass spectra in the $\mathbf\Phi$ and $\mathbf{d\Phi}$ theories within the scenario $\#1$.

We recall, see \cite{ch1, ch2}, that this scenario implies that, at the appropriate conditions, the quarks are not higgsed (\,i.e. $\langle{\ov Q}\rangle=\langle Q\rangle=0$\,) but form the coherent colorless diquark condensate (DC) and acquire the non-perturbative dynamical mass $\mc^2=\langle{\ov Q}Q\rangle$, and there appear light pseudo-Goldstone bosons $\Pi$ (pions) with masses $\mu(\Pi)\ll \mc$\,.\\

\section {Mass spectra at $N_F<N_c-1$}

\hspace{1cm} {\bf 2.1. Unbroken flavor symmetry}\\

There is $N_{\rm unbrok}=(2N_c-N_F)$ such vacua and all quarks are higgsed in all of them, but the hierarchies in the mass spectrum are parametrically different depending on the value of $\mph$ (see below). In any case, all $N_F^2$ fions are very heavy and dynamically irrelevant in these vacua at scales $\mu<\mu^{\rm pole}_{1}(\Phi)$ (see the Appendix A) and can be integrated out from the beginning.

All quarks are higgsed at the high scale $\mu=\mu_{\rm gl},\, \la\ll\mu_{\rm gl}\ll\mu^{\rm pole}_1(\Phi)$,
\bq
\mu^2_{\rm gl}=N_c g^2(\mu=\mu_{\rm gl})z_Q(\la,\mu_{\rm gl})\langle\Pi\rangle,\quad \langle\Pi\rangle=\langle{\ov Q}Q(\mu=\la)\rangle, \quad g^2=4\pi\alpha,
\eq
where (in the approximation of leading logs, $C_F=(N_c^2-1)/2N_c\simeq N_c/2$)
\bq
\frac{2\pi}{\alpha(\mu_{\rm gl})}\simeq \bo\ln{\frac{\mu_{\rm gl}}{\la}}\,\,,\,\, z_Q(\la,\mu_{\rm gl})\sim\Bigl (\frac
{\alpha(\la)}{\alpha(\mu_{\rm gl})}\Bigr)^{2C_F/\bo}\sim \Bigl (\ln\frac{\mu_{\rm gl}}{\la}\Bigr )^{N_c/\bo}\gg 1\,,
\quad \bo=3N_c-N_F\,.
\eq

Hence, after integrating out all heavy higgsed gluons and their superpartners at $\mu<\mu_{\rm gl}$ one remains with the $SU(N_c-N_F)$ pure Yang-Mills theory. Finally, after integrating out remained gluons at $\mu<\lym$ via the Veneziano-Yankielowicz (VY) procedure \cite{VY,TVY} (see section 2 in \cite{ch1} for more details), one obtains the Lagrangian of $N_F^2$ pions
\bq
K=z_Q(\la,\mu_{\rm gl})2{\rm Tr}\,\sqrt {\Pi^{\dagger}\Pi}\,,\quad W=-\nd S+W_{\Pi}\,,
\eq
\bq
S=\Biggl (\,\frac{\la^{\bo}}{\det \Pi}\,\Biggr )^{\frac{1}{N_c-N_F}}\,,\quad
W_{\Pi}=m_Q{\rm Tr}\,\Pi -\frac{1}{2\mph}\Biggl [{\rm Tr}\, ({\Pi}^2)- \frac{1}{N_c}({\rm Tr}\, \Pi)^2  \Biggr ],\nonumber
\eq
\bq
\langle\Pi_{ij}\rangle=\delta_{ij}\,\langle\Pi\rangle=\delta_{ij}\,\langle{\ov Q}_1 Q_1(\mu=\la)\rangle,\quad i,j=1\,...\,N_F\,.\nonumber
\eq

It follows from (2.3) that depending on the value of $\mph/\la \gg 1$\, there are two different regimes.\\

{\bf i)}\,\, At $\la\ll\mph \ll \mo$ the term $m_Q{\rm Tr}({\ov Q}Q)$ in the superpotential (2.3) gives only a small correction and one obtains
\bq
\langle \Pi \rangle_{\rm o}\sim \la^2\Bigl (\frac{\mph}{\la}\Biggr )^{\frac{N_c-N_F}{2N_c-N_F}}\gg \la^2\,.
\eq

There are $(2N_c-N_F)$ such vacua, this agrees with \cite{CKM}.
\footnote{\,
To see that there are just $2N_c-N_F$ vacua and not less, one has to separate slightly all quark masses, $m_Q\ra m_Q^{(i)},\,i=1...N_F,\, 0<(\delta m_Q)^{ij}=(m_Q^{(i)}-m_Q^{(j)})\ll {\ov m}_Q$. All quark mass terms give only small power corrections to (2.4), but just these corrections show the $Z_{2N_c-N_F}$ multiplicity of vacua.
}
The masses of heavy gluons and their superpartners are given in (2.1) while from (2.3) the pion masses are
\bq
\mu_{\rm o}(\Pi)\sim \frac{\langle \Pi \rangle_{\rm o}}{z_Q(\la,\mu_{\rm gl})\mph}\sim \frac{\la}{z_Q(\la,\mu_{\rm gl})}\Bigl (\frac{\la}{\mph}\Biggr )^{\frac{N_c}{2N_c-N_F}}\gg m_Q\,.
\eq
Besides, the scale of the gluino condensate of $SU(N_c-N_F)$ is
\bq
\lym=\langle S\rangle^{1/3}\sim \Biggl (\frac{\la^{\bo}}{\det \langle\Pi\rangle_{\rm o}}\Biggr )^{\frac{1}{3(N_c-N_F)}}\sim \la\Bigl(\frac{\la}{\mph}\Biggr )^{\frac{N_F}{3(2N_c-N_F)}}\,,\quad \mu_{\rm o}(\Pi)\ll \lym\ll \la\ll \mu_{\rm gl}\,,
\eq
and there is a large number of gluonia with the mass scale $\sim \lym$ (except for the case $N_F=N_c-1$ when the whole gauge group is higgsed and the non-perturbative superpotential in (2.3) originates from the instanton contribution).\\

{\bf ii)}\,\,\, $(2N_c-N_F)$ vacua split into two groups of vacua with parametrically different mass spectra at $\mph\gg \mo$. There are $N_c$\,\, SQCD vacua with $\langle \Pi \rangle_{\rm SQCD}\sim\la^2(\la/m_Q)^{(N_c-N_F)/N_c}$ differing by $Z_{N_c}$ phases (in these, the last term $\sim \Pi^2/\mph$ in the superpotential (2.3) can be neglected), and $(N_c-N_F)$ of nearly degenerate classical vacua with parametrically larger condensates $\langle \Pi \rangle_{\rm cl}\sim m_Q\mph$ (in these, the first non-perturbative quantum term $\sim S$ in the superpotential (2.3) gives only small corrections with $Z_{N_c-N_F}$ phases, but the multiplicity of vacua originates just from these small corrections). The properties of SQCD vacua have been described in detail in chapter 2 of \cite{ch1}, the pion masses are $\mu_{\rm SQCD}(\Pi)\sim m_Q/z_Q(\la,\mu^{SQCD}_{\rm gl})$ therein. In $(N_c-N_F)$ classical vacua the gluon and pion masses are given in (2.1) and (2.5) but now
\bq
\langle \Pi \rangle_{\rm cl}\sim  m_Q\mph \gg \la^2\,,\quad
\mu_{\rm cl}(\Pi)\sim \frac{m_Q}{z_Q(\la,\mu^{\rm cl}_{\rm gl})}\,,
\eq
and in all vacua (except for the case $N_F=N_c-1$ ) there is a large number of gluonia with the mass scale
\bq
\sim \lym=\langle S \rangle^{1/3}\sim\Biggl (\frac{\la^{\bo}}{\det \langle\Pi\rangle_{\rm SQCD}}\Biggr )^{\frac{1}{3(N_c-N_F)}}\sim \la\Biggl (\frac{m_Q}{\la}\Biggr)^{N_F/3N_c}\quad {\rm in\,\, N_c \,\,\,SQCD \,\, vacua}\,,
\eq
\bq
\sim \lym\sim\Biggl (\frac{\la^{\bo}}{\det \langle\Pi\rangle_{\rm cl}}\Biggr )^{\frac{1}{3(N_c-N_F)}}\sim \la\Biggl (\frac{\la^2}{m_Q\mph}\Biggr )^{\frac{N_F}{3(N_c-N_F)}}\,\, {\rm in\,\, (N_c-N_F) \,\,classical\,\, vacua}.
\eq

Finally, the change of regimes ${\bf i}\leftrightarrow {\bf ii}$ occurs at
\bq
\Biggl (\frac{\mo}{\la}\Biggr )^{\frac{N_c-N_F}{2N_c-N_F}}\sim \frac{m_Q \mo}{\la^2}\gg 1\quad \ra \quad
\mo \sim \la\Biggl (\frac{\la}{m_Q}\Biggr )^{\frac{2N_c-N_F}{N_c}}\,.
\eq

\hspace{1cm} {\bf 2.2\,\, Broken flavor symmetry\,: $U(N_F)\ra U(n_1)\times U(n_2)$}\\

The quark condensates $\langle{\ov Q}_j Q_i\rangle\sim C_i\delta_{ij}$ split into two groups in these vacua with the spontaneously broken flavor symmetry\,: there are $1\leq n_1\leq [N_F/2]$ equal values $\langle\Pi_1\rangle=\langle{\ov Q}_1 Q_1\rangle$ and $n_2=(N_F-n_1)\geq n_1$ equal values $\langle\Pi_2\rangle=\langle{\ov Q}_2 Q_2\rangle\neq \langle{\ov Q}_1 Q_1\rangle$ (unless stated explicitly, here and everywhere below in the text it is implied that $1-(n_1/N_c),\,\, 1-(n_2/N_c)$ and $(2N_c-N_F)/N_c$ are all $O(1)$\,). And there will be two different phases, depending on the value of $\mph/\la \gg 1$\, (see below).\\

{\,\,\,\bf 2.2.1}\,\,\, At $\la\ll\mph\ll\mo$ all qualitative properties are similar to those for an unbroken symmetry. All quarks are higgsed at high scales $\mu_{\rm gl, 1}\sim \mu_{\rm gl, 2}\gg \la$ and the low energy Lagrangian has the form (2.3). The term $m_Q{\rm Tr} ({\ov Q}Q)$ in the superpotential in (2.3) gives only small corrections, while (1.5) can be rewritten here in the form
\bq
\langle\Pi_1+\Pi_2\rangle_{\rm br}=\frac{1}{N_c}{\rm Tr}\,\langle \Pi\rangle_{\rm br}+m_Q\mph\simeq \frac{1}{N_c}\langle n_1\Pi_1+n_2\Pi_2\rangle_{\rm br}\,\, \ra \,\, \Bigl (1-\frac{n_1}{N_c}\Bigr )\langle\Pi_1\rangle_{\rm br}\simeq -\Bigl (1-\frac{n_2}{N_c}\Bigr )\langle\Pi_2\rangle_{\rm br},\nonumber
\eq
\bq
\langle S\rangle_{\rm br}=\Biggl (\frac{\la^{\bo}}{\langle\Pi_1\rangle^{n_1}_{\rm br}\langle\Pi_2\rangle^{n_2}_{\rm br}}\Biggr )^
{\frac{1}{N_c-N_F}}=\frac{\langle\Pi_1\rangle_{\rm br}\langle\Pi_2\rangle_{\rm br}}{\mph}\,,
\eq
\bq
\mu^2_{\rm gl, 1}\sim\mu^2_{\rm gl, 2}\sim  g^2(\mu=\mu_{\rm gl})z_Q(\la,\mu_{\rm gl})
\langle\Pi_{1,2}\rangle_{\rm br},
\,\, \langle \Pi_1 \rangle_{\rm br}\sim\langle\Pi_2\rangle_{\rm br}\sim \la^2\Biggl (\frac{\mph}{\la}
\Biggr )^{\frac{N_c-N_F}{2N_c-N_F}}\,.
\eq

The pion masses in this regime look as follows, see (2.3)\,:\, a) due to the spontaneous breaking of the flavor symmetry, $U(N_F)\ra U(n_1)\times U(n_2)$, there always will be $2n_1 n_2$ exactly massless particles and in this case these are the hybrids $\Pi_{12}$ and $\Pi_{21}$; \,b) other $n_1^2+n_2^2$\,\, `normal' pions have masses as in (2.5).

There are
\bq
N^{\rm tot}_{\rm brok}=\sum_{n_1=1}^{n_1=[N_F/2]}N_{\rm brok}(n_1)=
\sum_{n_1=1}^{n_1=[N_F/2]}(2N_c-N_F){\ov C}^{\, n_1}_{N_F}\,,\quad C^{\, n_1}_{N_F}=\frac{N_F!}{n_1!\, n_2!}
\eq
such vacua (the factor $2N_c-N_F$ originates from $Z_{2N_c-N_F}$ (see the footnote 1)\,, for even $N_F$ the last term with $n_1=N_F/2$ enters (2.13) with the additional factor $1/2$, i.e. ${\ov C}^{\, n_1}_{N_F}$ differ from the standard $C^{\,n_1}_{N_F}$ in (2.13) only by ${\ov C}^{\,n_1={\rm k}}_{N_F=2{\rm k}}=C^{\,n_1={\rm k}}_{N_F=2{\rm k}}/2$ ), so that the total number of vacua
\footnote {\,
By convention, we ignore the continuous multiplicity of vacua due to the spontaneous flavor symmetry breaking. Another way, one can separate slightly all quark masses (see the footnote 1), so that all Nambu-Goldstone bosons will acquire small masses $O(\delta m_Q)\ll {\ov m}_Q$.
}
is
\bq
N_{\rm tot}=\Bigl ( N_{\rm unbrok}=2N_c-N_F\Bigr )+N^{\rm tot}_{\rm brok}\,,
\eq
this agrees with \cite{CKM}.

{\,\,\,\bf 2.2.2}\,\,\, The change of the regime in these vacua with broken symmetry occurs at $\mo\ll\mph\ll{\tilde\mu}_{\Phi}$\,, see (2.10),(2.20), when all quarks are still higgsed but there appears
a large hierarchy between the values of quark condensates at $\mph\gg\mo$\,, see (1.5). Instead of $\langle \Pi_1 \rangle\sim \langle \Pi_2 \rangle $, they look now as: \\

{\bf a)}\,\, $\rm{br}1$ - vacua
\bq
\hspace*{-4mm}\langle \Pi_1 \rangle_{\rm br1}\simeq \Biggl (\rho_1=\frac{N_c}{N_c-n_1}\Biggr ) m_Q\mph\gg \la^2,\,\, \langle \Pi_2 \rangle_{\rm br1}\simeq\la^2\Bigl (\frac{\la}{m_Q\rho_1} \Bigr )^{\frac{N_c-n_2}{N_c-n_1}}\Bigl (\frac{\la}{\mph} \Bigr )^{\frac{n_1}{N_c-n_1}}\ll \langle \Pi_1 \rangle_{\rm br1}.
\eq
Unlike the mainly quantum $\langle\Pi\rangle_{\rm o}$ or mainly classical $\langle\Pi\rangle_{\rm cl}$ vacua with unbroken symmetry, these vacua are pseudo-classical\,: the largest value of the condensate $\langle \Pi_1 \rangle_{\rm br1}\sim m_Q\mph$ is classical while the smaller value of $\langle \Pi_2\rangle_{\rm br1}\sim\langle S\rangle_{\rm br1}/m_Q$ is of quantum origin, see (1.5). There are $N_{\rm br1}(n_1)=(N_c-n_1){\ov C}_{N_F}^{\, n_1}$ such vacua at given values of $n_1$ and $n_2$.\\
{\bf b)}\,\, $\rm{br2}$ - vacua. These are obtained from (2.15) by $n_1\leftrightarrow n_2$ and there are $N_{\rm br2}(n_1)=(N_c-n_2){\ov C}_{N_F}^{\, n_1}$ such vacua. Of course, the total number of vacua, $N_{\rm brok}(n_1)=N_{\rm br1}(n_1)+N_{\rm br2}(n_1)=(2N_c-N_F){\ov C}_{N_F}^{\, n_1}$ remains the same at $\mph\lessgtr\mo$.\\

We consider $\rm br1$ vacua (all results in $\rm br2$ vacua can be obtained by $n_1\leftrightarrow n_2$). In the range
$\mo\ll\mph\ll {\tilde\mu}_{\Phi}$ (see below) where all quarks are higgsed finally, the masses of higgsed gluons look now as
\bq
\mu^2_{\rm gl,1}\sim  g^2(\mu=\mu_{\rm gl,1})z_Q(\la,\mu_{\rm gl,1})\langle\Pi_1\rangle\gg
\mu^2_{\rm gl,2 }\,.
\eq
The superpotential in the low energy Lagrangian of pions loooks as in (2.3), but the Kahler term of pions is different.
We write it in the form \,:\, $K\sim z_Q(\la,\mu_{\rm gl,1}){\rm Tr}\sqrt{\Pi^{\dagger}_z\Pi_z}$\,. The $N_F\times N_F$ matrix $\Pi_z$ of pions looks as follows. Its $n_2\times n_2$ part consists of fields $z^{\prime}_Q(\mu_{\rm gl,1},\mu_{\rm gl,2})\Pi_{22}$, where $z^{\prime}_Q\ll 1$ is the perturbative logarithmic renormalization factor of ${\oq}_2,\, {\sq}_2$ quarks with unhiggsed colors which appears due to their additional RG evolution in the range of scales $\mu_{\rm gl,2}<\mu<\mu_{\rm gl,1}$, while at $\mu=\mu_{\rm gl,2}$ they are also higgsed. All other pion fields $\Pi_{11}, \Pi_{12}$ and $\Pi_{21}$ are normal. As a result, the pion masses look as follows. $2n_1n_2$ hybrid pions $\Pi_{12}$ and $\Pi_{21}$ are massless, while the masses of $n_1^2$\, $\Pi_{11}$ and $n_2^2$\, $\Pi_{22}$ are
\bq
\mu(\Pi_{11})\sim \frac{m_Q}{z_Q(\la,\mu_{\rm gl,1})}\,,\quad \mu(\Pi_{22})\sim \frac{m_Q}{z_Q(\la,\mu_{\rm gl,1})z^{\prime}_Q(\mu_{\rm gl,1},\mu_{\rm gl,2})}\gg \mu(\Pi_{11})\,.
\eq
Finally, the mass scale of gluonia from the unhiggsed $SU(N_c-N_F)$ group is $\sim \lym^{\rm (br1)}$\,, where
\bq
(\lym^{\rm (br1)})^3=\langle S\rangle_{\rm br1}=\frac{\langle \Pi_1\rangle\langle \Pi_2\rangle}{\mph}\sim m_Q\langle\Pi_2
\rangle \sim\la^3\Biggl (\frac{\la}{\mph}\Biggr)^{\frac{n_1}{N_c-n_1}}\Biggl (\frac{m_Q}{\la}\Biggr)^{\frac{n_2-n_1}{N_c-n_1}}\,.
\eq

{\,\,\,\bf 2.2.3}\,\,\,At scales $\la\ll\mu<\mu_{\rm gl,1}\sim \langle \Pi_1 \rangle^{1/2}\sim (m_Q\mph)^{1/2}$ (ignoring logarithmic factors) the light degrees of freedom include the $SU(N_c-n_1)$ gluons and active quarks ${\oq}_2,\, {\sq}_2$ with unhiggsed colors and  $n_2<(N_c-n_1)$ flavors, $n_1^2$ pions $\Pi_{11}$ and $2n_1 n_2$ hybrid pions $\Pi_{12}$ and $\Pi_{21}$ (in essence, these are ${\ov Q}_2,\, Q_2$ quarks with higgsed colors in this case). The scale factor ${\Lambda_1}$ of the gauge coupling in this lower energy theory is
\bq
\Lambda^{{\rm b}^{\prime}_{\rm o}}_1\sim\la^{\bo}/\det \Pi_{11}\,,\quad {\rm b}^{\prime}_{\rm o}=3(N_c-n_1)-n_2\,,\quad  \bo=3N_c-N_F\,.
\eq
The scale of the pole mass of ${\oq}_2,\, {\sq}_2$ quarks is $m_Q^{\rm pole}\sim m_Q$\,, while the scale of $\mu_{\rm gl,2}$ is $\mu_{\rm gl,2}\sim \langle{\oq}_2{\sq}_2\rangle^{1/2}=\langle\Pi_2\rangle^{1/2}$\,, with $\langle\Pi_2\rangle\ll \langle\Pi_1\rangle$ given in (2.15). Hence, the hierarchy at $\mo\ll\mph\ll{\tilde\mu}_{\Phi}$ looks as\,
$m_Q\ll {\Lambda_1}\ll\mu_{\rm gl, 2}\sim\langle\Pi_2\rangle^{1/2}$ and active ${\oq}_2,\, {\sq}_2$ quarks are also higgsed, while at $\mph\gg {\tilde\mu}_{\Phi}$ the hierarchy looks as $\langle\Pi_2\rangle^{1/2}\ll {\Lambda_1}\ll m_Q$ and the active quarks ${\oq}_2,\, {\sq}_2$ become too heavy and are in the $\rm HQ_2$ (heavy quark) phase. The phase changes at
\bq
\langle \Pi_2 \rangle^{1/2} \sim m_Q\sim {\langle\Lambda_1\rangle}\sim\lym^{(\rm br1)} \,\,\ra\,\, {\tilde\mu}_{\Phi}\sim \la \Biggl (\frac{\la}{m_Q}\Biggr )^{\frac{ \bo-n_1}{n_1}}\gg \mo\,.
\eq

Hence, we consider now this $Higgs_1-HQ_2$ phase realized at $\mph>{\tilde\mu}_{\Phi}$. For this it is convenient to retain all fields $\Phi$ although, in essence, they are too heavy and dynamically irrelevant. After integrating out all heavy higgsed gluons and ${\ov Q}_1, Q_1$ quarks, we write the Lagrangian at $\mu^2=\mu^2_{\rm gl,1}\sim N_c g^2(\mu=\mu_{\rm gl,1})z_Q(\la,\mu_{\rm gl,1})\langle\Pi_1\rangle$  in the form (see the Appendix A)
\bq
K=\Bigl [\,\frac{1}{f^2}{\rm Tr}(\Phi^\dagger\Phi)+z_Q(\la,\mu^2_{\rm gl,1})\Bigl (K_{\Pi}+K_{{\sq}_2}\Bigr )
\,\Bigr ],
\eq
\bq
K_{{\sq}_2}={\rm Tr}\Bigl ({\sq}^{\dagger}_2 {\sq}_2 +({\sq}_2\ra
{\oq}_2 )\Bigr )\,, \quad K_{\Pi}= 2{\rm Tr}\sqrt{\Pi^{\dagger}_{11}\Pi_{11}}+K_{\rm hybr},\nonumber
\eq
\bq
K_{\rm hybr}={\rm Tr}\Biggl (\Pi^{\dagger}_{12}\frac{1}{\sqrt{\Pi_{11}\Pi^{\dagger}_{11}}}\Pi_{12}+
\Pi_{21}\frac{1}{\sqrt{\Pi^{\dagger}_{11}\Pi_{11}}}\Pi^\dagger_{21}\Biggr ),\nonumber
\eq
\bq
W=\Bigl [-\frac{2\pi}{\alpha(\mu_{\rm gl,1})}{\textsf S}\Bigr ]+\frac{\mph}{2}\Biggl [{\rm Tr}\, (\Phi^2) -\frac{1}{\nd}\Bigl ({\rm Tr}\,\Phi\Bigr)^2\Biggr ]+{\rm Tr}\Bigl ({\oq_2}m^{\rm tot}_{{\sq}_2}{\sq}_2\Bigr )+W_{\Pi},\nonumber
\eq
\bq
W_{\Pi}= {\rm Tr}\Bigl (m_Q\Pi_{11}+m^{\rm tot}_{{\sq}_2}\,\Pi_{21}\frac{1}{\Pi_{11}}\Pi_{12}\Bigr )-
{\rm Tr}\Bigl (\Phi_{11}\Pi_{11}+\Phi_{12}\Pi_{21}+\Phi_{21}\Pi_{12} \Bigr ),
\quad m^{\rm tot}_{{\sq}_2}=(m_Q-\Phi_{22}).\nonumber
\eq
In (2.21): $\oq_2,\, \sq_2$ and $\textsf V$ are the active ${\ov Q}_2, Q_2$ guarks and gluons with unhiggsed colors ($\textsf S$ is their field strength squared), $\Pi_{12}, \Pi_{21}$ are the hybrid pions (in essence, these are the ${\ov Q}_2, Q_2$ guarks with higgsed colors), $z_Q(\la,\mu^2_{\rm gl,1})$ is the corresponding perturbative logarithmic renormalization factor of massless quarks, see (2.2). Evolving now down in the scale and integrating $\oq_2,\, \sq_2$ quarks as heavy ones at $\mu<m^{\rm pole}_{\sq_2}$ and then unhiggsed gluons at $\mu<\lym^{(\rm br1)}$ one obtains the Lagrangian of pions and fions
\bq
K=\Bigl [\frac{1}{f^2}{\rm Tr}(\Phi^\dagger\Phi)+z_Q(\la,\mu^2_{\rm gl,1})K_{\Pi}\Bigr ],\,
\eq
\bq
W=(N_c-n_1)S+\frac{\mph}{2}\Biggl [{\rm Tr} (\Phi^2) -\frac{1}{\nd}\Bigl ({\rm Tr}\,\Phi\Bigr)^2\Biggr ]+W_{\Pi}\,,
\quad\quad S=\Biggl [\frac{\la^{\bo}\det m^{\rm tot}_{{\sq}_2}}{\det \Pi_{11}}\Biggr ]^{\frac{1}{N_c-n_1}}\,,\nonumber
\eq

We start with determining the masses of hybrids $\Pi_{12}, \Pi_{21}$ and $\Phi_{12}, \Phi_{21}$. They are mixed and their kinetic and mass terms look as
\bq
K_{\rm hybr}={\rm Tr}\Bigl [\phi^{\dagger}_{12}\phi_{12}+\phi^{\dagger}_{21}\phi_{21}+\pi^{\dagger}_{12}
\pi_{12}+\pi^{\dagger}_{21}\pi_{21} \Bigr ],
\eq
\bq
W_{\rm hybr}={\rm Tr}\Bigl (m_{\phi}\phi_{12}\phi_{21}+m_{\pi}\pi_{12}\pi_{21}-m_{\rm \phi\pi}(\phi_{12}\pi_{21}+\phi_{21}\pi_{12})\Bigr )\,,\nonumber
\eq
\bq
m_{\phi}=f^2\mph,\quad m_{\pi}=\frac{m_Q-\langle\Phi_{2}\rangle}{z_Q}=\frac{\langle\Pi_1\rangle}{\mph z_Q}\sim\frac{m_Q}{z_Q}\ll m_{\phi}\,,\quad z_Q=z_Q(\la,\mu_{\rm gl,1})\,,\nonumber
\eq
\bq
m_{\rm \phi\pi}=\Bigl (\frac{f^2\langle\Pi_1\rangle}{z_Q}\Bigr )^{1/2},\quad m_{\rm \phi\pi}^2=m_{\phi}m_{\pi}\,.
\eq

Hence, the scalar potential looks as
\bq
V_S=|m|^2 |\Psi^{(-)}_{12}|^2+0|\Psi^{(+)}_{12}|^2 +(12\rightarrow 21),\quad |m|=(|m_{\phi}|+|m_{\pi}|)\,,
\eq
\bq
\Psi^{(-)}_{12}=\Bigl (c\,\phi_{12}-s\,\pi_{12} \Bigr ),\quad \Psi^{(+)}_{12}=\Bigl (c\,\pi_{12}+s\,\phi_{12} \Bigr ),
\quad c=\Bigl (\frac{|m_{\phi}|}{|m|}\Bigr )^{1/2},\quad s=\Bigl (\frac{|m_{\pi}|}{|m|}\Bigr )^{1/2}\,.\nonumber
\eq
Therefore, the fields $\Psi^{(-)}_{12}$ and $\Psi^{(-)}_{21}$ are heavy, with the masses $|m|\simeq |m_{\phi}|$, while the fields  $\Psi^{(+)}_{12}$ and $\Psi^{(+)}_{21}$ are massless. But the mixing is really parametrically small, so that the heavy fields are mainly $\phi_{12}, \phi_{21}$ while the massless ones are mainly $\pi_{12}, \pi_{21}$.
\footnote{\,
Everywhere below in the text we neglect mixing when it is small.
}

And finally from (2.22), the pole mass of  pions $\Pi_{11}$ is
\bq
\mu(\Pi_{11})\sim \frac{\langle\Pi_1\rangle}{z_Q(\la,\mu_{\rm gl,1})\mph}\sim\frac{m_Q}{z_Q(\la,\mu_{\rm gl,1})}\,.
\eq

On the whole for this $Higgs_1-HQ_2$ phase the mass spectrum looks as follows at $\mph\gg{\tilde\mu}_{\Phi}$\,. a) The heaviest are $n_1(2N_c-n_1)$ massive gluons and the same number of their scalar superpartners with the masses $\mu_{\rm gl,1}$, see (2.16), these masses originate from the higgsing of ${\ov Q}_1, Q_1$ quarks. b) There is a large number of 22-flavored hadrons made of weakly interacting and weakly confined non-relativistic $\oq_2, \sq_2$ quarks with unhiggsed colors (the string tension is $\sqrt\sigma\sim\lym^{(\rm br1)}\ll m^{\rm pole}_{\sq,2}$, see (2.18)), the scale of their masses is $m^{\rm pole}_{\sq,2}\sim m_Q/[z_Q(\la,\mu_{\rm gl,1})z^{\prime}_Q(\mu_{\rm gl,1}, m^{\rm pole}_{\sq,2})]$, where $z_Q\gg 1$ and $z^{\rm \prime}_Q\ll 1$ are the corresponding massless perturbative logarithmic renormalization factors. c) There are $n_1^2$ pions $\Pi_{11}$ with the masses (2.26), $\mu(\Pi_{11})\ll m^{\rm pole}_{\sq,2}$. d) There is a large number of gluonia made of gluons with unhiggsed colors, the scale of their masses is $\sim\lym^{(\rm br1)}$, see (2.18). e) The hybrids $\Pi_{12}, \Pi_{21}$ are massless.

All $N^2_F$ fions $\Phi_{ij}$ remain too heavy and dynamically irrelevant (see the footnote 3), their pole masses are
$\mu^{\rm pole}_1(\Phi)\sim f^2\mph\gg\mu_{\rm gl,1}$.

\section{ Quark condensates and multiplicity of vacua\\
\hspace*{4cm} at $\mathbf{N_c<N_F<2N_c}$ }

\hspace{3mm}  To obtain the numerical values of the quark condensates $\langle{\ov Q}_j Q_i\rangle=\delta_{ij}\langle{\ov Q} Q\rangle_i$ at $N_F>N_c$ (but only for this purpose), the simplest way is to use the known {\it exact form} of the non-perturbative contribution to the superpotential in the standard SQCD with the quark superpotential $m_Q{\rm Tr}({\ov Q}Q)$ and without the fions $\Phi$. It seems clear that at sufficiently large values of $\mph$ among the vacua of the $\Phi$-theory there should be $N_c$ vacua of SQCD in which, definitely, all fions $\Phi$ are too heavy and dynamically irrelevant. Therefore, they all can be integrated out and {\it the exact} superpotential can be written as ($\Pi_{ij}={\ov Q}
_j Q_i(\mu=\la),\, m_Q=m_Q(\mu=\la),\, \mph=\mph(\mu=\la)$, see Section 1 above and Sections 3 and 7 in \cite{ch1})
\bq
W=-\nd\Bigl (\frac{\det \Pi}{\la^{\bo}}\Bigr )^{1/\nd}+m_Q{\rm Tr}\,\Pi -\frac{1}{2\mph}
\Biggl [{\rm Tr}\, ({\Pi}^2)- \frac{1}{N_c}({\rm Tr}\, \Pi)^2  \Biggr ]\,.
\eq

Indeed, at sufficiently large $\mph$, there are $N_c$ vacuum solutions in (3.1) with the unbroken $SU(N_F)$ flavor symmetry. In these, the last term in (3.1) gives a small correction only and can be neglected and one obtains
\bq
\langle\Pi_{ij}\rangle_{SQCD}\simeq\delta_{ij}\frac{1}{m_Q}\Bigl (\lym^{(\rm SQCD)}\Bigr )^3=\delta_{ij}\frac{1}{m_Q}\Bigl (\la^{\bo}m_Q^{N_F}\Bigr)^{1/N_c}\,.
\eq

Now, using the holomorphic dependence of the exact superpotential on the chiral fields $({\ov Q}_j Q_i)$ and the chiral parameters $m_Q$ and $\mph$, the exact form (3.1) can be used to find the values of the quark condensates $\langle{\ov Q}_j Q_i\rangle$ in all other vacua of the $\Phi$ - theory and at all other values of $\mph>\la$. It is worth recalling only that, in general, as in the standard SQCD \cite{ch1,ch2,ch3}: a) (3.1) is not the superpotential of the genuine low energy Lagrangian describing lightest particles, it determines only the values of the vacuum condensates $\langle{\ov Q}_j Q_i\rangle$,\, b) and therefore, the notations in (3.1) do not imply literally that quarks are higgsed or form the diquark condensates, $\Pi_{ij}$ in (3.1) only have to be understood as convenient abbreviations for $({\ov Q}_j Q_i)$. (The genuine low energy Lagrangians in different vacua will be obtained below in sections 5-10, both in the direct and dual theories).  \\

{\bf 3.1}\,\, {\bf  Vacua with the unbroken flavor symmetry} $U(N_F)$. One obtains from (3.1) that at $\la\ll\mph\ll \mo$ there are two groups of such vacua with parametrically different values of condensates, $\langle{\ov Q}_j Q_i\rangle_L
=\delta_{ij}\langle{\ov Q} Q\rangle_L$ and $\langle{\ov Q}_j Q_i\rangle_S=\delta_{ij}\langle{\ov Q} Q\rangle_S$.

{\bf a}) There are $(2N_c-N_F)$ L - vacua (see also the footnote 1) with
\bq
\langle\qq(\mu=\la)\rangle_L=\langle\Pi_L\rangle\sim \la^2\Biggl (\frac{\la}{\mph}\Biggr )^{\frac{\nd}{2N_c-N_F}}\ll \la^2\,.
\eq
In these L (large) quantum vacua, the second term in the superpotential (3.1) gives numerically only a small correction.

{\bf b}) There are $(N_F-N_c)$ classical S - vacua with
\bq
\langle\qq(\mu=\la)\rangle_S=\langle\Pi_S\rangle\simeq -\frac{\nd}{N_c}\, m_Q\mph\,.
\eq
In these S (small) vacua, the first non-perturbative term in the superpotential (3.1) gives only small corrections with $Z_{N_F-N_c}$ phases, but just these corrections determine the multiplicity of these $(N_F-N_c)$ nearly degenerate vacua. On the whole, there are
\bq
N_{\rm unbrok}=(2N_c-N_F)+(N_F-N_c)=N_c
\eq
vacua with the unbroken flavor symmetry at $N_c<N_F<2N_c$.\\

One obtains from (3.1) that at $\mph\gg \mo$ the above $(2N_c-N_F)$ L - vacua and $(N_F-N_c)$ S - vacua degenerate into $N_c$ SQCD vacua (3.2).

The value of $\mo$ is determined from the matching
\bq
\Biggl [\langle\Pi\rangle_L\sim \la^2\Biggl (\frac{\la}{\mo}\Biggr )^{\frac{\nd}{2N_c-N_F}}\Biggr ]\sim \Biggl [\langle\Pi\rangle_S\sim m_Q\mo\Biggl ]\sim \Biggl [\langle\Pi\rangle_{\rm SQCD}\sim \la^2\Bigl (\frac{m_Q}{\la}\Bigr )^{\frac{\nd}{N_c}}\Biggl ]\quad\ra \nonumber
\eq
\bq
\ra \mo\sim \la\Bigl (\frac{\la}{m_Q}\Bigr )^{\frac{2N_c-N_F}{N_c}}\gg \la\,.
\eq

{\bf 3.2}\,\, {\bf  Vacua with the broken flavor symmetry} $U(N_F)\ra U(n_1)\times U(n_2),\, n_1\leq [N_F/2]$. In these, there are $n_1$ equal condensates $\langle{\ov Q}_1Q_1(\mu=\la)\rangle=\langle\Pi_1
\rangle$ and $n_2\geq n_1$ equal condensates $\langle{\ov Q}_2 Q_2(\mu=\la)\rangle=\langle\Pi_2\rangle\neq
\langle\Pi_1\rangle$. The simplest way to find the values of quark condensates in these vacua is to use (1.5). We rewrite it here for convenience
\bq
\langle\Pi_1+\Pi_2-\frac{1}{N_c}{\rm Tr}\Pi\rangle_{\rm br}=m_Q\mph\,,\quad\langle S\rangle_{\rm br}=\Bigl
(\frac{\det \langle\Pi\rangle_{\rm br}=\langle\Pi_1\rangle^{\rm {n}_1}_{\rm br}\langle\Pi_2\rangle^{\rm {n}_2}_{\rm br}}{\la^{\bo}}\Bigr )^{1/\nd}=\frac{\langle\Pi_1\rangle_{\rm br}\langle\Pi_2\rangle_{\rm br}}{\mph}\,.
\eq
Besides, the multiplicity of vacua will be shown below at given values of $n_1$ and $n_2\geq n_1$.\\

{\bf 3.2.1} The region $\la\ll\mph\ll\mo$.\\

{\bf a)} At $n_2\lessgtr N_c$, including $n_1=n_2=N_F/2$ for even $N_F$ but excluding $n_2=N_c$\,, there are $(2N_c-N_F){\ov C}^{\,n_1}_{N_F}$ L - type vacua with the parametric behavior of condensates (see the footnote 1)
\bq
(1-\frac{n_1}{N_c})\langle\Pi_1\rangle_{\rm Lt}\simeq -(1-\frac{n_2}{N_c})\langle\Pi_2\rangle_{\rm Lt}\sim \la^2\Biggl (\frac{\la}{\mph}\Biggr )^{\frac{\nd}{2N_c-N_F}},
\eq
i.e. as in the L - vacua above but $\langle\Pi_1\rangle_{\rm Lt}\neq\langle\Pi_2\rangle_{\rm Lt}$ here.

{\bf b)} At $n_2>N_c$ there are $(n_2-N_c)C^{n_1}_{N_F}$ $\rm br2$ - vacua with, see (3.7),
\bq
\langle\Pi_2\rangle_{\rm br2}\sim m_Q\mph\,,\quad \langle\Pi_1\rangle_{\rm br2}\sim \la^2\Bigl (\frac{\mph}{\la}\Bigr )^{\frac{n_2}{n_2-N_c}}\Bigl (\frac{m_Q}{\la}\Bigr )^{\frac{N_c-n_1}{n_2-N_c}},\quad
\frac{\langle\Pi_1\rangle_{\rm br2}}{\langle\Pi_2\rangle_{\rm br2}}\sim \Bigl (\frac{\mph}{\mo}\Bigr )^{\frac{N_c}{n_2-N_c}}\ll 1\,.
\eq

{\bf c)} At $n_1=\nd,\, n_2=N_c$ there are $(2N_c-N_F)\cdot C^{n_1=\nd}_{N_F}$ `special' vacua with, see (3.7),
\bq
\langle\Pi_1\rangle_{\rm spec}=\frac{N_c}{2N_c-N_F}(m_Q\mph)\,,\,\, \langle\Pi_2\rangle_{\rm spec}\sim \la^2\Bigl (\frac{\la}{\mph}\Bigr )^{\frac{\nd}{2N_c-N_F}},\,\, \frac{\langle\Pi_1\rangle_{\rm spec}}{\langle\Pi_2\rangle_{\rm spec}}\sim\Bigl (\frac{\mph}{\mo}\Bigr )^{\frac{N_c}{2N_c-N_F}}\ll 1.
\eq

On the whole, there are (\,$\theta(z)$ is the step function\,)
\bq
N_{\rm brok}(n_1)=\Bigl [(2N_c-N_F)+\theta(n_2-N_c)(n_2-N_c)\Bigr ]{\ov C}^{\,n_1}_{N_F}=
\eq
\bq
=\Bigl [(N_c-\nd)+\theta(\nd-n_1)(\nd-n_1)\Bigr ]{\ov C}^{\,n_1}_{N_F}\,,\nonumber
\eq
( ${\ov C}^{\,n_1}_{N_F}$ differ from the standard $C^{\,n_1}_{N_F}$ only by ${\ov C}^{\,n_1={\rm k}}_{N_F=2{\rm k}}=C^{\,n_1={\rm k}}_{N_F=2{\rm k}}/2$, see (2.13)\, ) vacua with the broken flavor symmetry $U(N_F)\ra U(n_1)\times U(n_2)$, this agrees with \cite{CKM}.\\ 

{\bf 3.2.2} The region $\mph\gg\mo$.\\

{\bf a)} At all values of $n_2\lessgtr N_c$, including $n_1=n_2=N_F/2$ at even $N_F$ and the `special' vacua with $n_1=\nd,\, n_2=N_c$, there are $(N_c-n_1){\ov C}^{\,n_1}_{N_F}$ $\rm br1$ - vacua with, see (3.7),
\bq
\langle\Pi_1\rangle_{\rm br1}\sim m_Q\mph\,,\quad \langle\Pi_2\rangle_{\rm br1}\sim \la^2\Bigl (\frac{\la}{\mph}\Bigr )^{\frac{n_1}{N_c-n_1}}\Bigl (\frac{\la}{m_Q}\Bigr )^{\frac{N_c-n_2}{N_c-n_1}}\,,\quad
\frac{\langle\Pi_2\rangle_{\rm br1}}{\langle\Pi_1\rangle_{\rm br1}}\sim \Bigl (\frac{\mo}{\mph}\Bigr )^{\frac{N_c}{N_c-n_1}}\ll 1\,.
\eq

{\bf b)} At $n_2<N_c$, including $n_1=n_2=N_F/2$, there are also $(N_c-n_2){\ov C}^{\,n_2}_{N_F}=(N_c-n_2){\ov C}^{\,n_1}_
{N_F}$\,\, $\rm br2$ - vacua with, see (3.7),
\bq
\langle\Pi_2\rangle_{\rm br2}\sim m_Q\mph\,,\quad \langle\Pi_1\rangle_{\rm br2}\sim \la^2\Bigl (\frac{\la}{\mph}\Bigr )^{\frac{n_2}{N_c-n_2}}\Bigl (\frac{\la}{m_Q}\Bigr )^{\frac{N_c-n_1}{N_c-n_2}}\,,\quad
\frac{\langle\Pi_1\rangle_{\rm br2}}{\langle\Pi_2\rangle_{\rm br2}}\sim \Bigl (\frac{\mo}{\mph}\Bigr )^{\frac{N_c}{N_c-n_2}}\ll 1\,.
\eq

On the whole, there are
\bq
N_{\rm brok}(n_1)=\Bigl [(N_c-n_1)+\theta (N_c-n_2)(N_c-n_2)\Bigr ]{\ov C}^{\,n_1}_{N_F}=
\eq
\bq
=\Bigl [(N_c-\nd)+\theta (\nd-n_1)(\nd-n_1)\Bigr ]{\ov C}^{\,n_1}_{N_F} \nonumber
\eq
vacua. As it should be, the number of vacua at $\mph\lessgtr \mo$ is the same.\\

As one can see from the above, all quark condensates become parametrically the same at $\mph\sim\mo$. Clearly, this region $\mph\sim\mo$ is very special and most of the quark condensates change their parametric behavior and hierarchies at $\mph\lessgtr\mo$. For example, the br2 - vacua with $n_2<N_c\,,\,\,\langle\Pi_2 \rangle\sim m_Q\mph\gg\langle\Pi_1\rangle$ at $\mph\gg\mo$ evolve into the L - type vacua with $\langle
\Pi_2\rangle\sim\langle\Pi_1\rangle\sim \la^2 (\la/\mph)^{\nd/(2N_c-N_F)}$ at $\mph\ll\mo$, while the br2 - vacua with $n_2>N_c\,,\,\,\langle\Pi_2\rangle\sim m_Q\mph\gg\langle\Pi_1\rangle$ at $\mph\ll\mo$ evolve into the br1 - vacua with $\langle\Pi_1\rangle\sim m_Q\mph\gg\langle\Pi_2\rangle$ at $\mph\gg\mo$, etc. The exception is the special vacua with $n_1=\nd,\, n_2=N_c$\,. In these, the parametric behavior $\langle\Pi_1 \rangle\sim m_Q\mph, \,\langle\Pi_2\rangle\sim \la^2(\la/\mph)^{\nd/(2N_c-N_F)}$ remains the same but the hierarchy is reversed at $\mph\lessgtr\mo\, :\, \langle\Pi_1\rangle/\langle\Pi_2\rangle\sim (\mph/\mo)^{N_c/(2N_c-N_F)}$.\\

The total number of all vacua at $N_c<N_F<2N_c$ is
\bq
N_{\rm tot}=\Bigl ( N_{\rm unbrok}=N_c \Bigr )+\Bigl ( N_{\rm brok}^{\rm tot}=\sum_{n_1=1}^{[N_F/2]}N_{\rm brok}(n_1)
\Bigr )=\sum_{k=0}^{N_c}(N_c-k)C^{\,k}_{N_F}\,,
\eq
this agrees with \cite{CKM}
\footnote{\,
But we disagree with their `derivation' in section 4.3. There is no their ${\cal N}_2$ vacua with $\langle M_{ii}\rangle\langle{\ov q}_i q_i\rangle/\la=\langle S\rangle=0,\,\, i=1,...N_F$ (no summation over $i$) in the dual theory at $m_Q\neq 0$. In all $N_{\rm tot}$ vacua in both direct and dual theories\,:\, $\langle\det M/\la^{\bo}\rangle^{1/\nd}=\langle\det {\ov Q}Q/\la^{\bo}\rangle^{1/\nd}=\langle S\rangle\neq 0$ at $m_Q\neq 0$ (see sections 5-10 below and the Appendix B). Really, the superpotential (4.48) contains all $N_{\rm tot}={\cal N}_1+{\cal N}_2$ vacua.
}
\,.

Comparing this with the number of vacua (2.13),(2.14) at $N_F<N_c$ it is seen that, for both $N_{\rm unbrok}$ and $N_{\rm brok}^{\rm tot}$ separately, the multiplicities of vacua at
$N_F<N_c$ and $N_F>N_c$  are not analytic continuations of each other.\\

The analog of (3.1) in the dual theory with $|\Lambda_q|=\la$, see (1.7), is obtained by the replacement $\Pi={\ov Q} Q(\mu=\la)\ra M(\mu=\la)$, so that $\langle M(\mu=\la)\rangle=\langle {\ov Q} Q(\mu=\la)\rangle$ in all vacua and multiplicities of vacua are the same.

\section{Fions: one or three generations}

\hspace{3mm} At $N_c<N_F<2N_c$ and in the interval of scales $\mu_H<\mu<\la$ ( $\mu_H$ is the largest physical mass in the quark-gluon sector), the quark and gluon fields are effectively massless. Because the quark renormalization factor $z_Q(\la,\mu\ll\la)=(\mu/\la)^{\gamma_Q>0}\ll 1$ decreases in this case in a {\it power fashion} with  lowering energy due to the perturbative RG evolution, it is seen  from (1.3) that the role of the 4-quark term $({\ov Q}Q)^2/\mph$ increases with lowering energy. Hence, while it is irrelevant at the scale $\mu\sim\la$ because $\mph\gg \la$, the question is whether it becomes dynamically relevant in the range of energies $\mu_H\ll\mu\ll \la$. For this, we estimate the scale $\mu_o$ where it becomes relevant in the massless theory (see section 7 in \cite{ch1} for the perturbative strong coupling regime at $N_c<N_F<3N_c/2$\,)
\bq
\frac{\mu_o}{\mph}\frac{1}{z^2_Q(\la,\mu_o)}=\frac{\mu_o}{\mph}\Bigl (\frac{\la}{\mu_o}\Bigr )^{2\gamma_Q}\sim 1\quad\ra \quad \frac{\mu_o}{\la}\sim \Bigl (\frac{\la}{\mph}\Bigr )^{\frac{1}{(2\gamma_Q-1)}}\,\,,
\eq
\bq
\gamma^{\rm conf}_Q=\frac{\bo}{N_F}\,\ra\,\frac{\mu^{\rm conf}_o}{\la}\sim \Bigl (\frac{\la}{\mph}\Bigr )^{\frac{N_F}{3(2N_c-N_F)}}\,\,,  \quad \gamma^{\rm strong}_Q=\frac{2N_c-N_F}{\nd}\,\ra\,
\frac{\mu^{\rm strong}_o}{\la}\sim \Bigl (\frac{\la}{\mph}\Bigr )^{\frac{N_F}{(5N_c-3N_F)}}\,\,.\nonumber
\eq

Hence, if $\mu_H\ll\mu_o$, then at scales $\mu<\mu_o$ the four-quark terms in the superpotential (1.3) cannot be neglected any more and we have to account for them. For this, we have to reinstate the fion fields $\Phi$ and to use the Lagrangian (1.1) in which the Kahler term at $\mu_H<\mu\ll\la$ looks as
\bq
K=\Bigl [\frac{z_{\Phi}(\la,\mu)}{f^2}{\rm Tr}\,(\Phi^\dagger \Phi)+z_Q(\la,\mu){\rm Tr}\Bigl (Q^\dagger Q+(Q\ra {\ov Q})\Bigr )\Bigr ],\,\, z_Q(\la,\mu)=\Bigl (\frac{\mu}{\la}\Bigr )^{\gamma_Q}\ll 1.
\eq

We recall that even at those scales $\mu$ that the running perturbative mass of fions $\mu_{\Phi}(\mu)\gg \mu$ and so they are too heavy and dynamically irrelevant, the quarks and gluons remain effectively massless and active. Therefore, due to the Yukawa interactions of fions with quarks, the loops  of still active light quarks (and gluons interacting with quarks)
still induce the running renormalization factor $z_{\Phi}(\la,\mu)$ of fions at all those scales where quarks are effectively massless, $\mu>\mu_H$. But, in contrast with a very slow logarithmic RG evolution at $N_F<N_c$ in section 2, the perturbative running mass of fions decreases now at $N_c<N_F<2N_c$ and $\mu<\la$ monotonically and {\it very quickly} with diminishing scale (see below), $\mph(\mu\ll \la)=\mph/f^2 z_\Phi(\la,\mu)\sim\mph(\mu/\la)^{|\gamma_{\Phi}|>1}\ll \mph$. Nevertheless, until $\mph(\mu)\gg \mu$, the fields $\Phi$ remain heavy and do not influence the RG  evolution. But,  when $\mu_H\ll\mu_o$ and $\mph(\mu)\sim\mph/z_{\Phi}(\la,\mu)$ is the main contribution to the fion mass
\footnote{\,
the cases when the additional contributions to the masses of fions from other perturbative or non-perturbative terms in the superpotential are not small in comparison with $\sim\mph/z_{\Phi}(\la,\mu)$ have to be considered separately
}
,
the quickly decreasing mass $\mph(\mu)$ becomes $\mu^{\rm pole}_2(\Phi)=\mph(\mu=\mu^{\rm pole}_2(\Phi))$ and $\mph(\mu<\mu^{\rm pole}_2(\Phi))< \mu$, so that\,: 1) there is a pole in the fion propagator at $p=\mu^{\rm pole}_2(\Phi)$, this is a second generation of fions (the first one is at $\mu^{\rm pole}_1(\Phi)\gg\la$, see Appendix A)\,; 2) the fields $\Phi$ {\it become effectively massless at $\mu<\mu^{\rm pole}_2(\Phi)$ and begin to influence the perturbative RG evolution}. In other words, the seemingly `heavy' fields $\Phi$ {\it return back}, they become effectively massless and dynamically {\it relevant}. Here and below the terms `relevant' and `irrelevant' (at a given scale $\mu$\,) will be used in the sense of whether the running mass $\sim\mph/z_{\Phi}(\la,\mu\ll\la)$ of fions at a given scale $\mu$ is $<\mu$, so that they are effectively massless and participate actively in interactions at this scale\,, or they remain too heavy with the running mass $>\mu$ whose interactions at this scale give only small corrections.

It seems clear that {\it the physical reason why the $4$-quark terms in the superpotential $(1.3)$ become relevant at scales $\mu<\mu_o$ is that the fion field $\Phi$ which was too heavy and so dynamically irrelevant at $\mu>\mu_o,\, \mph(\mu>\mu_o)>\mu$\,, becomes effectively massless at $\mu<\mu_o,\, \mph(\mu<\mu_o)<\mu$\,, and begins to participate actively in the RG evolution, i.e. it becomes relevant}. In other words, the four quark term in (1.3) `remembers' about fions and signals about the scale below which the fions become effectively massless, $\mu_o=\mu^{\rm pole}_2(\Phi)$. This allows us to find the value of $z_{\Phi}(\la,\mu_o)$,
\bq
\frac{f^2\mph}{z_{\Phi}(\la,\mu_o)}=\mu_o\,,\quad \frac{1}{f^{2}} z_{\Phi}(\la,\mu_o)=\frac{\mph}
{\mu_o}=\Bigl (\frac{\mu_o}{\la}\Bigr )^{\gamma_{\Phi}}\,,\quad \gamma_{\Phi}=-2\gamma_Q\,.
\eq

The perturbative running mass $\mph(\mu)\sim\mph/z_{\Phi}(\la,\mu\ll\la)\ll\mph$ of fions continues to decrease strongly with diminishing $\mu$ at all scales $\mu_H<\mu<\la$ until quarks remain effectively massless, and becomes frozen only at
scales below the quark physical mass, when the heavy quarks decouple.

Hence, if $\mu_H\gg\mu_o$\,, there is no pole in the fion propagator at the momenta $p<\la$ because the running fion mass is too large in this range of scales, $\mph(p>\mu_o)>p$. The fions remain dynamically irrelevant in this case at all momenta $p<\la$.

But when $\mu_H\ll\mu_o$, {\it there will be not only the second generation of fions at $p=\mu^{\rm pole}_2(\Phi)=\mu_o$
but also a third generation at $p\ll\mu_o$}. Indeed, after the heavy quarks decouple at momenta $p<\mu_H\ll\mu_o$ and the renormalization factor $z_{\Phi}(\la,\mu)$ of fions becomes frozen, $z_{\Phi}(\la,\mu<\mu_H)\sim z_{\Phi}(\la,\mu\sim
\mu_H)$, the frozen value $\mph(\mu<\mu_H)$ of the running fion mass is now $\mph(\mu\sim\mu_H)\ll p_H=\mu_H$. Hence, {there is one more pole in the fion propagator} at $p=\mu^{\rm pole}_3(\Phi)\sim \mph(\mu\sim\mu_H)\ll \mu_H$.

On the whole, in a few words for the direct theory (see the footnote 5 for reservations).\\
{\bf a)} The fions remain dynamically irrelevant and there are no poles in the fion propagator at momenta $p<\la$ if $\mu_H\gg\mu_o$.\\
{\bf b)} If $\mu_H\ll\mu_o\ll\la$, there are two poles in the fion propagator at momenta $p\ll\la$,\, $\mu^{\rm pole}_2(\Phi)\sim \mu_o$ and $\mu^{\rm pole}_3(\Phi)\sim \mph/z_{\Phi}(\la,\mu_H)\ll\mu^{\rm pole}_2(\Phi)$
(here and everywhere below in similar cases, - up to corrections due to possible nonzero decay widths of fions). In other words, the fions appear in three generations in this case (we recall that there is always the largest pole mass of fions $\mu^{\rm pole}_1(\Phi)\gg\la$, see the appendix A). Hence, the fions are effectively massless and dynamically relevant in the range of scales $\mu^{\rm pole}_3(\Phi)<\mu<\mu^{\rm pole}_2(\Phi)$.

Moreover, once the fions become effectively massless and dynamically relevant with respect to internal interactions, they begin to contribute simultaneously to the external anomalies ( the 't Hooft triangles in the external background fields).

The case $\mu_H\sim\mu_o$ requires additional information. The reason is that at scales $\mu\lesssim\mu_H$, in addition to the canonical kinetic term $\Phi^{\dagger}_R p^2\Phi_R$ (R=renormalized) of fions, there are also terms $\sim \Phi^{\dagger}_R p^2(p^2/\mu_H^2)^k\Phi_R$ with higher powers of momenta induced by loops of heavy quarks (and gluons). If $\mu_H\ll\mu_o$, then the pole in the fion propagator at $p=\mu^{\rm pole}_2(\Phi)=\mu_o$ is definitely there and, because $\mph(\mu=\mu_H)\ll\mu_H$, these additional terms are irrelevant in the region $p\sim\mph(\mu=\mu_H)\ll\mu_H$ and the pole in the fion propagator at $p=\mu^{\rm pole}_3(\Phi)=\mph(\mu=\mu_H)\ll\mu_H$ is also guaranteed.  But $\mph(\mu\sim\mu_H)\sim\mu_H$ if $\mu_H\sim\mu_o$, and these additional terms become relevant. Hence, whether there is pole in the fion propagator in this case or not depends on all these terms.\\

Now, if $\mu_H<\mu_o$ so that the fions become relevant at $\mu<\mu_o$, the question is\,: what are the values of the quark and fion anomalous dimensions, $\gamma_Q$ and $\gamma_\Phi$, in the massless perturbative regime at $\mu_H<\mu<\mu_o$\,?
To answer this question, we use the approach used in \cite{ch1} (see section 7). For this, we introduce first the corresponding massless Seiberg dual theory \cite{S2}. Our direct theory includes at $\mu_H<\mu<\mu^{\rm conf}_o$ not only the original effectively massless in this range of scales quark and gluon fields, but also the active $N_F^2$ fion fields $\Phi^j_i$ as they became now also effectively massless, so that the effective superpotential becomes nonzero and includes the Yukawa term ${\rm Tr}\,(\Phi{\ov Q}Q)$. Then, the massless dual theory with the same 't Hooft triangles includes only the massless qual quarks ${\ov q},\, q$ with $N_F$ flavors and the dual $SU(\nd=N_F-N_c)$ gluons. Further, one equates two NSVZ $\,{\widehat\beta}_{ext}$ - functions of the external baryon and $SU(N_F)_{L,R}$ - flavor vector fields in the direct and dual theories,
\bq
\frac{d}{d\,\ln \mu}\,\frac{2\pi}{\alpha_{ext}}={\widehat\beta}_{ext}= -\frac{2\pi}{\alpha^2_{ext}}\,\beta_{ext}= \sum_i T_i\,\bigl (1+\gamma_i\bigr )\,,
\eq
where the sum runs over all fields which are effectively massless at scales $\mu_H<\mu<\mu_o$, the unity in the brackets is due to one-loop contributions while the anomalous dimensions $\gamma_i$  of fields represent all higher-loop effects. It is worth noting that these general NSVZ forms (4.4) of the external `flavored' $\beta$-functions are independent of the kind of massless perturbative regime of the internal gauge theory, i.e. whether it is conformal, or the strong coupling or the IR free one.

The effectively massless particles in the direct theory here are the original quarks $Q,\,{\ov Q}$ and gluons and, in addition, the fions $\Phi^j_i$, while in the dual theory these are the dual quarks $q,\, {\ov q}$ and dual gluons only.
For the baryon currents one obtains
\bq
N_F N_c\,\Bigl ( B_Q=1 \Bigr )^2\,(1+\gamma_Q)=N_F \nd \,\Bigl ( B_q=\frac{N_c}{\nd}
\Bigr )^2\,(1+\gamma_q)\,,
\eq
while for the $SU(N_F)$  flavor currents one obtains
\bq
N_c\,(1+\gamma_Q)+N_F\,(1+\gamma_\Phi)=\nd \,(1+\gamma_q)\,.
\eq
Here, the left-hand sides are from the direct theory while the right-hand sides are from the dual one, $\gamma_Q$ and $\gamma_\Phi$ are the anomalous dimensions of the quark $Q$ and fion $\Phi$\,, while $\gamma_q$ is the anomalous dimension of the dual quark $q$.

The massless dual theory is in the conformal regime at $3N_c/2<N_F<2N_c$\,, so that $\gamma^{\rm conf}_q=\rm{{\ov b}_o}/N_F=(3\nd-N_F)/N_F$. Therefore, one obtains from (4.5),(4.6) that $\gamma^{\rm conf}_Q=\bo/N_F=(3N_c-N_F)/N_F$ and $\gamma^{\rm conf}_\Phi=-2\gamma_Q$, i.e. while only the quark-gluon sector of the direct theory behaves conformally at scales $\mu^{\rm conf}_o<\mu< \la$ where the fion fields $\Phi$ remain heavy and irrelevant, the whole theory including the fields $\Phi$ becomes conformal at scales $\mu_H<\mu< \mu^{\rm conf}_o$ where fions become effectively massless and relevant.
\footnote{\,
This does not mean that nothing changes at all after the fion field $\Phi$ begins to participate actively in the perturbative RG evolution at $\mu_H<\mu<\mu^{\rm conf}_o$. In particular, the frozen fixed point values of the gauge and Yukawa couplings $a^*$ and $a_{f}^*$ will change, and various appropriate Green functions will change their behavior, etc.
}

In the region $N_c<N_F<3N_c/2$ where the massless direct theory is in the strong coupling regime $a(\mu\ll\la)\gg 1$ (see section 7 in \cite{ch1}), the massless dual theory is in the IR free logarithmic regime. Hence, $\gamma_q$ is logarithmically small at $\mu\ll\mu^{\rm strong}_o\ll\la,\, \gamma_q\ra 0$, see (4.1), and one obtains from (4.5),(4.6) in this case the values of $\gamma^{\rm strong}_Q(\mu_H\ll\mu\ll\mu^{\rm strong}_o\ll\la)$ and $\gamma^{\rm strong}_{\Phi}(\mu_H\ll\mu\ll\mu^{\rm strong}_o\ll\la)$
\bq
\gamma^{\rm strong}_Q=\frac{2N_c-N_F}{\nd}\,,\quad \gamma^{\rm strong}_{\Phi}=-(1+\gamma^{\rm strong}_Q)=-\,\frac{N_c}{\nd}\,.
\eq

Therefore (within this approach, and in the absence of any other known way to obtain the values of $\gamma^{\rm strong}_Q$
and $\gamma^{\rm strong}_{\Phi}$ in the strong coupling regime), the quark anomalous dimension is $\gamma^{\rm strong}_Q=(2N_c-N_F)/\nd$ in the whole range $\mu_H<\mu<\la$, while the fion anomalous dimension $\gamma^{\rm strong}_{\Phi}$ changes when it becomes relevant, $\gamma^{\rm strong}_{\Phi}=-2\gamma^{\rm strong}_Q$ at $\mu^{\rm strong}_o<\mu<\la$ and $\gamma^{\rm strong}_{\Phi}=-(1+\gamma^{\rm strong}_Q)$ at $\mu_H<\mu<\mu^{\rm strong}_o$.\\

In the rest of this paper the mass spectra of the direct and dual theories will be considered within the conformal window $3N_c/2<N_F<2N_c$ only.

\section{Direct theory. Unbroken flavor symmetry}

\hspace{3mm} As in the standard ${\cal N}=1$ SQCD with the superpotential $W=m_Q{\rm Tr}({\ov Q}Q)$, the results for the mass spectrum at $N_F>N_c$ of the theory with the Lagrangian (1.2) depend essentially on the used dynamical scenario. Two different scenarios $\#1$ \cite{ch1, ch2} and $\#2$ \cite{ch3} have been considered previously. In this paper the mass spectra of the theory (1.2) will be calculated within the scenario $\#1$.

We recall that this scenario implies that when the scale of the quark condensate is in the range $\mq<\mc=|\langle{\ov Q}Q(\mu=\la)\rangle|^{1/2}<\la$ (and nothing prevents), the quarks are not higgsed but form the coherent colorless diquark condensate (DC) and acquire the dynamical mass $\mc$, and there appear light pseudo-Goldstone bosons $\Pi$ (pions) with masses $\mu(\Pi)\ll \mc$, see \cite{ch1, ch2}.\\

\hspace{5mm}{\bf 5.1\,\,\, L - vacua}\\

In these $(2N_c-N_F)$ vacua at $\la\ll\mph\ll\mo$, we compare first the possible constituent and pole quark masses. As for the constituent mass, it looks as, see (3.3),
\bq
\Bigl (\mu^{(L)}_C\Bigr )^2=\langle\Pi\rangle_L=\langle{\ov Q}Q(\mu=\la)\rangle_L\sim \la^2\Biggl (\frac{\la}{\mph}\Biggr )^{\frac{\nd}{2N_c-N_F}}\ll \la^2\,,\quad \la\ll \mph\ll\mo\,.
\eq

In vacua with unbroken flavor symmetry all quark masses are equal, see (1.2),
\bq
\langle m^{\rm tot}_Q\rangle\equiv\langle m^{\rm tot}_Q\rangle(\mu=\la)=m_Q+\frac{\nd}{N_c\mph}\langle\Pi\rangle ,\quad \langle\Pi\rangle=\langle{\ov Q}_1 Q_1(\mu=\la)\rangle\,.
\eq
Then, the quark pole mass looks as
\bq
m^{\rm pole}_Q=\frac{\langle m^{\rm tot}_Q\rangle}{z_Q(\la,\mu=m^{\rm pole}_Q)}\,\,,\,\,\, z_Q(\la,\mu\ll\la)=\Bigl (\frac{\mu}{\la}\Bigr )^{\gamma_Q=(\bo/N_F)}\ll 1\,.
\eq

Now, in L-vacua with $\langle\Pi\rangle_L$\,, see (5.2),
\bq
\frac{\langle m^{\rm tot}_Q\rangle_L}{\la}\sim \frac{\langle\Pi\rangle_L}{\la\mph}\sim \Bigl (\frac{\la}{\mph}\Bigr )^{\frac{N_c}{2N_c-N_F}},\quad \frac{m^{\rm pole}_{Q,L}}{\la}=\Bigl (\frac{\langle m^{\rm tot}_Q\rangle_L}{\la}\Bigr )^{\frac{N_F}{3N_c}}\sim \nonumber
\eq
\bq
\sim\Bigl (\frac{\la}{\mph}\Bigr )^{\frac{N_F}{3(2N_c-N_F)}}\sim\lym^{(L)}\,,\quad\frac{m^{\rm pole}_{Q,L}}
{\mu^{(L)}_C}\sim \Bigl (\frac{\la}{\mph}\Bigr )^{\frac{\bo}{6(2N_c-N_F)}}\ll 1\,.
\eq

Therefore, all quarks are in the $DC$ (diquark condensate) phase. We check now whether the fions will be relevant in these L-vacua. For this, we have to compare $\mu^{\rm conf}_o$ in (4.1) with $\mu^{(L)}_C$ in (5.1),
\bq
\frac{\mu^{\rm conf}_o}{\mu^{(L)}_C}\sim \Bigl (\frac{\la}{\mph}\Bigr )^{\frac{\bo}{6(2N_c-N_F)}}\ll 1\,.
\eq

Therefore, the fion fields remain dynamically irrelevant in these L-vacua, they can be integrated out and, at $\mu<\la$, we can start directly with the Lagrangian (1.3). Proceeding now as in section 3 of \cite{ch1}, i.e. integrating out first the heaviest constituent quarks at $\mu<\mu^{(L)}_C$ and then gluons at $\mu<\lym^{(L)}$, one obtains the Lagrangian of $N^2_F$ pions,
\bq
K={\rm Tr}\,\sqrt {\Pi^{\dagger}\Pi}\,\,,\quad
W= - \nd\Biggl (\frac{\det \Pi}{\la^{\bo}}\Biggr )^{1/\nd}+m_Q{\rm Tr}\,\Pi -\frac{1}{2\mph}
\Biggl [{\rm Tr}\, (\Pi^2)- \frac{1}{N_c}({\rm Tr}\, \Pi)^2  \Biggr ]\,.
\eq
The term with $m_Q$ in (5.6) can be neglected in these L-vacua at $\la\ll\mph\ll\mo$, and one obtains for the pion masses
\bq
\mu_L(\Pi)\sim\frac{\langle\Pi\rangle_L}{\mph}\sim \la\Bigl (\frac{\la}{\mph}\Bigr )^{\frac{N_c}{2N_c-N_F}},\quad \frac{\mu_L(\Pi)}{\lym^{(L)}}\sim \Bigl (\frac{\la}{\mph}\Bigr )^{\frac{\bo}{3(2N_c-N_F)}}\ll 1\,.
\eq

On the whole, the mass spectrum in these $(2N_c-N_F)$ L-vacua with the unbroken flavor symmetry looks as follows
at $\la\ll\mph\ll\mo$.\\
{\bf a)}\, There is a large number of heaviest flavored hadrons made of  non-relativistic and weakly confined (the string tension is ${\sqrt\sigma}\sim \lym^{(L)}\ll \mu^{(L)}_C$\,) constituent quarks with the masses $\mu^{(L)}_C=\langle{\ov Q}Q\rangle^{1/2}_L$ (5.1).\\
{\bf b)}\, There is a large number of gluonia with the mass scale $\sim \lym^{(L)}=\langle S\rangle^{1/3}_L\ll \mu^{(L)}_C$\, (5.4).\\
{\bf c)}\, The lightest are $N_F^2$ pions with the masses $\mu_L(\Pi)$ in (5.7). The fions are dynamically irrelevant in these vacua.

The term with $m_Q$ cannot be neglected any more in (5.6) at $\mph\gtrsim \mo$  and all condensates and masses evolve to those in the standard SQCD, see section 3 in \cite{ch1}.\\

\hspace{5mm}{\bf 5.2\,\,\, S - vacua}\\

In these $(N_F-N_c)$ vacua at $\la\ll\mph\ll\mo$, the possible constituent and pole quark masses look as, see (1.2),(3.4),
\bq
\Bigl (\mu^{(S)}_C\Bigr )^2=\langle\Pi\rangle_S\simeq \frac{\nd}{N_c}\, (m_Q\mph)\,,
\eq
\bq
\frac{\langle m^{\rm tot}_Q\rangle_S}{\la}\sim \frac{\langle S\rangle_S}{\la\langle\Pi\rangle_S}\sim \Bigl (\frac{\langle\Pi\rangle_S}{\la^2}\Bigr )^{N_c/\nd}\sim \Bigl (\frac{m_Q\mph}{\la^2}\Bigr )^{N_c/\nd}\,,\nonumber
\eq
\bq
\frac{m^{\rm pole}_{Q,S}}{\la}\sim \Bigl (\frac{m_Q\mph}{\la^2}\Bigr )^{N_F/3\nd},\quad \frac{m^{\rm pole}_{Q,S}}{\mcs}
\sim \Bigl (\frac{m_Q\mph}{\la^2}\Bigr )^{\bo/6\nd}\ll 1\,,\quad \la\ll \mph\ll\mo\,.\nonumber
\eq
Therefore, all quarks are in the $DC$-phase also. But the analog of (5.5) looks now as
\bq
\frac{\mu^{\rm conf}_o}{\mcs}\sim \Bigl (\frac{\la}{m_Q}\Bigr )^{1/2}\Bigl (\frac{\la}{\mph}\Bigr )^{\frac{6N_c-N_F}{6(2N_c-N_F)}}\,.
\eq
Hence, in these S-vacua, there appear two additional generations of fions at $\la\ll\mph<\mph^{(S,{\,\rm conf})}$,
\bq
\mph^{(S,{\,\rm conf})}=\la\Bigl(\frac{\la}{m_Q}\Bigr )^{\frac{3(2N_c-N_F)}{6N_c-N_F}}\,,
\quad \la\ll\mph^{(S,{\,\rm conf})}\ll\mo\,,
\eq
while the fions remain too heavy and dynamically irrelevant at $\mph^{(S,{\,\rm conf})}<\mph<\mo$.

The conformal regime continues here down to the scale $\mu_H$, where $\mu_H$ is the largest physical mass in the quark-gluon sector and here it is $\mu_H=\mcs\sim (m_Q\mph)^{1/2}$. The RG evolution of the quark and fion fields becomes frozen at scales $\mu<\mcs$ because the heavy constituent quarks decouple. Proceeding as in \cite{ch1} (see section 3), i.e. integrating out first constituent quarks as heavy ones at $\mu<\mcs$ and then gluons at $\mu<\lym^{(S)}$, one obtains the Lagrangian of the fion and pion fields
\footnote{\,
from now on we omit the constant $f=O(1)$ for simplicity
}\,,
\bq
K={\rm Tr}\,\sqrt{\Pi^\dagger\Pi}+z^{(S)}_\Phi(\la,\mcs){\rm Tr}\,(\Phi^\dagger \Phi)\,,\quad z^{(S)}_\Phi(\la,\mu)\sim\Bigl (\frac{\la}{\mu}\Bigr )^{2\bo/N_F}\gg 1\,,
\eq
\bq
W=\,-\nd\, S+\frac{\mph}{2}\Biggl [{\rm Tr}\,(\Phi^2)-\frac{1}{\nd}\Bigl ({\rm Tr}\,\Phi\Bigr )^2\Biggr ]+{\rm Tr}\,  (m_Q-\Phi)\Pi\,,\quad S=\Biggl (\frac{\det \Pi}{\la^\bo}\Biggr )^{1/\nd}.\nonumber
\eq

The non-perturbative term with the determinant in (5.11) can be neglected in these S-vacua with $\langle \Pi\rangle_S
\sim m_Q\mph$ at $\la\ll\mo$, and one obtains for the masses of $\Phi$ and $\Pi$ (see the footnote 3)
\bq
\mu_S(\Phi)\sim \frac{\mph}{z^{(S)}_\Phi(\la,\mcs)}\,,\quad \mu_S(\Pi)\sim m_Q\,,\quad \frac{\mu_S(\Phi)}{\mcs}
\sim \Bigl (\frac{\mph}{\mph^{(S,{\,\rm conf})}}\Bigr )^{\frac{6N_c-N_F}{2N_F}}\,.
\eq

Another way, because $\mu_S(\Phi)\gg\mu_S(\Pi)$, one can integrate out first in (5.11) the heavier fields $\Phi$ at scales $\mu<\mu_S(\Phi)$ and obtains
\bq
K_{\Pi}={\rm Tr}\,\sqrt{\Pi^\dagger\Pi}\,, \quad W_{\Pi}=\,-\nd\, S -\frac{1}{2\mph}\Biggl [{\rm Tr}\,(\Pi^2)-\frac{1}{N_c}\Bigl ({\rm Tr}\,\Pi\Bigr )^2\Biggr ]+m_Q{\rm Tr}\,\Pi\,.
\eq
Neglecting the term $\sim S$ with the determinant in these S-vacua, one obtains from (5.13) that the pion mass is $\mu_S(\Pi)\sim m_Q$.

We discuss now in more detail the fion masses in these S-vacua.

{\bf i)} At $\mph>\mph^{(S,{\,\rm conf})}$ the running mass
$\mu_{\Phi}(\mu)=\mph/z_{\Phi}^{(S)}(\la,\mu)$ of fions remains larger than the scale $\mu$ in the whole range $\mcs<\mu<
\la$, until it becomes frozen at lower scales $\mu<\mcs$ at its value $\mph/z_{\Phi}^{(S)}(\la,\mcs)>\mcs$. Hence, {\it there are no poles in the propagators of fions at all momenta} $p<\la\ll\mu^{\rm pole}_1(\Phi)$ and the `mass' $\mu_S(\Phi)$ in (5.12) is not the observable pole mass but simply the limiting value of the mass term in the fion propagator at low momenta.

{\bf ii)} The situation is qualitatively different at $\la<\mph<\mph^{(S,{\,\rm conf})}$. The running mass of fions
$\mu_{\Phi}(\mu)$ is $\mu_{\Phi}(\mu)>\mu$ at $\mu>\mu^{\rm conf}_o$ but it becomes $\mu_{\Phi}(\mu)<\mu$ at $\mu<\mu^{\rm conf}_o$. Therefore, {\it there is a second pole in the fion propagator at $p=\mu^{\rm conf}_o$}
( the first pole is at $\mu^{\rm pole}_1(\Phi)\sim\mph\gg\la$, see the Appendix A). Moreover, because the frozen value $\mu_{\Phi}(\mu=\mcs)$ of $\mu_{\Phi}(\mu)$ is $\mu_{\Phi}(\mu=\mcs)<\mcs$, {\it there is a third pole in the fion propagator at $p=\mu_{\Phi}(\mu=\mcs)<\mcs$}. Therefore,  there are now {\it three generations} of fions at $\la<\mph<
\mph^{(S,{\,\rm conf})}$ with three different observable pole masses. In addition to the first universal generation with the very large pole mass $\mu^{\rm pole}_1(\Phi)\gg \la$, there are in this case {\it two additional generations} with parametrically smaller pole masses
\bq
\mu^{\rm pole}_2(\Phi)=\mu^{\rm conf}_o\sim \la \Bigl (\frac{\la}{\mph}\Bigr )^{\frac{N_F}{3(2N_c-N_F)}}>\mcs\,,
\eq
\bq
\mu^{\rm pole}_3(\Phi)=\mu_S(\Phi)\sim \la \Bigl (\frac{\mph}{\la}\Bigr )^{\frac{3N_c}{N_F}}\Bigl (\frac{m_Q}{\la} \Bigr )^{{\frac{\bo}{N_F}}}<\mcs<\mu^{\rm pole}_2(\Phi)\ll\la\,.\nonumber
\eq

On the whole, the mass spectrum looks in these S-vacua as follows at $\la\ll\mph\ll\mo$. \\
{\bf\,a)}\,  There is a large number of flavored hadrons made of non-relativistic (and weakly confined, the string tension is $\sqrt{\sigma}\sim\lym^{(S)}\ll\mcs$ ) constituent quarks with the masses $\mcs\sim (m_Q\mph)^{1/2}$.\\
{\bf b)}\, There is a large number of gluonia with the mass scale $\sim \lym^{(S)}\sim\la(m_Q\mph/\la^2)^{N_F/3\nd}
\ll\mcs$\,.\\
{\bf c)}\, The lightest are $N^2_F$ pions with the masses $\sim m_Q$.\\
{\bf d)}\, Besides, at $\la<\mph<\mph^{(S,{\,\rm conf})}$, there are two additional generations of $N^2_F$ fions with the pole masses (5.14). The fions are effectively massless and dynamically relevant in the range of scales $\mu^{\rm pole}_3(\Phi)<\mu<\mu^{\rm pole}_2(\Phi)$. At $\mph>\mph^{(S,{\,\rm conf})}$ these additional poles in the fion propagator are absent and fions are dynamically irrelevant at $\mu<\la$.

Finally, all condensates and observable masses evolve to those in the standard SQCD at $\mph>\mo$, see section 3 in \cite{ch1}.

\section{Dual theory. Unbroken flavor symmetry}

\hspace{3mm} The Lagrangian of the dual theory \cite{S1,S2} at the scale $\mu=\la$ is given in (1.6).
From (1.6), the running mass of mions at $\mu\sim \la$ (and not too large $\mph$, see below) is $\sim \la^2/\mph\ll \la$ and only decreases at lower energies, so that mions are effectively massless at least in the interval of scales $\mu_H<\mu<\la$, where $\mu_H$ is the largest physical mass. Therefore, the regime is conformal at $\mu_H<\mu<\la$
and, with $|\Lambda_q|=\la$, both direct and dual theories enter the conformal regime simultaneously at $\mu<\la$.\\

\hspace {5mm} {\bf 6.1\,\,\, L - vacua}\\

The running mass $\mu_{q,L}\equiv \mu_{q,L}(\mu=\la)$ of dual quarks ${\ov q},\,q$ and their pole mass in these $(2N_c-N_F)$ dual L - vacua look as (we recall that $\langle M\rangle=\langle\Pi\rangle$ in all vacua, see section 1
and (3.3)\,)
\bq
\frac{\mu_{q,L}}{\la}=\frac{\langle M\rangle_L}{\la^2}\sim\Bigl (\frac{\la}{\mph}\Bigr )^{\frac{\nd}{2N_c-N_F}}\,\,,\,\, \frac{\mu^{\rm pole}_{q,L}}{\la}=\Bigl (\frac{\mu_{q,L}}{\la}\Bigr )^{N_F/3\nd}\sim\Bigl (\frac{\la}{\mph}\Bigr )^{\frac{N_F}{3(2N_c-N_F)}}\,,\,\, \la\ll\mph\ll\mo\,,
\eq
\bq
\frac{\mu_{q,L}}{\la}=\frac{\langle M\rangle_L}{\la^2}\sim\Bigl (\frac{m_Q}{\la}\Bigr )^{\frac{\nd}{N_c}}\,,\quad \frac{\mu^{\rm pole}_{q,L}}{\la}=\Bigl (\frac{\mu_{q,L}}{\la}\Bigr )^{N_F/3\nd}\sim\Bigl (\frac{m_Q}{\la}\Bigr )^{\frac{N_F}{3N_c}}\,,\quad \mph\gg\mo.\nonumber
\eq

The value of the dual quark constituent mass can be found from the Konishi anomaly (1.8)
\bq
\frac{(\dl)^2}{\la^2}=\frac{\langle N(\mu=\la)\rangle_L=\langle{\ov q}q(\mu=\la)\rangle_L}{\la^2}=\frac{\langle{ S}\rangle_L}{\langle M\rangle_L\la}\sim\Bigl (\frac{\la}{\mph}\Bigr )^{\frac{N_c}{(2N_c-N_F)}}\,,
\,\,\la\ll\mph\ll\mo\,,
\eq
\bq
\frac{(\dl)^2}{\la^2}=\frac{\langle N(\mu=\la)\rangle_L=\langle{\ov q}q(\mu=\la)\rangle_L}{\la^2}=
\frac{m_Q}{\la}\,,\quad \mph\gg \mo\,.\nonumber
\eq
Therefore,
\bq
\frac{\mu^{\rm pole}_{q,L}}{\dl}\sim\Bigl (\frac{\la}{\mph}\Bigr )^{\frac{2N_F-3N_c}{6(2N_c-N_F)}}\ll 1\,,\quad
\la\ll\mph\ll\mo\,,\quad \frac{3}{2}N_c<N_F<2N_c\,,
\eq
and this shows that the dual quarks ${\ov q},\,q$ are in the DC phase where they acquire the large constituent mass
$\dl$ and $N_F^2$ lighter dual pions $N^i_j$ (nions) are formed, see section 5 in \cite{ch1}. Hence, the regime is conformal at $\mu_H=\dl<\mu<\la$\,, while the heavy constituent quarks decouple and the RG flow of mion fields becomes frozen at $\mu<\dl$. Therefore, proceeding as in \cite{ch1} and integrating out the dual constituent quarks as heavy ones at $\mu<\dl$ and then the dual gluons at $\mu<\lym^{(L)}$\,, one obtains the Lagrangian of mions and nions (all fields are always normalized at $\mu=\la,\,\,{\rm \bd}=3\nd-N_F=2N_F-3N_c$\,)
\bq
K=\Biggl [ \frac{z^{(L)}_M}{\la^2}\,{\rm Tr\,\Bigl ( M^{\dagger} M\Bigr )}\,+ \rm {Tr\,\sqrt {N^\dagger N}}\Biggr ]\,,
\quad {\it z^{(L)}_M}=\Bigl (\frac{\la}{\dl}\Bigr )^{2\bd/N_F}\gg 1\,,\quad \la\ll\mph\ll\mo\,,
\eq
\bq
W=\Biggl [\frac{1}{\la}{\rm Tr \Bigl (- M N \Bigr )}+ N_c\Bigl ( \frac{\det\,{\rm N}}{\Lambda_{\rm Q}^{\rm\bd} }\Bigr)
^{1/N_c}+m_Q\rm {Tr\, M}-\frac{1}{2\mph}\Bigl ({\rm Tr}\, (M^2)- \frac{1}{N_c}({\rm Tr}\, M)^2\Bigr ) \Biggr ]\,. \nonumber
\eq

The main contribution to the mion mass at $\la\ll\mph\ll \mo$ originates from the term $\sim M^2/\mph$ in (6.4)
\bq
{\mu}_L(M)\sim \frac{\la^2}{\mph {\it z^{(L)}_M}}\sim \la\Bigl (\frac{\la}{\mph} \Bigr )^{\frac{\nd\bo}{N_F(2N_c-N_F)}}\,,
\quad \frac{{\mu}_L(M)}{\dl}\sim\Bigl (\frac{\la}{\mph}\Bigr )^{{\rm \bd}/2N_F}\ll 1\,.
\eq

The running mion mass $\mu_{M}(\mu)$ at the scale $\mu\sim \la$ is $\mu_{M}(\mu\sim\la)\sim\la^2/\mph\ll\la$, i.e. {\it it is effectively massless and so dynamically relevant}. With decreasing scale $\mu_{M}(\mu)$ decreases but more slowly than the scale $\mu$ itself, because $\gamma_M= - (2{\rm\bd}/N_F),\,\,|\gamma_M|<1$ at $3/2<N_F/N_c<2$\,. Hence, if nothing prevents, the mion becomes too heavy and {\it dynamically irrelevant} at scales (compare with (4.1)\,)
\bq
\mu_{M}(\mu)=\frac{\la^2}{\mph z_M(\mu)}=\frac{\la^2}{\mph}\Bigl (\frac{\mu}{\la}\Bigr )^{2\bd/N_F}>\mu\,,\,\,{\rm i.\,\,e.\,\, at}\,\, \mu<{\ov\mu}^{\,\rm conf}_o=\mu^{\rm conf}_o=\la\Bigl (\frac{\la}{\mph}\Bigr )^{\frac{N_F}{3(2N_c-N_F)}}\,,
\eq
\bq
\frac{\mu^{\rm conf}_o}{{\ov\mu}^{L}_C}\sim \Bigl (\frac{\la}{\mph}\Bigr )^{\frac{\bd}{6(2N_c-N_F)}}\ll 1\,,
\quad\frac{\mu^{\rm conf}_o}{\mu_{L}(M)}\sim \Bigl (\frac{\la}{\mph}\Bigr )^{\frac{(2N_F-3N_c)^2}{3N_F(2N_c-N_F)}}\ll 1\,,
\quad \la<\mph<\mo\,.\nonumber
\eq

Therefore, the hierarchy of masses at $\la<\mph<\mo$ looks as\,: ${\ov\mu}^L_C>{\mu}_L(M)>\mu^{\rm conf}_o$. Now, at scales $\mu<{\ov\mu}^L_C$ the mass ${\mu}_L(M)>\mu^{\rm conf}_o$ does not run any more. Hence, {\it there is no pole} in the mion propagator at $p\sim \mu^{\rm conf}_o$ but {\it there is one} at $p\sim {\mu}_L(M)\gg\mu^{\rm conf}_o$, so that ${\mu}_L(M)$ in (6.5) is the only pole mass of mions in this range of $\mph$ in these dual L - vacua.

The mions $M$ are much heavier than nions $N$ at $\la\ll\mph\ll\mo$\,. Hence, they can be integrated out in (6.4) and one obtains
\bq
K=\rm {Tr\,\sqrt {N^\dagger N}}\,,\quad W=N_c\Bigl ( \frac{\det\,{\rm N}}{\Lambda_{\rm Q}^{\rm\bd} }\Bigr)
^{1/N_c}+\frac{\mph}{2}\Bigl ({\rm Tr}\,({\widehat N}^2)-\frac{1}{\nd}({\rm Tr}\, {\widehat N})^2\Bigr ),\,\, {\widehat N}=(m_Q-\frac{N}{\la})\,.
\eq
From (6.7), the nion mass looks in these dual L - vacua as
\bq
{\mu}_L(N)\sim \mph\langle N \rangle_L \sim\left\{\begin{array}{l l}
\la\Bigl (\la/\mph\Bigr)^{\frac{\nd}{(2N_c-N_F)}} &   \,{\rm at}\quad \la\ll \mph\ll \mo
\\
\vspace*{1mm}
\\
 m_Q\mph/\la &   \, {\rm at}\quad \mo\ll\mph\ll {\ov\mu}^{\,\rm dSQCD}_{\Phi}
\end{array}\right.
\eq

At $\mph\gtrsim\mo\sim \la(\la/m_Q)^{(2N_c-N_F)/N_c}$ the term $m_Q{\rm Tr}M$ cannot be neglected in (6.4) any more and the vacuum values $\langle M\rangle_L$ and $\langle N\rangle_L$ match those in the dual SQCD (\,dSQCD\,)\,:
$\langle M\rangle_{dSQCD}\sim \la^2(m_Q/\la)^{\nd/N_c}$\,,\,\, $\langle N \rangle_{dSQCD}\sim m_Q\la$, and $z_M^{(L)}\sim z_M^{\rm dSQCD}\sim (\la/m_Q)^{\bd/N_F}$. But the mion and nion masses are not matched yet\,: ${\mu}^{\,\rm pole}_{L} (M)\gg {\mu}_{\rm dSQCD}(M)$\,, while ${\mu}_{L} (N)\ll {\mu}_{\rm dSQCD}(N)$ at $\mph\sim\mo$. From (6.4), with increasing $\mph>\mo$\,, the increasing mass ${\mu}_{L} (N)$  and decreasing mass ${\mu}^{\,\rm pole}_{L} (M)$ both match those in the dSQCD at $\mph\sim{\ov\mu}^{\,\rm dSQCD}_{\Phi}$ and stay then at their dSQCD values at $\mph>{\ov\mu}^{\,\rm dSQCD}_{\Phi}$\,, see section 5 in \cite{ch1} (and the hierarchies $\mu^{\rm conf}_o/{\ov\mu}^{L}_C\ll 1$ and $\mu^{\rm conf}_o/\mu^{\rm pole}_{L}(M)\ll 1$ are maintained at $\mph>\mo$)\,, where
\bq
{\mu}_L (N)\sim \frac{m_Q{\ov\mu}^{\,\rm dSQCD}_{\Phi}}{\la}\sim {\mu}_{\,\rm dSQCD}(N)\sim \la\Bigl (\frac{m_Q}{\la}\Bigr )^{\frac{3\nd}{2N_F}}\quad\ra\quad {\ov\mu}^{\,\rm dSQCD}_{\Phi}\sim \la \Bigl(\frac{\la}{m_Q}\Bigr)^{\frac{\bo}{2N_F}},
\eq
\bq
{\mu}^{\,\rm pole}_L (M)\sim \frac{\la^2}{{\ov\mu}^{\,\rm dSQCD}_{\Phi}}\Biggl [\frac{1}{z_M^{(L)}}\sim \Bigl (\frac{m_Q}{\la}\Bigr )^{\frac{\rm\bd}{N_F}}\Biggr ]\sim {\mu}_{\,\rm dSQCD}(M)\sim \la\Bigl (\frac{m_Q}{\la}\Bigr )^{\frac{3\nd}{2N_F}}\,\ra\,{\ov\mu}^{\,\rm dSQCD}_{\Phi}\sim \la \Bigl(\frac{\la}{m_Q}\Bigr )_{.}^{\frac{\bo}{2N_F}}\nonumber
\eq

On the whole, the mass spectrum in these $(2N_c-N_F)$ L - vacua of dSQCD look as follows.\\
{\bf a)}\, There is a large number of heaviest flavored hadrons, mesons and baryons, made of non-relativistic and weakly confined (the string tesion is $\sqrt\sigma\sim\lym^{(L)}\ll \dl$) constituent dual quarks with the masses\,: $\dl\sim\la(\la/\mph)^{N_c/2(2N_c-N_F)}$ at $\la\ll \mph\ll\mo$ and $\dl\sim (m_Q\la)^{1/2}$ at $\mph\gg\mo$.\\
{\bf b)}\, A large number of gluonia with the mass scale\,: $\lym^{(L)}=\langle S\rangle^{1/3}_L\sim\la(\la/\mph)
^{N_F/3(2N_c-N_F)}$ at $\la\ll \mph\ll\mo$, and $\lym^{(L)}\sim (\la^{\bo} m_Q^{N_F})^{1/3N_c}\sim \lym^{(\rm SQCD)}$ at
$\mph\gg\mo$.\\
{\bf c)}\, $N_F^2$ mions $M$ with the pole masses\,: $\mu^{\rm pole}_L(M)\sim\la(\la/\mph)^{\nd\bo/N_F(2N_c-N_F)}\ll\dl$ at $\la\ll \mph\ll{\ov\mu}^{\,\rm dSQCD}_{\Phi}$\,, and $\mu^{\rm pole}_L(M)\sim \la\Bigl (m_Q/\la\Bigr )^{3\nd/2N_F}\sim \mu^{\,\rm pole}_{\rm dSQCD}(M)\ll\dl$ at $\mph\gg{\ov\mu}^{\,\rm dSQCD}_{\Phi}\gg\mo$.\\
{\bf d)}\, $N_F^2$ nions $N$ with the masses (6.8) at $\la\ll\mph\ll{\ov\mu}^{\,\rm dSQCD}_{\Phi}$, and $\mu_L(N)\sim
\la\Bigl (m_Q/\la\Bigr )^{3\nd/2N_F}\\ \sim \mu_{\,\rm dSQCD}(N)$ at $\mph\gg{\ov\mu}^{\,\rm dSQCD}_{\Phi}$.

Therefore, the masses of constituent quarks and gluonia match those in dSQCD at $\mph>\mo$, while the masses of mions and nions match those in dSQCD at $\mph>{\ov\mu}^{\,\rm dSQCD}_{\Phi}\gg\mo$ only. The mions are dynamically relevant in the range of scales ${\mu}^{\,\rm pole}_L (M)<\mu<\la$ in these L-vacua at all values $\mph\gg \la$.\\

\hspace{1cm} {\bf 6.2\,\,\, S - vacua}\\

Proceeding as in the previous section 6.1, the running mass $\mu_{q,S}\equiv \mu_{q,S}(\mu=\la)$ of dual quarks ${\ov q},\,q$ and their pole mass in these $(N_F-N_c)$ dual S-vacua are
\bq
\mu_{q,S}=\frac{\langle M\rangle_S}{\la}\sim\frac{m_Q\mph}{\la}\,\,,\quad \frac{\mu^{\rm pole}_{q,S}}{\la}=\Bigl (\frac{\mu_{q,S}}{\la}\Bigr )^{N_F/3\nd}\sim\Bigl (\frac{m_Q\mph}{\la^2}\Bigr )^{N_F/3\nd}\,,\,\, \la\ll\mph\ll\mo\,.
\eq
\bq
\frac{\mu_{q,S}}{\la}=\frac{\langle M\rangle_S}{\la^2}\sim\Bigl (\frac{m_Q}{\la}\Bigr )^{\frac{\nd}{N_c}}\,,\quad \frac{\mu^{\rm pole}_{q,S}}{\la}=\Bigl (\frac{\mu_{q,S}}{\la}\Bigr )^{N_F/3\nd}\sim\Bigl (\frac{m_Q}{\la}\Bigr )^{N_F/3N_c}\,,\quad \mph\gg\mo.\nonumber
\eq
The value of the dual quark constituent mass can be found from the Konishi anomaly (1.8)
\bq
\frac{(\ds)^2}{\la^2}=\frac{\langle N\rangle_S=\langle{\ov q}q(\mu=\la)\rangle_S}{\la^2}=\frac{\langle S\rangle_S}{\langle M\rangle_S\la}\sim\Bigl (\frac{m_Q\mph}{\la^2}\Bigr )^{N_c/\nd}\,,\,\,\la\ll\mph\ll\mo\,,
\eq
\bq
\frac{(\ds)^2}{\la^2}=\frac{\langle N\rangle_S=\langle{\ov q}q(\mu=\la)\rangle_S}{\la^2}=
\frac{m_Q}{\la}\,,\quad \mph\gg \mo\,.\nonumber
\eq
Therefore,
\bq
\frac{\mu^{\rm pole}_{q,S}}{\ds}\sim\Bigl (\frac{m_Q\mph}{\la^2}\Bigr )^{(2N_F-3N_c)/6\nd}\ll 1\,,
\quad \la\ll\mph\ll\mo\,,\quad\frac{3}{2}N_c<N_F<2N_c\,,
\eq
and this shows that dual quarks ${\ov q},\,q$ are here also in the DC phase where they acquire the constituent mass
$\ds$ and $N_F^2$ lighter dual pions $N^i_j$ (nions) are formed, see  section 5 in \cite{ch1}.

The regime is conformal at $\ds<\mu<\la$, while the constituent dual quarks decouple at $\mu<\ds$ and the RG evolution of the mion renormalization factor $z_M^{(S)}(\mu)$ becomes frozen, $z_M^{(S)}=z_M^{(S)}(\mu<\ds)=z_M^{(S)}(\mu=\ds)=
(\la/\ds)^{2\bd/N_F}$. Hence, after integrating out dual constituent quarks at $\mu<\ds$ and dual gluons at $\mu<\lym^{(S)}$ one obtains the Lagrangian (6.4) with a replacement $z_M^{(L)}\ra z_M^{(S)}$\,.

As for the low energy value of the mion mass in these $(N_F-N_c)$ dual S-vacua, it looks at $\la\ll\mph\ll\mo$ as
\bq
{\mu}_S(M)\sim \frac{\la^2}{\mph z_M^{(S)}}\sim \frac{\la^2}{\mph}\Bigl (\frac{m_Q\mph}{\la^2}\Bigr )^{N_c\bd/N_F\nd}\,.
\eq

But comparing the quark constituent mass $\ds$ with $\mu^{\rm conf}_o$ from (6.6) one obtains (compare with (5.10)\,)
\bq
\frac{\ds}{\mu^{\rm conf}_o}\gg 1\quad \ra\quad \mph\gg {\ov\mu}^{\,(S,{\,\rm conf})}_{\Phi}\sim\la\Bigl (\frac{\la}{m_Q}\Bigr )^{\frac{3(2N_c-N_F)}{N_F\bd+2N_c\bo}}\gg \mu_{\Phi}^{(S,{\,\rm conf})}\,,
\eq
while ${\mu^{\rm conf}_o}\gg\ds$ at $\la\ll\mph\ll{\ov\mu}^{\,(S,{\,\rm conf})}_{\Phi}\ll\mo$.

Hence, the hierarchy of scales and masses looks here as $\la>\ds>\mu_S(M)>\mu^{\rm conf}_o$ at $\mph>{\ov\mu}^{\,(S,{\,\rm conf})}_{\Phi}$ only. This means that only in this region, as in the preceding section 6.1, the mass $\mu_S(M)$ in (6.13) is {\it the mion pole mass} and the mions are dynamically relevant at scales $\mu^{\rm pole}_S(M)<\mu<\la$.

The hierarchy of scales and masses looks as $\mu^{\rm conf}_o>\mu_S(M)>\ds$ at $\la\ll\mph\ll {\ov\mu}^{\,(S,{\,\rm conf})}_{\Phi}$. Therefore, there {\it is no pole} in the mion propagator at this frozen low energy value $p\sim\mu_S(M)$ of the mion mass term. But, because the running mion mass $\mu^{(S)}_M(\mu)$ behaves in this range of $\mph$ as
\bq
\mu^{(S)}_M(\mu)\sim \frac{\la^2}{\mph z^{(S)}_M(\mu)}\sim \frac{\la^2}{\mph}\Bigl (\frac{\mu}{\la}\Bigr )^{2\bd/N_F}
\eq
in the vicinity of $\mu^{\rm conf}_o$, it is $\mu^{(S)}_M(\mu)<\mu^{\rm conf}_o$ at $\mu>\mu^{\rm conf}_o$ and $\mu^{(S)}
_M(\mu)>\mu^{\rm conf}_o$ at $\mu<\mu^{\rm conf}_o$, so that {\it there is a pole} in the mion propagator at $p\sim \mu^{\rm conf}_o$ and, besides, the mions are effectively massless and dynamically relevant in the range of scales $\mu^{\rm pole}_S(M)=\mu^{\rm conf}_o<\mu<\la$.

In any case, the mions $M$ are much heavier than nions $N$ at $\la\ll\mph\ll{\ov\mu}^{\,\rm dSQCD}_{\Phi}$\,. Hence, they can be integrated out and one obtains (6.7). From this, the mass of nions looks in these dual S - vacua as
\bq
{\mu}_S(N)\sim\frac{m_Q\mph}{\la}\quad {\rm at}\quad \la\ll\mph\ll{\ov\mu}^{\,\rm dSQCD}_{\Phi}\,.
\eq

On the whole, the mass spectrum in these $(N_F-N_c)$  S - vacua of dSQCD looks as follows.\\
{\bf a)}\, There is a large number of flavored hadrons, mesons and baryons, made of non-relativistic and weakly confined (the string tesion is $\sqrt\sigma\sim\lym^{(S)}\ll \ds$) constituent dual quarks with the masses $\ds\sim\la
(m_Q\mph/\la^2)^{N_c/2\nd}$ at $\la\ll \mph\ll\mo$.\\
{\bf b)}\, A large number of gluonia with the mass scale\,: $\lym^{(S)}=\langle S\rangle^{1/3}_S\sim\la(m_Q\mph/\la^2)
^{N_F/3\nd}$ at $\la\ll \mph\ll\mo$ and $\lym^{(S)}\sim (\la^{\bo} m_Q^{N_F})^{1/3N_c}\sim\lym^{(\rm SQCD)}$ at
$\mph\gg\mo$.\\
{\bf c)}\, $N_F^2$ mions $M$ with the pole masses\,: $\mu^{\rm pole}_S(M)\sim \mu^{\rm conf}_o\sim\la\Bigl (\la/\mph\Bigr )^{N_F/3(2N_c-N_F)}$ at $\la\ll\mph\ll{\ov\mu}^{\,(S,{\,\rm conf})}_{\Phi}\ll\mo$ and (6.13) at ${\ov\mu}^{\,(S,{\,\rm conf})}_{\Phi}\ll\mph\ll\mo$.\\
{\bf d)}\, $N_F^2$ dual pions $N$ (nions) with the masses (6.16).

All condensates and mass spectra in these S - vacua and in the L - vacua from the preceding section 6.1 degenerate and become the same at $\mph\gg\mo\sim \la(\la/m_Q)^{(2N_c-N_F)/N_c}$.

\section{Direct theory. Broken flavor symmetry.\\
\hspace*{2cm} The region $\mathbf {\la<\mph<\mo}$}

\hspace{5mm} {\bf 7.1}\,\,\,{\bf L - type vacua}\\

In these vacua
\footnote{\,
It will be implied everywhere below in the text that the numbers $n_1$ and $n_2$ are such that $1-(n_1/N_c),\, 1-(n_2/N_c)$ and $1-(N_F/2N_c)$ are all $O(1)$. The only exception considered explicitly below will be the special vacua with $n_2=N_c,\,n_1=\nd$.
}
and in this range of $\mph$, the term $m_Q{\ov Q}Q$ in the superpotential can be neglected and the parametric behavior is the same as in L - vacua in section 5.1. Therefore, all quarks are in the DC phase. The difference with previous calculations in section 5 in this region is that the flavor symmetry is broken spontaneously in a large number of these vacua with $\langle\Pi_1\rangle\neq\langle\Pi_2\rangle\,,\,\,\langle\Pi_1\rangle=\langle{\ov Q}_1Q_1\rangle\sim\langle
\Pi_2\rangle=\langle{\ov Q}_2 Q_2\rangle\gg (m_Q\mph)$, while the classical S - type vacua are absent.

Hence, proceeding as before and integrating out first all heaviest constituent quarks with the masses $\mu_{C,1}=\langle
\Pi_1\rangle^{1/2}\sim\mu_{C,2}=\langle\Pi_2\rangle^{1/2}\sim\langle\Pi\rangle_L^{1/2}\sim\la(\la/\mph )^{\nd/2(2N_c-N_F)}\gg (m_Q\mph)^{1/2}$ and then gluons at $\mu<\lym^{\rm (br)}\sim\lym^{(L)}$, one obtains the Lagrangian of $N_F^2$ pions
\bq
K={\rm Tr}\,\sqrt {\Pi^{\dagger}\Pi}\,\,,\quad
W= -\nd\,S+m_Q{\rm Tr}\,\Pi -\frac{1}{2\mph}\Biggl [{\rm Tr}\, {(\Pi}^2)- \frac{1}{N_c}({\rm Tr}\, \Pi)^2
\Biggr ]\,,\nonumber
\eq
\bq
S=\Biggl (\frac{\det \Pi}{\la^{\bo}}\Biggr )^{1/\nd},\quad \Bigl (\lym^{\rm (br)}\Bigr )^3=\langle S\rangle_{\rm br}=\Biggl (\frac{\langle\Pi_1\rangle^{n_1}\langle\Pi_2\rangle^{n_2}}{\la^{\bo}}\Biggr )^{1/\nd}
=\frac{\langle\Pi_1\rangle\langle\Pi_2\rangle}{\mph}\,,
\eq
\bq
\langle\Pi_1\rangle+\langle\Pi_2\rangle=m_Q\mph+\frac{1}{N_c}{\rm Tr}\,\langle\Pi\rangle\simeq\frac{1}{N_c}
\Biggl (n_1\langle\Pi_1\rangle+n_2\langle\Pi_2\rangle\Biggr ).\nonumber
\eq
From (7.1)
\bq
m_Q\mph\ll\langle\Pi_1\rangle_{Lt}\sim\langle\Pi_2\rangle_{Lt}\sim \la^2\Biggl (\frac{\la}{\mph}\Biggr )^{\frac{\nd}{2N_c-N_F}}\ll \la^2\,,\quad \la\ll\mph\ll\mo\sim \la\Bigl (\frac{\la}{m_Q}\Bigr )^{\frac{2N_c-N_F}{N_c}}\,.
\eq

As a check of self-consistency, we have now instead of (5.2)
\bq
\langle m^{\rm tot}_{Q,1}\rangle=\frac{\langle\Pi_2\rangle}{\mph}\,,\quad \langle m^{\rm tot}_{Q,2}\rangle=\frac
{\langle\Pi_1\rangle}{\mph}\sim \langle m^{\rm tot}_{Q,1}\rangle\,,
\eq
so that (5.4) is parametrically the same.

On the whole, the main qualitative difference in the mass spectra in comparison with the L - vacua $\langle\Pi\rangle_L$
with unbroken flavor symmetry in section 5.1 is that the hybrid pions $\Pi_{12}$ and $\Pi_{21}$ are  Nambu-Goldstone particles here and are exactly massless.\\

\hspace{5mm} {\bf 7.2\,\,\,  br2 - vacua}\\

In these vacua with $n_2>N_c,\, n_1<\nd<N_c$\,, see (3.9),
\bq
\langle\Pi_2\rangle_{\rm br2}\simeq \Bigl (\rho_2=-\frac{N_c}{n_2-N_c}\Bigr )m_Q\mph\,,\quad \langle\Pi_1\rangle_{\rm br2}\sim \la^2\Bigl (\frac{\mph}{\la}\Bigr )^{\frac{n_2}{n_2-N_c}}\Bigl (\frac{m_Q}{\la}\Bigr )^{\frac{N_c-n_1}{n_2-N_c}}
\,,
\eq
\bq
\frac{\langle\Pi_1\rangle_{\rm br2}}{\langle\Pi_2\rangle_{\rm br2}}\sim \Bigl (\frac{\mph}{\mo}\Bigr )^{\frac{N_c}{n_2-N_c}}\ll 1\,,\quad \la\ll\mph\ll\mo\sim\la\Bigl (\frac{\la}{m_Q}\Bigr )^{\frac{2N_c-N_F}{N_c}}\,.\nonumber
\eq
To see what is the phase in this case we look at hierarchies of possible masses. The pole masses of quarks $Q_1$
and $Q_2$ look as
\bq
\langle m^{\rm tot}_{Q,1}\rangle=\frac{\langle\Pi_2\rangle}{\mph}\sim m_Q\,,\quad \frac{\langle m^{\rm tot}_{Q,2}\rangle}{\langle m^{\rm tot}_{Q,1}\rangle}=\frac{\langle\Pi_1\rangle}{\langle\Pi_2\rangle}\ll 1\,,
\eq
\bq
\quad m_{Q,1}^{\rm pole}\sim \frac{\langle m^{\rm tot}_{Q,1}\rangle}{z_Q(\la, m^{\rm pole}_{Q,1})}\sim
\la\Bigl (\frac{m_Q}{\la}\Bigr )^{\frac{N_F}{3N_c}}\gg m_{Q,2}^{\rm pole}\,,\quad z_Q(\la,\mu\ll\la)=\Bigl (\frac{\mu}
{\la}\Bigr )^{\bo/N_F}\,,\nonumber
\eq
while their constituent masses look as
\bq
\mu_{C,2}\sim \langle\Pi_2\rangle^{1/2}\sim (m_Q\mph)^{1/2}\gg \mu_{C,1}\sim \langle\Pi_1\rangle^{1/2}\,.
\eq
The masses of gluons due to possible higgsing of quarks look as
\bq
\mu_{gl,2}\sim \Bigl [\langle\Pi_2\rangle z_Q(\la,\mu_{gl,2})\Bigr ]^{1/2}\sim \la\Bigl (\frac{\langle\Pi_2\rangle}{\la}\Bigr )^{N_F/3\nd}\gg \mu_{gl,1}\,.
\eq
Because from (7.5),(7.6)
\bq
\frac{m_{Q,1}^{\rm pole}}{\mu_{C,2}}\sim \Bigl (\frac{\la}{\mph}\Bigr )^{\frac{1}{2}}\Bigl (\frac{m_Q}{\la}\Bigr )^{\frac{2N_F-3N_c}{6N_c}}\ll 1\quad  {\rm and}\quad \frac{\mu_{gl,2}}{m_{Q,1}^{\rm pole}}\sim\Bigl (\frac{\mph}{\mo}\Bigr )^{N_F/3\nd}\ll 1\,,
\eq
this shows that the quarks ${\ov Q}_2, Q_2$ are in the $DC_2$ - phase in the whole region $\la\ll\mph\ll\mo$.
But, as for the quarks ${\ov Q}_1, Q_1$,
\bq
\frac{\mu_{C,1}}{m_{Q,1}^{\rm pole}}\sim\Bigl (\frac{\mph}{\la}\Bigr )^{\frac{n_2}{2(n_2-N_c)}}\Bigl (\frac{m_Q}{\la}\Bigr )^{\frac{\delta}{6N_c(n_2-N_c)}}\,,\,\,   \delta=[\,n_1\bd -(2 N_F\nd-3N_c^2)\,]\,,\,\, \bd=2N_F-3N_c\,.
\eq
It follows from (7.9) that \\
a1)\,\, if $(2N_F\nd-3N_c^2)<0$, then $\delta>0$ and
\bq
\frac{\mu_{C,1}}{ m_{Q,1}^{\rm pole}}\gg 1\quad\ra\quad\mph\gg {\widehat\mu}_{\Phi}\,,
\quad {\widehat\mu}_{\Phi}=\Bigl (\frac{\la}{m_Q}\Bigr )^{\frac{\delta>0}{3n_2N_c}}\ll\mo\,,
\eq
so that at all $0<n_1<\nd$ the overall phase is $DC_1-DC_2$ at ${\widehat\mu}_{\Phi}\ll\mph\ll\mo$ and $HQ_1-DC_2$ at $\la\ll\mph\ll{\widehat\mu}_{\Phi}$\,;\\
a2)\,\,if $(2N_F\nd-3N_c^2)>0$, then $\mu_{C,1}\gg m_{Q,1}^{\rm pole}$ at $\delta<0$, i.e. the overall phase is $DC_1-DC_2$ at $0<n_1<n^o_1=(2N_F\nd-3N_c^2)/\bd$ only, but in the whole region $\la\ll\mph\ll\mo$\,;\\
a3)\,\,$\delta>0$ at $(2N_F\nd-3N_c^2)>0$, so that the phases is as in 'a1' above but now at $n^o_1<n_1<\nd$ only.\\

We start with the $DC_1-DC_2$ phase and recall that the largest constituent mass $\mu_{C,2}$ of ${\ov Q}_2, Q_2$ quarks is formed in this phase not at the scale $\mu\sim\mu_{C,2}$, but both constituent masses $\mu_{C,2}$ and $\mu_{C,1}$ are formed at the smaller scale $\mu\sim\mu_{C,1}$ \cite{ch2}. Therefore, the RG flow of quarks and gluons is conformal down to $\mu\sim\mu_{C,1}\ll\mu_{C,2}$. , integrating out all constituent quarks at $\mu<\mu_{C,1}$ and then all gluons at $\mu<\lym^{(\rm br2)}\sim (m_Q\langle \Pi_1\rangle)^{1/3}$, the lower energy Lagrangian looks as
\bq
K={\rm Tr}\,\Bigl [\,z_{\Phi}(\la,\mu_{C,1})(\Phi^{\dagger}\Phi)+\sqrt{\Pi^\dagger\Pi}\,\,\Bigr ],\quad z_{\Phi}=
\Bigl (\frac{\la}{\mu_{C,1}}\Bigr )^{2\bo/N_F}=\Bigl (\frac{\la^2}{\langle\Pi_1\rangle}\Bigr )^{\bo/N_F}\gg 1\,,\nonumber
\eq
\bq
W=-\nd S+\frac{\mph}{2}\Biggl [{\rm Tr}\,(\Phi^2) -\frac{1}{\nd}\Bigl ({\rm Tr}\,\Phi\Bigr)^2\Biggr ]+{\rm Tr}\,(m_Q-\Phi)\Pi\,,\quad S=\Bigl (\frac{\det\Pi}{\la^{\bo}}\Bigr )^{1/\nd}\,,
\eq
and one has to choose the br2 - vacua in (7.11).

To see whether fions are relevant or not in this $DC_1-DC_2$ phase, we compare $\mu_{C,1}$ and $\mu_o^{\rm conf}$, see (4.1),
\bq
\frac{\mu_o^{\rm conf}}{\mu_{C,1}}\sim \Bigl (\frac{\la}{\mph}\Bigr )^{\frac{N_F}{3(2N_c-N_F)}}\frac{\la}{\langle\Pi_1
\rangle^{1/2}}>1\quad\ra\quad \la\ll\mph<\mph^{\rm (relev)}=\la\Bigl (\frac{\la}{m_Q}\Bigr )^{\rho}\ll\mo\,,
\eq
\bq
\rho=\frac{3(N_c-n_1)(2N_c-N_F)}{2N_F(n_2-N_c)+3n_2(2N_c-N_F)}>0\,.\nonumber
\eq
Therefore, the fions are dynamically relevant in this $DC_1-DC_2$ phase at $\la\ll\mph<\mph^{\rm (relev)}$ and become irrelevant at $\mph>\mph^{\rm (relev)}$. Hence, at $\la\ll\mph\ll\mph^{\rm (relev)}$, there is the second generation of
all $N_F^2$ fions with the pole masses $\mu_{2}^{\rm pole}(\Phi_{ij})=\mu_o^{\rm conf}\gg\mu_{C,1}$ .\\

From (7.11), the mass terms of hybrids $\Phi_{12}, \Phi_{21}$ and $\Pi_{12}, \Pi_{21}$ look as in (2.23), but instead of (2.24) one has now
\bq
m_{\phi}=\frac{\mph}{z_{\Phi}(\la,\mu_{C,1})}\,,\quad m_{\pi}=\frac{\langle\Pi_1+\Pi_2\rangle}{\mph}\,,\quad m^2_{\phi\pi}= m_{\phi}m_{\pi}\,.
\eq

The exact equality  $m^2_{\phi\pi}= m_{\phi}m_{\pi}$ ensures that one of the two eigenvalues is zero. As one can check, the mixing $\phi_{12}\leftrightarrow\pi_{12}$ is parametrically small, so that the massless particles are mainly pions $\pi_{12},\pi_{21}$, while the heavy hybrids are mainly $\phi_{12},\phi_{21}$,
\bq
\mu_3^{\rm pole}(\Phi_{12})=\mu_3^{\rm pole}(\Phi_{21})\sim \frac{\mph}{z_{\Phi}(\la,\mu_{C,1})}\sim \mph
\Bigl (\frac{\langle\Pi_1\rangle}{\la^2}\Bigr )^{\bo/N_F}\,.
\eq
Further, the mixings $\phi_{11}\leftrightarrow\pi_{11}$ and $\phi_{22}\leftrightarrow\pi_{22}$ are also parametrically small and fions are much heavier than pions, with their third generation pole masses
\bq
\mu_3^{\rm pole}(\Phi_{11})\simeq\mu_3^{\rm pole}(\Phi_{22})\sim \frac{\mph}{z_{\Phi}(\la,\mu_{C,1})}\sim \mph
\Bigl (\frac{\langle\Pi_1\rangle}{\la^2}\Bigr )^{\bo/N_F}\,.
\eq
Hence, after integrating all fions one obtains the superpotential of pions
\bq
W= -\nd\,S+m_Q{\rm Tr}\,\Pi -\frac{1}{2\mph}\Biggl [{\rm Tr}\, {(\Pi}^2)- \frac{1}{N_c}({\rm Tr}\, \Pi)^2
\Biggr ]\,,
\eq
and one has to choose br2 - vacua in (7.16). Then one obtains for the pion masses
\bq
\mu(\Pi_{11})\simeq\mu(\Pi_{22})\sim \frac{\langle\Pi_2\rangle}{\mph}\sim m_Q\,.
\eq

On the whole for the mass spectrum in this $DC_1-DC_2$ phase.\\
1)\,The heaviest are 22-flavored hadrons made of the constituent quarks ${\ov Q}_2, Q_2$ with the masses $\mu_{C,2}\sim (m_Q\mph)^{1/2}$.\\
2)\, The next mass scale is due to 11-flavored hadrons made of the constituent ${\ov Q}_1, Q_1$ quarks with the masses $\mu_{C,1}\sim\langle\Pi_1\rangle^{1/2}\ll\mu_{C,2}$.\\
3)\, The gluonia with the mass scale $\lym^{\rm (br2)}\sim ( m_Q\langle\Pi_1\rangle)^{1/3}\ll \mu_{C,1}$.\\
4)\, $n_1^2$ pions $\Pi_{11}$ and $n_2^2$ pions $\Pi_{22}$ with the masses $\sim m_Q\ll\lym^{\rm (br2)}$.\\
5)\, $2 n_1 n_2$ massless hybrids $\Pi_{12},\Pi_{21}$.

And finally, as for the fions.\\
a1)\, At $(2N_F\nd-3N_c^2)<0,\,n_1>0,\, \delta>0$ and when $\mph$ is in the interval ${\widehat\mu}_{\Phi}\ll\mph\ll\mu^{\rm (relev)}_{\Phi}$, see (7.10),(7.12), all $N_F^2$ fions appear at scales $\mu<\la$ in the two generations with the pole masses $\mu_{2}^{\rm pole}(\Phi_{ij})\sim\mu_o^{\rm conf}\gg\mu_{C,1}$ and $\mu_{3}^{\rm pole}(\Phi_{ij})\ll\mu_{C,1}$, see (7.14),(7.15). They are dynamically relevant then in the range of scales $\mu_{3}^{\rm pole}(\Phi_{ij})<\mu<\mu_{2}^{\rm pole}(\Phi_{ij})$. But there are no poles in the fion propagators at $\mu<\la$ and they become dynamically irrelevant at $\mu^{\rm (relev)}_{\Phi}\ll\mu_{\Phi}\ll\mo$.\\
a2)\, At $(2N_F\nd-3N_c^2)>0,\,n_1<n^o_1,\,\delta<0$ - the same as in 'a1' above but the fions are relevant now in a much wider interval $\la\ll\mph\ll\mu^{\rm (relev)}_{\Phi}$.\\
a3)\, At $(2N_F\nd-3N_c^2)>0,\, \delta>0$ - the same as in 'a1' above but at $ n_1>n^o_1$ only.\\

We consider now the $HQ_1-DC_2$ phase and proceed in this case similarly to that in section 4 of \cite{ch2} where the standard SQCD in this phase was considered. The difference is due to fions and their Yukawa interactions with quarks. We recall also that the conformal regime is maintained in this phase in the range of scales $m^{\rm pole}_{Q,1}<\mu<\la$. After integrating out the constituent quarks ${\ov Q}_2, Q_2$ and the quarks ${\ov Q}_1, Q_1$ as heavy ones at $\mu<m^{\rm pole}_{Q,1}$, the Lagrangian takes the form
\bq
K=\Bigl [\,z_{\Phi}(\la, m^{\rm pole}_{Q,1}){\rm Tr}\,(\Phi^{\dagger}\Phi)+{\rm Tr} \,\sqrt{\Pi^{\dagger}_{22}\Pi_{22}}\,\,\Bigr ]\,,\quad
W=\Bigl [\,-\frac{2\pi}{\alpha(\mu)}S\,\Bigr ]+W_{\Phi}+W_{\Pi}\,,
\eq
\bq
W_{\Phi}=\frac{\mph}{2}\Biggl [{\rm Tr}\, (\Phi^2) -\frac{1}{\nd}\Bigl ({\rm Tr}\,\Phi\Bigr)^2\Biggr ],\,\,
W_{\Pi}=-n_2\Bigl (\frac{\det\Pi_{22}}{\la^{\bo}\det m^{\rm tot}_{Q,1}}\Bigr )^{\frac{1}{n_2-N_c}}+{\rm Tr}\,\Pi_{22}\Bigl (m^{\rm tot}_{Q,2}-\Phi_{21}\frac{1}{m^{\rm tot}_{Q,1}}\Phi_{12}\Bigr ),\nonumber
\eq
\bq
z_{\Phi}(\la, m^{\rm pole}_{Q,1})=\Bigl (\frac{\la}{m^{\rm pole}_{Q,1}}\Bigr )^{2\bo/N_F}\,,\quad
m^{\rm tot}_{Q,1}=(m_Q-\Phi_{11})\,,\quad m^{\rm tot}_{Q,2}=(m_Q-\Phi_{22})\,,\nonumber
\eq
with pions $\Pi_{22}$ and $m^{\rm tot}_{Q,1}$ sitting on $\lym^{(\rm br2)}$ inside $\alpha(\mu)$ in (7.18). The pions $\Pi_{22}$ and all fions are frozen and do not evolve any more at $\mu<m^{\rm pole}_{Q,1}$. Further, after integrating out gluons at $\mu<\lym^{(\rm br2)}$ through the VY - procedure \cite{VY}, the superpotential looks as
\bq
W=W_{\Phi}-(n_2-N_c)\Bigl (\frac{\det\Pi_{22}}{\la^{\bo}\det m^{\rm tot}_{Q,1}}\Bigr )^{\frac{1}{n_2-N_c}}+{\rm Tr}\,\Pi_{22}\Bigl (m^{\rm tot}_{Q,2}-\Phi_{21}\frac{1}{m^{\rm tot}_{Q,1}}\Phi_{12}\Bigr )\,.
\eq

One obtains from the above for the masses of fions and pions in this $HQ_1-DC_2$ phase.\\
1)\, Because $m^{\rm pole}_{Q,1}\ll\mu_o^{\rm conf}$ in the region $\la\ll\mph\ll {\widehat\mu}_{\Phi}\ll\mo$
with this phase, there is the second generation of all $N_F^2$ fions with the pole masses $\mu_2^{\rm pole}(\Phi_{ij})\sim\mu_o^{\rm conf}$.\\
2)\, There is the third  generation of $\Phi_{11}$ and $\Phi_{22}$ fions with the pole masses (the mixing of $\Phi_{22}$ and $\Pi_{22}$ is parametrically small)
\bq
\mu_3^{\rm pole}(\Phi_{11})\sim \mu_3^{\rm pole}(\Phi_{22})\sim \frac{\mph}{z_{\Phi}(\la,m^{\rm pole}_{Q,1})}\sim
\mph\Bigl (\frac{m_Q}{\la}\Bigr )^{2\bo/3N_c}\ll m^{\rm pole}_{Q,1}\sim\la\Bigl (\frac{m_Q}{\la}\Bigr )^{N_F/3N_c}\,,
\eq
and so the fions $\Phi_{11}$ and $\Phi_{22}$ are dynamically relevant in the range of scales $\mu_3^{\rm pole}(\Phi_{11})<\mu<\mu_o^{\rm conf}$.\\
3)\, The mass of $\Pi_{22}$ pions is $\mu(\Pi_{22})\sim m_Q\ll \mu_3^{\rm pole}(\Phi_{22})$.\\
4) The third generation hybrids are massless, $\mu_3^{\rm pole}(\Phi_{12})=\mu_3^{\rm pole}(\Phi_{21})=0$.

In addition, there are in a mass spectrum\,: the heaviest 22-flavored hadrons made of the constituent quarks ${\ov Q}_2, Q_2$ with the masses $\mu_{C,2}\sim (m_Q\mph)^{1/2}$, the next mass scale is due to 11-flavored hadrons made of ${\ov Q}_1, Q_1$ quarks with the masses $m^{\rm pole}_{Q,1}\ll\mu_{C,2}$ and, finally, there are gluonia with the mass scale $\sim\lym^{\rm (br2)}\sim (m_Q\langle\Pi_1\rangle)^{1/3}\ll m^{\rm pole}_{Q,1}$.\\

\hspace{5mm} {\bf 7.3}\,\,\,{\bf Special vacua, $\mathbf{n_2=N_c,\, n_1=\nd}$ }\\

In these vacua at $\la\ll\mph\ll\mo$, see (3.7),(3.10),
\bq
\langle\Pi_1\rangle_{\rm spec}=\frac{N_c}{2N_c-N_F}(m_Q\mph)\,,\,\, \langle\Pi_2\rangle_{\rm spec}=\la^2\Bigl (\frac{\la}{\mph}\Bigr )^{\frac{\nd}{2N_c-N_F}}\,,\,\, \frac{\langle\Pi_1\rangle_{\rm spec}}{\langle\Pi_2\rangle_{\rm spec}}\sim\Bigl (\frac{\mph}{\mo}\Bigr )^{\frac{N_c}{2N_c-N_F}}\ll 1
\eq

The most important possible masses look here as follows
\bq
\mu_{C,2}\sim\langle\Pi_2\rangle^{1/2}_{\rm spec}\,,\quad \mu_{C,1}\sim\langle\Pi_1\rangle^{1/2}_{\rm spec}\sim (m_Q\mph)^{1/2}\ll\mu_{C,2}\,,
\eq
\bq
\langle m^{\rm tot}_{Q,1}\rangle=\frac{\langle\Pi_2\rangle_{\rm spec}}{\mph}\sim\la\Bigl (\frac{\la}{\mph}\Bigr )^{\frac{N_c}{2N_c-N_F}}\,\,\ra\,\, m^{\rm pole}_{Q,1}\sim\la\Bigl (\frac{\la}{\mph}\Bigr )^{\frac{N_F}{3(2N_c-N_F)}}
\gg m^{\rm pole}_{Q,2}\,,\nonumber
\eq
\bq
\mu^2_{\rm gl,2}\sim (a_*\sim 1)\langle\Pi_2\rangle_{\rm spec}\Bigl (\frac{\mu_{\rm gl,2}}{\la}\Bigr )^{\frac{\bo}{N_F}}
\,\,\ra\,\, \mu_{\rm gl,2}\sim \la\Bigl (\frac{\la}{\mph}\Bigr )^{\frac{N_F}{3(2N_c-N_F)}}\sim  m^{\rm pole}_{Q,1}\gg\mu_{\rm gl,1}\,,\nonumber
\eq
\bq
\frac{m^{\rm pole}_{Q,1}}{\mu_{C,2}}\sim\Bigl (\frac{\la}{\mph}\Bigr )^{\frac{\bo}{6(2N_c-N_F)}}\ll 1\,,\nonumber
\eq
\bq
\Bigl (\frac{\mu_{C,1}}{m^{\rm pole}_{Q,1}}\Bigr )^2\sim\frac{m_Q}{\la}\Bigl (\frac{\mph}{\la}\Bigr )^
{\frac{6N_c-N_F}{3(2N_c-N_F)}}> 1\quad\ra\quad \mph>\mu^{(\rm DC)}_{\Phi}\sim\Bigl (\frac{\la}{m_Q}\Bigr )^{\frac{3(2N_c-N_F)}{6N_c-N_F}}\ll\mo\,,
\eq
where $\mu_{C,2}$ is the possible constituent mass of ${\ov Q}_2, Q_2$ quarks and $\mu_{\rm gl,2}$ is the gluon mass due to their possible higgsing. Because $\mu_{\rm gl,2}\sim m^{\rm pole}_{Q,1}$ it is unclear beforehand whether the phase is $DC_2-HQ_1$ or $Higgs_2-HQ_1$. But an attempt to write the standard superpotential for the $DC_2-HQ_1$ phase shows that it will be singular at $n_2=N_c$ \cite{ch2} and so the phase $DC_2-HQ_1$ cannot be realized in these special vacua (at least in a standard way). We assume here that the overall phase is $HQ_1-Higgs_2$ and the whole gauge group is higgsed at $\la\ll\mu\ll\mu^{(\rm DC)}_{\Phi}$, while the phase will be $DC_1-DC_2$ at $\mu^{(\rm DC)}_{\Phi}\ll\mu\ll\mo$.

We start with the $HQ_1-Higgs_2$ phase. Supposing that $m^{\rm pole}_{Q,1}=(\rm several)\mu_{\rm gl,2}$ and integrating out first the quarks ${\ov Q}_1, Q_1$ as heavy ones at $\mu<m^{\rm pole}_{Q,1}$ and then all higgsed gluons and their superpartners at $\mu<\mu_{gl,2}$, the Lagrangian takes the form
\bq
K={\rm Tr}\,\Biggl [\, z_{\Phi}(\Phi^\dagger\Phi)+ z_Q\Biggl ( 2\sqrt {\Pi^\dagger_{22}\Pi_{22} }+B^{\dagger}_2 B_2+
{\ov B}^{\,\dagger}_2{\ov B}_2\, \Biggr )\,
\Biggr ]\,,
\eq
\bq
z_Q=z_Q(\la,m^{\rm pole}_{Q,1})=\Bigl (\frac{m^{\rm pole}_{Q,1}}{\la}\Bigr )^{\bo/N_F}\,,\,\, z_{\Phi}=z_{\Phi}(\la,m^{\rm pole}_{Q,1})=1/z^2_Q\,,\nonumber
\eq
\bq
W=W_{\rm non-pert}+W_{\Phi}+{\rm Tr}\,\Pi_{22}\Bigl (m^{\rm tot}_{Q,2}-\Phi_{21}\frac{1}{m^{\rm tot}_{Q,1}}\Phi_{12}
\Bigr ),\quad W_{\Phi}=\frac{\mph}{2}\Biggl [{\rm Tr}\, (\Phi^2) -\frac{1}{\nd}\Bigl ({\rm Tr}\,\Phi\Bigr)^2\Biggr ],
\nonumber
\eq
\bq
m^{\rm tot}_{Q,1}=m_Q-\Phi_{11}\,,\,\,  m^{\rm tot}_{Q,2}=m_Q-\Phi_{22},\nonumber
\eq
where for the non-perturbative term we use the form proposed in \cite{S1}
\bq
W_{\rm non-pert}=A\Bigl [\,1-\frac{\det\Pi_{22}}{\lambda^{2N_c}}+\frac{{\ov B}_2 B_2}{\lambda^2}\, \Bigr ]\,,\,\,
\langle A\rangle=\langle S\rangle\,,\,\, \lambda^2=\Bigl (\la^{\bo}\det m^{\rm tot}_{Q,1}\Bigr )^{\frac{1}{N_c}},\,\, \langle\lambda^2\rangle=\langle\Pi_2\rangle,
\eq
in which $A$ is the auxiliary field.

From (7.24),(7.25), the hybrids $\Phi_{12}, \Phi_{21}$ are massless, the baryons ${\ov B}_2,\, B_2$ are light
\bq
\mu(B_2)=\mu({\ov B}_2)\sim \frac{m_Q}{z_Q}\sim m_Q\Bigl (\frac{\mph}{\la}\Bigr )^{\frac{\bo}{3(2N_c-N_F)}}\ll \mu_{\rm gl,2}\,,
\eq
while all other masses are parametrically $\sim\mu_{\rm gl,2}\sim m^{\rm pole}_{Q,1}$ (the pion masses increased due to their mixing with the fions). Besides, in particular, because $\mu_o^{\rm conf}\sim m^{\rm pole}_{Q,1}$ in these special vacua, there is no warranty that these nonzero masses of fions $\Phi_{11}$ and $\Phi_{22}$ are the pole masses. Maybe yes, but maybe not (see section 4).

On the whole, there are three scales in the mass spectrum\,:\, the hybrid fions $\Phi_{12}, \Phi_{21}$ are massless, the baryons have small masses (7.26), while all other masses are $\mu_{\rm gl,2}\sim m^{\rm pole}_{Q,1}\sim\la(\la/\mph)^{N_F/3(2N_c-N_F)}$ in these special vacua at $\la\ll\mph\ll\mu^{(\rm DC)}_{\Phi}$.\\

Now, we consider the phase $DC_1-DC_2$ with $\mu_{C,2}\gg\mu_{C,1}\gg\mu_o^{\rm conf}\sim m^{\rm pole}_{Q,1}$ in these special vacua at $\mu^{(\rm DC)}_{\Phi}\ll\mph\ll\mo$. We can proceed then as in this phase in section 7.2 above and to start directly with (7.11). But just because $\mu_{C,1}\gg\mu_o^{\rm conf}$, there are no poles in the fion propagators at all scales $\mu<\la$ and all fions are too heavy and dynamically irrelevant here. Hence, after integrating them out in (7.11), one obtains the superpotential (7.16). From this, the masses of $\Pi_{11}$ and $\Pi_{22}$ pions are as in (7.17), while the hybrids $\Pi_{12}$ and $\Pi_{21}$ are massless. We recall finally that the masses of constituent quarks are here $\mu_{C,2}\sim \langle\Pi_2\rangle_{\rm spec}^{1/2}\gg\mu_{C,1}$ and $\mu_{C,1}\sim\langle\Pi_1\rangle^{1/2}_{\rm spec}\gg m^{\rm pole}_{Q,1}$, and the mass scale of gluonia is $\lym^{(\rm spec)}=[\langle\Pi_1\rangle_{\rm spec}\langle\Pi_2\rangle_{\rm spec}/\mph]^{1/3}\sim\la\,[(m_Q/\la)(\la/\mph)^{\nd/(2N_c-N_F)}]^{1/3}\ll\mu_{C,1}$.

\section{Direct theory. Broken flavor symmetry.\\
\hspace*{1.5cm} The region $\mathbf {\mo\ll\mph\ll\la^2/m_Q}$ }

\hspace{5mm} {\bf 8.1\,\,\,  br1 - vacua, $\mathbf{DC_1-DC_2}$ phase}\\

In all L - type, br2 and special vacua the theory enters the region $\mph\sim (\rm several)\mo$ with all quarks in the $DC_1-DC_2$ phase and $\langle\Pi_1\rangle\sim\langle\Pi_2\rangle\sim\la^2(m_Q/\la)^{\nd/N_c}$, see sections 3 and 7. But there appears then a large hierarchy between the quark condensates with increasing $\mph$ at $\mph\gg\mo$ in these br1 - vacua
\bq
\langle\Pi_1\rangle\simeq \Bigl (\rho_1=\frac{N_c}{N_c-n_1}\Bigr )\, m_Q\mph\gg\langle\Pi_2\rangle\sim \la^2\Bigl (\frac{\la}{\mph}\Bigr )^{\frac{n_1}{N_c-n_1}}\Bigl (\frac{m_Q}{\la}\Bigr )^{\frac{n_2-N_c}{N_c-n_1}}\,,
\eq
i.e. the constituent masses $\mu_{C,1}$ of ${\ov Q}_1, Q_1$ quarks become parametrically larger than $\mu_{C,2}$ of ${\ov Q}_2, Q_2$ quarks.

We recall now once more the important feature of the $DC$ phase \cite{ch2}\,: {\it the $DC_1$ phase cannot be formed separately}, i.e. the largest constituent mass $\mu_{C,1}=\langle\Pi_1\rangle^{1/2}\sim (m_Q\mph)^{1/2}$ of ${\ov Q}_1, Q_1$ quarks is not formed at the scale $\mu\sim\mu_{C,1}$\,, but only at the appropriate lower scale $\mu_{\rm lower}$ below which {\it all quark flavors become massive}. Therefore, if nothing prevents, this lower scale in the case considered may be either the constituent mass $\mu_{C,2}=\langle\Pi_2\rangle^{1/2}$ of ${\ov Q}_2, Q_2$ quarks or their pole mass $m_{Q,2}^{\rm pole}$ (and both are parametrically smaller now than $\mu_{C,1}$ at $\mph\gg\mo$)\,, $\mu_{\rm lower}={\rm max}(\,\mu_{C,2}\,,\,m_{Q,2}^{\rm pole}\,)$.

But there is another competing mass $\mu_{\rm gl,1}$\,, i.e. the mass of gluons due to possible higgsing of ${\ov Q}_1, Q_1$ quarks,
\bq
\mu^2_{\rm gl,1}\sim a(\mu=\mu_{\rm gl,1})\langle\Pi_1\rangle z_Q(\la,\mu_{\rm gl,1}),\quad z_Q(\la, \mu_{\rm gl,1})=\Bigl ( \frac{\mu_{\rm gl,1}}{\la}\Bigr )^{\bo/N_F}\,,
\eq
\bq
\frac{\mu_{\rm gl,1}}{\la}\sim \Bigl (\frac{\langle\Pi_1\rangle}{\la^2}\Bigr )^{\frac{1}{2-\gamma_Q}}\sim
\Bigl (\frac{m_Q\mph}{\la^2}\Bigr )^{N_F/3\nd}, \,\,
a(\mu)\equiv\frac{N_c\alpha(\mu)}{2\pi},\,\, a_{\rm conf}(\mu=\mu_{\rm gl,1}\ll \la)=a_*=O(1)<1\,.\nonumber
\eq

It is seen from
\bq
\frac{m_{Q,2}^{\rm pole}}{\la}\sim \Bigl (\frac{m_Q}{\la}\Bigr )^{N_F/3N_c}\,,\quad \frac{m_{Q,2}^{\rm pole}}{\mu_{\rm gl,1}}\sim \Bigl (\frac{\mo}{\mph}\Bigr )^{N_F/3\nd}\ll 1\,,
\eq
that $\mu_{\rm gl,1}\gg m_{Q,2}^{\rm pole}$ at $\mph\gg \mo$\,, but as for $\mu_{C,2}$\,, one obtains
\bq
\mu_{C,2}=\langle\Pi_2\rangle^{1/2}\,, \quad \frac{\mu_{\rm gl,1}}{\mu_{C,2}}>1 \quad {\rm at}\quad \mph>\mu^{\rm (higgs)}_{\Phi}\,, \quad \frac{\mo}
{\la}\ll\frac{\mu^{\rm (higgs)}_{\Phi}}{\la}\sim \Bigl (\frac{\la}{m_Q}\Bigr )^{\sigma}\ll \frac{\la}{m_Q}\,,
\eq
\bq
\sigma=\frac{\bo(N_c-n_1)+3\nd(2N_c-N_F)}{2N_F(N_c-n_1)+3n_1\nd}>0\,.
\eq

Therefore, when $\mu_{C,2}\gg\mu_{\rm gl,1}$ at $\mo\ll\mph\ll\mu^{\rm (higgs)}_{\Phi}$\,, the quarks still are in the $DC_1-DC_2$ phase where both dynamical constituent masses $\mu_{C,1}$ and $\mu_{C,2}$ are formed {\it simultaneously} in the threshold region $[(\rm several)\mu_{C,2}>\mu>\mu_{C,2}/(\rm several)]\gg\mu_{\rm gl,1}$\,. But at $\mph\gg\mu^{\rm (higgs)}_{\Phi}$, when $\mu_{\rm gl,1}\gg\mu_{C,2}$\,, the quarks ${\ov Q}_1, Q_1$ {\it are higgsed at the higher scale} $\mu\sim \mu_{\rm gl,1}\gg\mu_{C,2}$\,, before reaching the lower scale $\mu\sim\mu_{C,2}$ where the constituent masses $\mu_{C,1}$ and $\mu_{C,2}$ are formed (this variant was not considered in \cite{ch2}). Hence, the phase will be either $Higgs_1-DC_2$ or $Higgs_1-HQ_2$, depending on the phase of ${\oq}_2, {\sq}_2$ quarks with unhiggsed colors (it is worth recalling here that, unlike the non-perturbative mechanism of the $DC_1$ condensate formation, the higgsing at $\mu\sim\mu_{\rm gl,1}$ {\it operates separately}, independently of what is going on at lower scales with ${\oq}_2, {\sq}_2$ quarks with unhiggsed colors).

Hence, the phase $DC_1-DC_2$ is maintained in the region $\mo<\mph<\mu^{\rm (higgs)}_{\Phi}$ in these br1 - vacua
and only the hierarchy of vacuum condensates $\langle\Pi_1\rangle$ and $\langle\Pi_2\rangle$ is different here in comparison with the section 7.1 above, $\langle\Pi_1\rangle\simeq\rho_1 m_Q\mph\gg \langle\Pi_2\rangle$. Therefore,
the Lagrangian of pions will be (7.1), $n_1^2$ pions $\Pi_{11}=({\ov Q}_1 Q_1)$ and $n_2^2$ pions $\Pi_{22}=({\ov Q}_2 Q_2)$ will have masses $\mu(\Pi_{11})= \mu(\Pi_{22})\sim m_Q$, while $2n_1n_2$ hybrids $\Pi_{12}=({\ov Q}_1 Q_2)$ and $\Pi_{21}=({\ov Q}_2 Q_1)$ will be massless. The masses of gluonia are $\sim \lym^{\rm (br1)}=\langle S\rangle^{1/3}_{\rm br1}$, as in (2.18). The hierarchy of nonzero masses looks here as\,: $\mu(\Pi_{11})\sim\mu(\Pi_{22})\ll\lym^{\rm (br1)}\ll\mu_{C,2}\ll\mu_{C,1}\ll \la$.

Finally, it remains to see that fions are dynamically irrelevant here. The renormalization factor $z_{\Phi}(\la,\mu)$ of fions becomes frozen at $z_{\Phi}(\la,\mu_{C,2})=(\la^2/\langle\Pi_2\rangle)^{\bo/N_F}\gg 1$, after all constituent quarks decouple at $\mu< \mu_{C,2}$\,. Hence, the running mass of fions stops at the value $\mu_{\Phi}(\mu=\mu_{C,2})= \mph/z_{\Phi}(\la,\mu_{C,2})$. To see that they are irrelevant at $\la\ll\mph\ll\mu^{\rm (higgs)}_{\Phi}$ it is sufficient to check that $\mu_{C,2}>\mu^{\rm conf}_o$, see (4.1). This is fulfilled as $\mu_{C,2}>m_{Q,2}^{\rm pole}\sim\la(m_Q/\la)^{N_F/3N_c}$ here and $m_{Q,2}^{\rm pole}>\mu^{\rm conf}_o$ at $\mph>\mo$. Hence, there are no poles in the fion propagators at $\mu<\la$ and all fions remain dynamically irrelevant in this case.\\

\hspace{5mm} {\bf 8.2\,\,\,  br1 - vacua, $\mathbf{Higgs_1-DC_2}$ phase}\\

At $\mo\ll\mu^{\rm higgs}_{\Phi}<\mph\ll\la^2/m_Q$ the quarks ${\ov Q}_1, Q_1$ are higgsed at $\mu\sim\mu_{\rm gl,1}\ll\la$ in the conformal regime at $a_{+}(\mu=\mu_{\rm gl,1})\sim a_{-}(\mu=\mu_{\rm gl,1})=a_*=O(1)<1$. The lower energy theory at $\mu<\mu_{\rm gl,1}$ has $N^{\,\prime}_c=(N_c-n_1)$ colors, $N^{\,\prime}_F=(N_F-n_1)=n_2$ flavors, ${\rm b}^{\prime}_o=(\bo-2n_1)$, and the new scale factor of its gauge coupling is
\bq
\Bigl [\Lambda^{\prime}_Q(\Pi_{11})\Bigr ]^{{\rm b}^{\prime}_o}=\frac{z^{n_2}_Q(\la, \mu_{\rm gl,1})\la^{\bo}}{\det \Pi_{11}}\,,\quad \Lambda^{\prime}_Q=\langle\Lambda^{\prime}_Q(\Pi_{11})\rangle\sim\mu_{\rm gl,1}\,.
\eq

We consider first the case $n_1<\bo/2$ when the lower energy theory with $N^{\prime}_c=(N_c-n_1)$ colors and $\,N^{\prime}_F=(N_F-n_1)=n_2$ flavors and with unbroken flavor symmetry $U(n_2)$ remains in the conformal window with $3/2<N^{\prime}_F/N^{\prime}_c<3$ at $\mu<\mu_{\rm gl,1}$. Then the phase of ${\oq}_2, \sq_2$ quarks with unhiggsed colors is $DC_2$. Indeed, the new values of the pole and constituent masses of ${\oq}_2, {\sq}_2$ quarks with unhiggsed colors look now as (\,$\gamma^{+}_Q=\bo/N_F>0,\, \gamma^{-}_Q={\rm b}^{\prime}_o/N^{\prime}_F>0$\,)
\bq
m^{\rm pole}_{\sq,2}=\Bigl (\langle m^{\rm tot}_{\sq,2}\rangle=\frac{\langle\Pi_1\rangle}{\mph}\sim m_Q \Bigr )\Bigl (z^{+}_Q(\la,\mu_{\rm gl,1})z^{-}_Q(\mu_{\rm gl,1}, m^{\rm pole}_{\sq,2}\Bigr )^{-1}\,,\quad (\mu^{\,\prime}_{C,2})^2=\langle\Pi_2\rangle z^+_Q(\la,\mu_{\rm gl,1})\,,\nonumber
\eq
\bq
z^{+}_Q=\Bigl (\frac{\mu_{\rm gl,1}}{\la}\Bigr )^{\gamma^{+}_Q}\ll 1,\,\, z^{-}_Q =\Bigl (\frac{ m^{\rm pole}_{\sq,2}}{\mu_{\rm gl,1}}\Bigr )^{\gamma^{-}_Q}\ll 1\,,\quad
\frac{m^{\rm pole}_{\sq,2}}{\mu^{\,\prime}_{C,2}}\sim \Biggl (\frac{\mo}{\mph}\Biggr )^{\frac{N_c(\bo-2n_1)}{6\nd(N_c-n_1)}}\ll 1\,.
\eq

Therefore, after the heaviest particles with the masses $\sim\mu_{\rm gl,1}$ have been integrated out, the Lagrangian at $\mu\sim \mu_{\rm gl,1}\ll \la$ takes the form (2.21), with a replacement of the logarithmic renormalization factor $z_Q(\la,\mu_{\rm gl,1}\gg\la)\gg 1$ and $z_{\Phi}\sim 1$ by the power-like ones, $z^+_Q(\la,\mu_{\rm gl,1}\ll\la)\ll 1$ and $z^+_{\Phi}(\la,\mu_{\rm gl,1}\ll\la)\gg 1$. Then, after integrating out the constituent $\oq_2, \sq_2$ quarks with the masses $\mu^{\,\prime}_{C,2}$ at $\mu<\mu^{\,\prime}_{C,2}$ and unhiggsed gluons at $\mu<\lym^{\rm (br1)}$\,, the Lagrangian looks as
\bq
K=\Bigl [z^{+}_{\Phi}(\la,\mu_{\rm gl,1})K_{\Phi}+z^{+}_Q(\la,\mu^2_{\rm gl,1})K_{\Pi}\Bigr ],\quad z^{+}_Q=\Bigl (\frac{\mu_{\rm gl,1}}{\la}\Bigr )^{\gamma^{+}_Q}\ll 1,\quad z^{+}_{\Phi}=\Bigl (\frac{1}{z^{+}_Q}\Bigr )^2\gg 1\,,
\eq
\bq
K_{\Phi}={\rm Tr}\Bigl [\,\Bigl (\Phi_{11}^{\dagger}\Phi_{11}+\Phi_{12}^{\dagger}\Phi_{12}+\Phi_{21}^{\dagger}
\Phi_{21}\Bigr )+z^{-}_{\Phi}(\mu_{\rm gl,1},\mu^{\,\prime}_{C,2} )\Phi_{22}^{\dagger}\Phi_{22}\,\Bigr ],\quad z^{-}_{\Phi}=\Bigl (\frac{\mu_{\rm gl,1})}{\mu^{\,\prime}_{C,2}}\Bigr )^{2\gamma^{-}_Q}\gg 1\,,\nonumber
\eq
\bq
K_{\Pi}={\rm Tr}\,\Bigl [2\,\sqrt{\Pi^\dagger_{11}\Pi_{11}}+\Biggl(\Pi^{\dagger}_{12}\frac{1}{\sqrt{\Pi_{11}
\Pi^{\dagger}_{11}}}\Pi_{12}+\Pi_{21}\frac{1}{\sqrt{\Pi^{\dagger}_{11}\Pi_{11}}}\Pi^\dagger_{21}\Biggr )+
\sqrt{\Pi^\dagger_{22}\Pi_{22}}\,\,\Bigr ]\,.\nonumber
\eq
\bq
W=-\nd S+\frac{\mph}{2}\Biggl [{\rm Tr}\, (\Phi^2) -\frac{1}{\nd}\Bigl ({\rm Tr}\,\Phi\Bigr)^2\Biggr ]+W_{\Pi}\,,
\quad S=\Bigl (\frac{\det\Pi_{11}\det\Pi_{22}}{\la^{\bo}} \Bigr )^{1/\nd}\,,
\eq
\bq
W_{\Pi}= {\rm Tr}\Bigl [ (m_Q-\Phi_{11})\Pi_{11}+(m_Q-\Phi_{22})\Bigl (\Pi_{22}+\Pi_{21}\frac{1}{\Pi_{11}}\Pi_{12}
\Bigr )-\Bigl (\Phi_{12}\Pi_{21}+\Phi_{21}\Pi_{12} \Bigr )\,\Bigr ]\,.\nonumber
\eq

We starts with the hybrids. The fions $\Phi_{12}$ are much heavier than $\Pi_{12}$ and mixing between them is
parametrically small. Neglecting it and integrating out $\Phi_{12}$, one obtains for the mass terms of $\Pi_{12}$ and
$\Pi_{21}$, see (1.5),
\bq
W_{(\Pi)}^{\rm hybr}=\Bigl (\,\frac{m_Q-\langle\Phi_{2}\rangle=\langle m_{Q,2}^{\rm tot}\rangle}{\langle\Pi_{1}\rangle}
-\frac{1}{\mph}\,\Bigr )\Pi_{21}\Pi_{12}=0\,.
\eq

Further, the fions $\Phi_{11}$ and $\Phi_{22}$ are also much heavier than the pions $\Pi_{11},\Pi_{22}$. After integrating them out the superpotential of pions $\Pi_{11},\Pi_{22}$ looks as
\bq
W=-\nd S+m_Q{\rm Tr}\,\Pi-\frac{1}{\mph}\,\Bigl [\,{\rm Tr}\,(\Pi^2_{11}+\Pi^2_{22})-\frac{1}{N_c}\Bigl ({\rm Tr}\,\Pi \Bigr )^2\,\Bigr ].
\eq
Hence, from (8.8) and (8.11), the pion masses look as
\bq
\mu(\Pi_{11})\sim\frac{\langle\Pi_{1}\rangle}{z^+_Q\mph}\sim\frac{m_Q}{z^+_Q}\sim\mu(\Pi_{22})\,.
\eq

On the whole, the mass spectrum looks as follows.\\
a) The heaviest are $(2n_1N_c-n_1^2)$ higgsed gluons and their superpartners with the masses $\mu_{\rm gl,1}\sim(m_Q\mph)^{1/2}\ll\la$.\\
b) There is a large number of 22-flavored hadrons made of non-relativistic and weakly confined constituent quarks
${\oq}_2, {\sq}_2$ with the masses $\mu^{\,\prime}_{C,2}\ll\mu_{\rm gl,1}$ (the string tension is ${\sqrt\sigma}\sim \lym^{\rm (br1)}\ll\mu^{\,\prime}_{C,2}$).\\
c) $n^2_1$ pions $\Pi_{11}$ and $n^2_2$ pions $\Pi_{22}$ have masses $\mu(\Pi_{11})\sim\mu(\Pi_{22})\sim m_Q/z^+_Q\ll\lym^{\rm (br1)}$\,,
\bq
\Biggl (\frac{\mu(\Pi_{11})}{\lym^{\rm (br1)}}\Biggr )^3\sim\Biggl (\frac{\mu(\Pi_{11})}{\mu^{\,\prime}_{C,2}}\Biggr )^2\sim
\Biggl (\frac{\mo}{\mph}\Biggr )^{\frac{N_c(\bo-2n_1)}{\nd(N_c-n_1)}}\ll 1,\quad
\lym^{\rm (br1)}\sim \Bigl (m_Q\langle\Pi_2\rangle_{\rm br1}\Bigr )^{1/3}\,.
\eq
d) The $2n_1n_2$ hybrids $\Pi_{12}$ and $\Pi_{21}$ are  massless.
\footnote{\,
As one  can check, all $N_F^2$ fions remain dynamically irrelevant in this region $\mph>\mo \,.$
}
\\

\hspace{5mm} {\bf 8.3\,\,\,  br1 - vacua, $\mathbf{Higgs_1-HQ_2}$ phase}\\

We consider now the case $\rm b^\prime_o<0$\,, i.e. $N^\prime_F/N^\prime_c>3,\,\,n_1>\bo/2$\,. Then (neglecting logarithmic factors), one has to replace $z^{-}_Q\ra 1$ in (8.7) at $\mu^{\rm (higgs)}_{\Phi}<\mph\ll\la^2/m_Q$ and obtains
\bq
\frac{\mu^{\,\prime}_{C,2}}{m^{\rm pole}_{\sq,2}}\sim \Biggl (\frac{\mo}{\mph}\Biggr )^{\frac{N_c(2n_1-\bo)}{2\nd(N_c-n_1)}}\ll 1\,.
\eq
This shows that ${\oq}_2, {\sq}_2$ quarks with unhiggsed colors are in the $HQ_2$ - phase and there is no lighter $\Pi_{22}$ pions. The lagrangian of pions takes now the form (2.22). This phase $Higgs_1-HQ_2$ is preserved also in the region $\mph\gg\la^2/m_Q$.\\

Finally, we comment in short on the behavior in the region $\mph\gg \la^2/m_Q$ when $\rm b^\prime_o>0$. Then
$\mu_{\rm gl, 1}\sim\langle\Pi_1\rangle^{1/2}\sim (m_Q\mph)^{1/2}\gg \la$\,, i.e. the quarks ${\ov Q}_1, Q_1$ will be higgsed in the weak coupling region with the logarithmic RG flow, and (neglecting logarithmic factors) $(\Lambda^\prime_Q)^{\bo^\prime}\sim\la^\bo/\det \langle\Pi_{11}\rangle,\,\, \Lambda^{\prime}_Q\ll\la$ now. Hence, while $\Lambda^{\prime}_Q\sim\mu_{\rm gl,1}$ increased with increasing $\mph$ at $\mu_{\rm gl,1}\ll\la$, when $\mph$ becomes larger than $\la^2/m_Q$ and increases, $\Lambda^{\prime}_Q$ begins to decrease in a power-like fashion while the ratios $\mu^{\,\prime}_{C,2}/\Lambda^{\prime}_Q\,,\,\,m^{\rm pole}_{\sq, 2}/\Lambda^{\prime}_Q\,,\,\, m^{\rm pole}_{\sq, 2}/\mu^{\,\prime}_{C,2}$ {\it are increasing} with $\mph$. Until $\mph$ is not too large, $\mph<\mu^{\prime}_{\Phi}$\,, the hierarchy $\Lambda^{\prime}_Q>\mu^{\,\prime}_{C,2}> m^{\rm pole}_{\sq, 2}$ is preserved and ${\oq}_2, {\sq}_2$ quarks stay in the $DC_2$ phase. But $m^{\rm pole}_{\sq,2}\sim\mu^{\,\prime}_{C,2}\sim\Lambda^{\prime}_Q\sim m_Q$ at $\mph\sim\mu^
{\prime}_{\Phi}\sim\la(\la/m_Q)^{(\bo-n_1)/n_1}\gg\la^2/m_Q$, and the hierarchy is reversed at $\mph>\mu^{\prime}_{\Phi}$, it becomes $m^{\rm pole}_{\sq, 2}>\mu^{\,\prime}_{C,2}>\Lambda^{\prime}_Q$\,.
\footnote{\,
In this region $\mu^{\,\prime}_{C,2}\sim \mu_{\rm gl, 2}$ has the meaning of the gluon mass due to possible higgsing of ${\oq}_2, {\sq}_2$ quarks.
}
The phase is changed when $\mph$ becomes larger than $\mu^{\prime}_{\Phi},\, DC_2\ra HQ_2$, the ${\oq}_2, {\sq}_2$ quarks will be in the $HQ_2$ (heavy quark) phase and there will be no lighter $\Pi_{22}$ pions. The Lagrangian of pions $\Pi_{11}$ and hybrids $\Pi_{12},\,\Pi_{21}$ has the form (2.22). With further increasing $\mph$ this phase $Higgs_1-HQ_2$ stays intact.\\

\hspace{5mm} {\bf 8.4\,\,\,  br2 and special vacua}\\

At $n_2<N_c$ there are also $\rm br2$ - vacua, see section 3. For these, their properties can be obtained by the replacement $n_1\leftrightarrow n_2$ in formulas of the preceding sections 8.1 and 8.3\,. The only difference is that, because $n_2\geq N_F/2$ and so $2n_2>\bo$, there is no analog of the conformal regime at $\mu<\mu_{\rm gl,1}$ with $2n_1<\bo$ in section 8.2. I.e., see (8.5), at $\mph>{\mu}^{\rm higgs}_{\Phi}(n_1\leftrightarrow n_2)$ the lower energy theory at $\mu<\mu_{\rm gl,2}$ will be always in the perturbative IR free logarithmic regime and the overall phase will be $Higgs_2-HQ_1$.

As for the special vacua (see section 3), their properties can also be obtained with $n_1=\nd,\, n_2=N_c$
in formulas of the preceding sections 8.1-8.3\,.\\

\section{Dual theory. Broken flavor symmetry.\\
\hspace*{1.6cm} The region $\mathbf {\la<\mph<\mo}$}

\hspace{5mm} {\bf 9.1\,\,\,  L - type vacua}.\\

\hspace{3mm} Below in this section\,: $M_{11}$ is the $n_1\times n_1$ matrix, $\langle M_1\rangle=\langle M_{11}\rangle=\langle
\Pi_1\rangle$,\, $M_{22}$ and $N_{22}=({\ov q}_2 q_2)$ are $n_2\times n_2$ matrices, $\langle M_2\rangle=\langle M_{22}\rangle=\langle\Pi_2\rangle$\,, and $M_{\rm hybr}$ includes $2n_1\times n_2$\, $M_{12}$ and $M_{21}$ mesons (and the same for $N_F^2$ nions $N$\,).\\

Although $\langle M_1\rangle_{Lt}\neq\langle M_2\rangle_{Lt}$ are not equal now, but their parametric behavior in the region $\la\ll\mph\ll\mo$ is the same here as in the L - vacua with the unbroken flavor symmetry in section 6.1, $\langle M_1=\Pi_1\rangle_{Lt}\sim\langle M_2=\Pi_2
\rangle_{Lt}\sim \langle M\rangle_L$. I.e., the phase is $DC_1-DC_2,\, N_F^2$ nions $N_{ij}$ are formed and the Lagrangian is (6.4). All $N_F^2$ mions have masses (6.5) and are much heavier than nions. Hence, the Lagrangian of nions is (6.7). The masses of $n^2_1$ nions $N_{11}$ and $n^2_2$ nions $N_{22}$ are still $\mu(N_{11})\sim \mu(N_{22})\sim \la(\la/\mph)^{\nd/(2N_c-N_F)}$. But, due to a spontaneous breaking of the flavor symmetry, the masses of hybrid nions $N_{12}$ and $N_{21}$ differ qualitatively now from those in section 6.1. They are massless Nambu-Goldstone particles here.\\

\hspace{5mm} {\bf 9.2\,\,\,  br2 - vacua}\\

The condensates of mions and dual quarks in these vacua with $n_2>N_c\,, n_1<\nd$ at $\la\ll\mph\ll\mo$ look as
\bq
\langle M_2\rangle\simeq \Bigl (\rho_2=-\frac{n_2-N_c}{N_c}\Bigr )m_Q\mph,\,\,\, \langle M_1\rangle\sim \la^2
\Bigl(\frac{\mph}{\la}\Bigr )^{\frac{n_2}{n_2-N_c}}\Bigl (\frac{m_Q}{\la}\Bigr )^{\frac{N_c-n_1}{n_2-N_c}},\,\,\,
\frac{\langle M_1\rangle}{\langle M_2\rangle}\sim \Bigl (\frac{\mph}{\mo}\Bigr )^{\frac{N_c}{n_2-N_c}}\ll 1\,,\nonumber
\eq
\bq
\langle N_1\rangle=\langle{\ov q}_1 q_1(\mu=\la)\rangle=\frac{\la\langle S\rangle}{\langle M_1\rangle}
=\frac{\la\langle M_2\rangle}{\mph}\sim m_Q\la\gg\langle N_2\rangle\,.
\eq
From these, some possible characteristic masses look as
\bq
\mu_{q,2}=\frac{\langle M_2\rangle}{\la}\sim\frac{m_Q\mph}{\la},\quad \mu^{\rm pole}_{q,2}\sim\la\Bigl (\frac{m_Q\mph}{\la^2}\Bigr )^{N_F/3\nd}\gg\mu^{\rm pole}_{q,1}\,,
\eq
\bq
{\ov\mu}^2_{C,1}\sim \langle N_1\rangle=\frac{\la\langle S\rangle}{\langle M_1\rangle}=\frac{\la\langle M_2\rangle}{\mph}\sim m_Q\la\,,\quad \frac{\mu^{\rm pole}_{q,2}}{{\ov\mu}_{C,1}}\ll 1\,,
\nonumber
\eq
\bq
{\ov\mu}^2_{C,2}\sim \langle N_2\rangle=\frac{\la\langle S\rangle}{\langle M_2\rangle}=\frac{\la\langle M_1\rangle}{\mph}\sim\la^2\Bigl (\frac{\mph}{\la}\Bigr )^{\frac{N_c}{n_2-N_c}}
\Bigl (\frac{m_Q}{\la}\Bigr )^{\frac{N_c-n_1}{n_2-N_c}}\ll {\ov\mu}^2_{C,1}\,,\nonumber
\eq
\bq
{\ov\mu}_{\rm gl,1}\sim\la\Bigl (\frac{\langle N_1\rangle}{\la^2}\Bigr )^{N_F/3N_c}\sim\la\Bigl(\frac{m_Q}{\la}\Bigr )^{N_F/3N_c}\gg{\ov\mu}_{\rm gl,2}\,,\nonumber
\eq
\bq
\frac{{\ov\mu}_{\rm gl,1}}{\mu^{\rm pole}_{q,2}}\sim \Bigl(\frac{\mo}{\mph}\Bigr )^{N_F/3\nd}\gg 1\,,\quad
\frac{{\ov\mu}_{\rm gl,1}}{{\ov\mu}_{C,1}}\sim \Bigl(\frac{m_Q}{\la}\Bigr )^{\bd/6N_c}\ll 1\,,\quad
\bd=2N_F-3N_c\,,
\eq
\bq
\frac{{\ov\mu}_{\rm gl,1}}{{\ov\mu}_{C,2}}\sim\Bigl(\frac{\la}{\mph}\Bigr )^{\frac{N_c}{2(n_2-N_c)}}
\Bigl(\frac{\la}{m_Q}\Bigr )^{\frac{\delta}{6N_c(n_2-N_c)}}\,,\quad \delta=n_1\bd-(2 N_F\nd-3N_c^2)\,,
\eq
where $\mu^{\rm pole}_{q,1}$ and $\mu^{\rm pole}_{q,2}$ are the pole masses of ${\ov q}_1, q_1$ and ${\ov q}_2, q_2$ quarks, ${\ov\mu}_{C,1}$ and ${\ov\mu}_{C,2}$ are their constituent masses and ${\ov\mu}_{\rm gl,1},\, {\ov\mu}_{\rm gl,2}$ are the gluon masses due to possible higgsing of these quarks.

It is seen from (9.1)-(9.4) that the largest mass is ${\ov\mu}_{C,1}$. But we recall that it does not work by itself
\cite{ch2}, it is important what is the next mass. It follows from (9.1)-(9.4) that\,:\\
a1)\, if $(2 N_F\nd -3N_c^2)<0$, then $\delta>0$ and at all $0<n_1<\nd$ the phase is $DC_1-DC_2$ in the region ${\widehat{\ov\mu}}_{\Phi}\ll\mph\ll\mo$ and $Higgs_1-HQ_2$ in the region $\la\ll\mph\ll {\widehat{\ov\mu}}_{\Phi}$, see (7.10),
\bq
{\widehat{\ov\mu}}_{\Phi}=\la\Bigl (\frac{\la}{m_Q}\Bigr )^{\frac{\delta\,>0}{3N_c^2}}\gg {\widehat\mu}_{\Phi}\,;
\eq
a2)\,  if $(2 N_F\nd -3N_c^2)>0$ and $0<n_1<{\ov n}^o_1=n^o_1=(2 N_F\nd -3N_c^2)/\bd$, then $\delta<0$ and the next mass
is ${\ov\mu}_{C,2}$\,, this means that the overall phase is $DC_1-DC_2$ in the whole region $\la\ll\mph\ll\mo$\,;\\
a3)\, if $(2 N_F\nd -3N_c^2)>0$ and $\delta>0$, then the phases is as in 'a1' but at $n^o_1<n_1<\nd$ only.

Besides, when ${\ov q}_1, q_1$ quarks are higgsed, the lower energy theory at $\mu<{\ov\mu}_{\rm gl,1}$ remains in the conformal regime at $0<n_1<\bd/2$ in the case 'a2' and at $n^o_1<n_1<\bd/2$ in the case 'a3', and enters the IR  free regime at $n_1>\bd/2$ in the case 'a2' and at $n_1>{\rm max}\,[\, n^o_1,\,\bd/2\,]$ in the case 'a3'.\\

We consider first the $DC_1-DC_2$ phase. The quarks acquire the constituent masses ${\ov\mu}_{C,1}$ and ${\ov\mu}_{C,2}$ and $N_F^2$ dual pions (nions) are formed at the scale $\mu\sim{\ov\mu}_{C,2}\ll{\ov\mu}_{C,1}$. After integrating out all constituent quarks at $\mu<{\ov\mu}_{C,2}$ and then all gluons at $\mu<\lym^{(\rm br2)}$, the Lagrangian of nions $N$ and mions $M$ looks as in (6.4), with a replacement ${\ov\mu}^{L}_{C}\ra{\ov\mu}_{C,2}$. All mions are much heavier here than nions and mixing between them is small (and is neglected). Hence, after integrating out all mions one obtains (6.7) and the nion masses look as ( at $\la\ll\mph\ll\mo$ for 'a2' and ${\widehat{\ov\mu}}_{\Phi}\ll\mph\ll\mo$ for 'a1' and 'a3')
\bq
\mu(N_{11})\sim \mu(N_{22})\sim \frac{m_Q\mph}{\la}\,,\quad \mu(N_{12})=\mu(N_{21})=0\,.
\eq

As for the mion running masses, they all become frozen at $\mu<{\ov\mu}_{C,2}$ at the value
\bq
\mu(M_{ij})\sim \frac{\la^2}{\mph z_M(\la,{\ov\mu}_{C,2})}\gg\mu(N)\,,\quad z_M(\la,{\ov\mu}_{C,2})=\Bigl (\frac{\la}{{\ov\mu}_{C,2}}\Bigr )^{2\bd/N_F}\,.
\eq

To see whether mions are dynamically relevant or not in this $DC_1-DC_2$ phase, we compare ${\ov\mu}_{C,2}$ and $\mu_o^{\rm conf}$, see (9.2),(6.6),
\bq
\Bigl (\frac{\mu_o^{\rm conf}}{{\ov\mu}_{C,2}}\Bigr )^2\sim \Bigl (\frac{\la}{m_Q}\Bigr )^{\frac{N_c-n_1}{n_2-N_c}}
\Bigl (\frac{\mph}{\la}\Bigr )^{\ov\rho}\gg 1\,,\quad \ov\rho=\frac{2n_1 N_F-(2N_F^2+N_c N_F-6 N_c^2)}
{3(n_2-N_c)(2N_c-N_F)}>0\,.
\eq
It is seen from (9.8) that $\mu(M)$ in (9.7) are not the pole masses of mions, the hierarchies look here as $\mu_o^{\rm conf}\gg\mu(M)\gg{\ov\mu}_{C,2}$. {\it The mion propagators have poles at $p=\mu^{\rm pole}(M)\sim\mu_o^{\rm conf}$ only} and so the mions are dynamically relevant only at scales $\mu_o^{\rm conf}<\mu<\la$.

On the whole, in addition to the mions and nions, the mass spectrum in this $DC_1-DC_2$ phase includes\,:\, a)\, a large number of hadrons made of the weakly confined (the string tension is ${\sqrt\sigma}\sim\lym^{(\rm br2)}\ll{\ov\mu}_{C,2}
\ll{\ov\mu}_{C,1}$) constituent dual quarks with the masses ${\ov\mu}_{C,1}$ and ${\ov\mu}_{C,2}$,\,
b)\, a large number of gluonia made of $SU(\nd)$ dual gluons with the mass scale $\sim\lym^{(\rm br2)}=(\langle M_1\rangle\langle M_2\rangle/\mph)^{1/3}$.\\

We consider now the $Higgs_1-HQ_2$ phase. The largest physical mass here is ${\ov\mu}_{\rm gl,1}$, see (9.2). The lower energy theory at $\mu<{\ov\mu}_{\rm gl,1}$ has $\nd^{\,\prime}=\nd-n_1$ dual colors and $N_F^{\prime}=N_F-n_1=n_2>N_c$ lighter $\ov{\textsf{q}}_2$, $\textsf{q}_2$  quarks with unhiggsed colors. It is in the conformal regime at $2<N_F^{\prime}/\nd^{\,\prime}<3$, i.e. at $n_1<\bd/2$, and in the IR  free one at $N_F^{\prime}/\nd^{\,\prime}>3,\, n_1>\bd/2$.

We start with $2<N_F^{\prime}/\nd^{\,\prime}<3$. After integrating out all heaviest higgsed gluons and their superpartners at $\mu<{\ov\mu}_{\rm gl,1}$, the Lagrangian of remained lighter degrees of freedom looks as
\bq
K=z_M{\rm Tr}\,\Bigl [\frac{M^\dagger M}{\la^2}\Bigr ]+z_q  {\rm Tr}\,\Bigl [2\sqrt
{N^{\dagger}_{11}N_{11}}+K_{\rm hybr}+ \textsf{q}^{\dagger}_2\textsf{q}_2+(\textsf{q}_2\ra
\ov{\textsf{q}}_2)\,\Bigr ]\,,
\eq
\bq
K_{\rm hybr}=\Biggl (N^{\dagger}_{12}\frac{1}{\sqrt{N_{11} N^{\dagger}_{11}}}N_{12}+
N_{21}\frac{1}{\sqrt{N^{\dagger}_{11}N_{11}}}N^\dagger_{21}\Biggr ),\nonumber
\eq
\bq
z_q=z_q (\la,{\ov\mu}_{\rm gl,1})=\Bigl (\frac{{\ov\mu}_{\rm gl,1}}{\la}\Bigr )^{\bd/N_F}\,,\quad
z_M=z_M (\la,{\ov\mu}_{\rm gl,1})=1/z^2_q\,,\nonumber
\eq
\bq
W=\Biggl [-\frac{2\pi}{{\ov\alpha}(\mu)}\ov{\textsf{S}}\Biggr ]+W_M -W_{MN}-\frac{1}{\la}{\rm Tr}\Bigl (\ov
{\textsf{q}}_2 M_{22}\textsf{q}_2\Bigr )\,,
\eq
\bq
W_M=m_Q\rm {Tr\, M}-\frac{1}{2\mph}\Bigl ({\rm Tr}\, (M^2)- \frac{1}{N_c}({\rm Tr}\, M)^2\Bigr )\,,\nonumber
\eq
\bq
W_{MN}=\frac{1}{\la}{\rm Tr}\, \Bigl ( M_{11}N_{11}+N_{12}M_{21}+N_{21}M_{12}+M_{22}N_{21}\frac{1}{N_{11}}N_{12}
\Bigr )\,,\nonumber
\eq
where $N_{11}$ are the dual pions (nions) due to higgsing of ${\ov q}_1, q_1$ quarks (besides, they are sitting inside ${\ov\alpha}(\mu)$), $\ov{\textsf{S}}$ is the gauge field strength squared of unhiggsed dual gluons, $\ov{\textsf{q}}_2$ and $\textsf{q}_2$ are the ${\ov q}_2, q_2$ quarks with unhiggsed colors and $N_{12}, N_{21}$ are the hybrid nions (in essence, these are the ${\ov q}_2, q_2$ quarks with higgsed colors).

The quarks $\ov{\textsf{q}}_2, \textsf{q}_2$ are in the $HQ_2$ phase. After integrating them out at $\mu<{\ov\mu}^{\rm pole}_{q,2}$ and then unhiggsed gluons at $\lym^{(\rm br2)}$ the lower energy Lagrangian is
\bq
K=\frac{z_M}{\la^2}{\rm Tr}\,\Bigl [\,M^\dagger_{11} M_{11}+M^{\dagger}_{12}M_{12}+M^{\dagger}_{21}M_{21}+z^{\prime}_M M^{\dagger}_{22} M_{22}\Bigr ]+z_q  {\rm Tr}\,\Bigl [2\sqrt{N^{\dagger}_{11}N_{11}}+K_{\rm hybr}\Bigr ]\,,\nonumber
\eq
\bq
W=(\nd-n_1)\Biggl (\frac{\la^{\bd}\det ( M_{22}/\la)}{\det N_{11}}\Biggr )^{1/(\nd-n_1)}+W_M - W_{MN}\,,
\eq
\bq
{\ov\mu}_{\rm gl,1} \sim\la\Bigl(\frac{m_Q}{\la}\Bigr )^{N_F/3N_c}\,,
\quad z^{\prime}_M=z^{\prime}_M({\ov\mu}_{\rm gl,1},{{\ov\mu}}^{\rm pole}_{q,2})=\Bigl (\frac{{\ov\mu}_{\rm gl,1}}
{{\ov\mu}^{\rm pole}_{q,2}}\Bigr )^{2{\bd}^\prime/N_F^\prime}\,.\nonumber
\eq
The factor $z^{\prime}_M$ in (9.11) appears due to the additional evolution of $\ov{\textsf{q}}_2, \textsf{q}_2$ quarks
and $M_{22}$ mions in the range of scales ${\ov\mu}^{\rm pole}_{q,2}<\mu<{\ov\mu}_{\rm gl,1}$, while ${\ov\mu}^{\rm pole}_{q,2}$ is the pole mass of $\ov{\textsf{q}}_2, \textsf{q}_2$ quarks in this $Higgs_1-HQ_2$ phase
\bq
{\ov\mu}^{\rm pole}_{q,2}=\frac{\mu_{q,2}}{z_q z^\prime_q}\,,\quad z^\prime_q=z^\prime_q({\ov\mu}_{\rm gl,1},
{\ov\mu}^{\rm pole}_{q,2})=\Bigl (\frac{{\ov\mu}^{\rm pole}_{q,2}}{{\ov\mu}_{\rm gl,1}}\Bigr )^{{\bd}^\prime/N_F^
\prime}\ll 1\,.
\eq

From (9.11)\,: i)\, the low energy values of the mion running masses $M_{11}$ are much larger than those of nions $N_{11}$ and the mixing among them is small;\, ii)\, the same for the hybrids $M_{12}$ and $N_{12}$. Hence, one obtains from (9.11) for the masses
\bq
\mu(M_{11})\sim\mu(M_{12})\sim\mu(M_{21})\sim\frac{\la^2}{\mph z_M}\,\,,\quad \mu(M_{22})\sim \frac{\la^2}{\mph z_M z^\prime_M}\,\,,
\eq
\bq
\mu(N_{11})\sim \frac{\mph\langle N_1\rangle}{z_q\la^2}\sim\la\Bigl (\frac{m_Q\mph}{\la^2}\Bigr )^{2\bo/3N_c}\,,\quad
\mu(N_{12})=\mu(N_{21})=0\,.
\eq

To see whether (9.13) are the mion poles masses or not it is sufficient to check the hierarchies $\mu_o^{\rm conf}\gg
\mu(M_{11})\gg{\ov\mu}_{\rm gl,1},\,\,\mu(M_{11})\gg\mu(M_{22})\gg{\ov\mu}^{\rm pole}_{q,2}$. Therefore, the values of the running mion 'masses' in (9.13) are not their poles masses, these are simply the low energy limiting values of mass terms in their propagators.  The propagators of all mions have poles at $p=\mu^{\rm pole}(M)\sim\mu_o^{\rm conf}$ only and so they are dynamically relevant only in the range of scales $\mu_o^{\rm conf}<\mu<\la$.

On the whole, in addition to the mions and nions, the mass spectrum in this $Higgs_1-HQ_2$ phase includes\,:\, a)\, the heaviest $n_1(2\nd-n_1)$ higgsed dual gluons and the same number of their superpartners,\, b)\, a large number of hadrons made of the unhiggsed $\ov{\textsf{q}}_2, \textsf{q}_2$ quarks with the mass scale $\sim {\ov\mu}^{\rm pole}_{q,2}$ (9.12),\, c)\, a large number of gluonia made of the unhiggsed $SU(\nd-n_1)$ dual gluons with the mass scale $\sim\lym^{(\rm br2)}=(\langle M_1\rangle\langle M_2\rangle/\mph)^{1/3}$.\\

We consider now the case $N_F^{\prime}/\nd^{\,\prime}>3$ where the theory at $\mu<{\ov\mu}_{\rm gl,1}$ is in the
logarithmic IR  free regime at scales ${\ov\mu}^{\rm pole}_{q,2}<\mu<{\ov\mu}_{\rm gl,1}$. The Lagrangians will be as in (9.9)-(9.11) and only $z^\prime_M$ in (9.11) and ${\ov\mu}^{\rm pole}_{q,2}$ in (9.12) will be different. Neglecting logarithmic factors, one can replace $z^\prime_M\ra 1$ in (9.11) and $z^\prime_q\ra 1$ in (9.12). The nion masses will be
as in (9.14) while $\mu(M_{22})$ in (9.13) will be $\mu(M_{22})\sim\mu(M_{11})$.\\

\hspace{5mm} {\bf 9.3\,\,\,  Special  vacua,\, $\mathbf{n_1=\nd,\, n_2=N_c}$}\\

The most important possible masses look here as follows,
\bq
\langle M_1\rangle_{\rm spec}=\frac{N_c}{2N_c-N_F}(m_Q\mph)\,,\quad \langle M_2\rangle_{\rm spec}=\la^2\Bigl (\frac{\la}{\mph}\Bigr )^{\nd/(2N_c-N_F)}\gg\langle M_1\rangle_{\rm spec}\,,
\eq
\bq
\mu_{q,2}=\frac{\langle M_2\rangle}{\la}\,,\quad \mu^{\rm pole}_{q,2}\sim\la\Bigl (\frac{\langle M_2\rangle}{\la^2}\Bigr )^{N_F/3\nd}\sim\la\Bigl (\frac{\la}{\mph}\Bigr )^{N_F/3(2N_c-N_F)}\gg \mu^{\rm pole}_{q,1}\,,\nonumber
\eq
\bq
{\ov\mu}^2_{C,1}\sim \langle N_1\rangle=\frac{\langle M_2\rangle\la}{\mph}\sim \la^2\Bigl (\frac{\la}{\mph}\Bigr )^{\frac{N_c}{2N_c-N_F}}\gg {\ov\mu}^2_{C,2}\sim m_Q\la\,,\quad\frac{\mu^{\rm pole}_{q,2}}{{\ov\mu}_{C,1}}\sim \Bigl (\frac{\la}{\mph}\Bigr )^{\frac{\bd}{6(2N_c-N_F)}}\ll 1\,,\nonumber
\eq
\bq
{\ov\mu}_{\rm gl,1}\sim \la\Bigl (\frac{\langle N_1\rangle}{\la^2}\Bigr )^{N_F/3N_c}\sim \mu^{\rm pole}_{q,2}\gg {\ov\mu}_{\rm gl,2}\,,\nonumber
\eq
and, see (7.23),
\bq
\Bigl (\frac{{\ov\mu}_{C,2}}{\mu^{\rm pole}_{q,2}}\Bigr )^{2}\sim\frac{m_Q}{\la}\Bigl (\frac{\mph}{\la}\Bigr )^{\frac{2N_F}{3(2N_c-N_F)}}> 1\quad \ra\quad \mph>{\ov\mu}^{(\rm DC)}_{\Phi}\sim \la\Bigl (\frac{\la}{m_Q}\Bigr )^{\frac{3(2N_c-N_F)}{2N_F}}\gg \mu^{(\rm DC)}_{\Phi}\,,
\eq
where ${\ov\mu}_{C,1}$ is the possible constituent mass of ${\ov q}_1, q_1$ quarks and ${\ov\mu}_{\rm gl,1}$ is the gluon mass due to their possible higgsing. Because ${\ov\mu}_{\rm gl,1}\sim \mu^{\rm pole}_{q,2}$ it is unclear beforehand whether the phase is $DC_1-HQ_2$ or $Higgs_1-HQ_2$.
\footnote{\,
But taking $\bd/\nd\ll 1$ and using the results from \cite{ch3} we obtain $\mu^{\rm pole}_{q,2}/{\ov\mu}_{\rm gl,1}\sim
\exp \{3\nd/14\bd\}\gg 1$.
}
But an attempt to write the standard superpotential for the $DC_1-HQ_2$ phase shows that it will be singular at $n_1=\nd$ \cite{ch2} and, similarly to the special vacua in the direct theory in section 7.3, we assume here that
the overall phase will be $Higgs_1-HQ_2$ and the whole dual gauge group will be higgsed at $\la\ll\mu\ll{\ov\mu}^{(\rm DC)}_{\Phi}$, while the phase will be $DC_1-DC_2$ at ${\ov\mu}^{(\rm DC)}_{\Phi}\ll\mu\ll\mo$.

We start with the $Higgs_1-HQ_2$ phase and proceed as in the section 7.3. I.e., after integrating out first the quarks ${\ov q}_2, q_2$ as heavy ones at $\mu<\mu^{\rm pole}_{q,2}$ and then all higgsed dual gluons and their superpartners at $\mu<{\ov\mu}_{\rm gl,1}$, the Lagrangian takes the form
\bq
K={\rm Tr}\,\Biggl [\, z_M\frac{M^\dagger M}{\la^2}+ z_q\Biggl (\sqrt{N^\dagger_{11} N_{11}}+b^{\dagger}_1 b_1+
{\ov b}^{\,\dagger}_1 {\ov b}_1\Biggr )\,\Biggr ]\,, \nonumber
\eq
\bq
z_q=z_q(\la,\mu^{\rm pole}_{q,2})=\Bigl (\frac{\mu^{\rm pole}_{q,2}}{\la}\Bigr )^{\bd/N_F}\,,\quad z_M=z_M(\la,\mu^{\rm pole}_{q,2})=1/z^2_q\,,\nonumber
\eq
\bq
W=W_{\rm non-pert}-W_M -\frac{1}{\la}{\rm Tr}\, N_{11}\Bigl (M_{11}-M_{12}\frac{1}{M_{22}}M_{21}\Bigr )\,,
\eq
\bq
W_M=\frac{1}{2\mph}\Biggl [{\rm Tr} (M^2) -\frac{1}{N_c}\Bigl ({\rm Tr}\,M\Bigr)^2\Biggr ]+m_Q{\rm Tr}\,M\,,\nonumber
\eq
where the non-perturbative term looks here as
\bq
W_{\rm non-pert}={\ov A}\,\Bigl [\,1-\frac{\det N_{11}}{{\ov\lambda}^{2\nd}}+\frac{{\ov b}_1 b_1}{{\ov\lambda}^2}\,
\Bigr ]\,,\quad \langle \ov A\rangle=\langle S\rangle=\frac{\langle M_1\rangle\langle M_2\rangle}{\mph}\,,
\eq
\bq
{\ov\lambda}^2=\Biggl (\la^{\bd}\det \frac{M_{22}}{\la}\Biggr )^{1/\nd}\,,\quad \langle{\ov\lambda}^2\rangle=
\langle N_1\rangle=\langle m^{\rm tot}_{Q,1}\rangle\la=\frac{\langle M_2\rangle\la}{\mph}\,,\nonumber
\eq
and $\ov A$ is the auxiliary field.

From (9.17),(9.18)\,:\, the hybrids $M_{12}, M_{21}$ are massless, the baryons ${\ov b}_1,\, b_1$ are light
\bq
\mu(b_1)=\mu ({\ov b}_1)\sim \frac{\langle M_1\rangle}{z_q\la}\sim m_Q\Bigl (\frac{\mph}{\la}\Bigr )^{\frac{\bo}{3(2N_c-N_F)}}\ll {\ov\mu}_{\rm gl,1}\,,
\eq
while all other masses are $\sim{\ov\mu}_{\rm gl,1}\sim \mu^{\rm pole}_{q,2}$ (the nion masses increased due to their mixing with the mions). Besides, in particular, because $\mu_o^{\rm conf}\sim \mu^{\rm pole}_{q,2}$ in these special vacua, there is no warranty that these nonzero masses of mions $M_{11}$ and $M_{22}$ are the pole masses. Maybe so but maybe not (see section 4).

On the whole, there are three scales in the mass spectrum\,:\, the hybrid mions $M_{12}, M_{21}$ are massless, the baryon masses are (9.19), while all other masses are $\sim {\ov\mu}_{\rm gl,1}\sim\mu^{\rm pole}_{q,2}\sim\\
\sim\la(\la/\mph)^{N_F/3(2N_c-N_F)}$ in these special vacua at $\la\ll\mph\ll{\ov\mu}^{(\rm DC)}_{\Phi}$.\\

Now, we consider the phase $DC_1-DC_2$ with ${\ov\mu}_{C,1}\gg{\ov\mu}_{C,2}\gg\mu_o^{\rm conf}\sim \mu^{\rm pole}_{q,2}$ in these special vacua at ${\ov\mu}^{(\rm DC)}_{\Phi}\ll\mph\ll\mo$. We can proceed then as in this phase in section 9.2 above and to start directly with the Lagrangian (6.4). Because ${\ov\mu}_{C,2}\gg\mu_o^{\rm conf}$, there are poles in the mion propagators at $ p=\mu^{\rm pole}(M_{ij})\ll {\ov\mu}_{C,2}\ll\la$,
\bq
\mu^{\rm pole}(M_{ij})\sim\frac{\la^2}{z_M(\la,{\ov\mu}_{C,2})\mph}\sim\frac{\la^2}{\mph}\Bigl (\frac{m_Q}{\la}\Bigr )^{\bd/N_F}\gg\mu_o^{\rm conf}\,,\quad z_M(\la,{\ov\mu}_{C,2})=\Bigl (\frac{\la}{m_Q}\Bigr )^{\bd/N_F}\,,
\eq
and all mions are dynamically relevant here in the range of scales $\mu^{\rm pole}(M_{ij})<\mu<\la$. Hence, after integrating them out in (6.4) one obtains the Lagrangian (6.7). From this, the masses of $N_{11}$ and $N_{22}$ nions are $\mu(N_{11})\sim\mu(N_{22})\sim\la(\la/\mph)^{\nd/(2N_c-N_F)}$, while the hybrids $N_{12}$ and $N_{21}$ are massless. We recall finally that the masses of constituent quarks are here ${\ov\mu}_{C,1}\sim (\langle M_2\rangle_{\rm spec}/\mph)^{1/2}\gg{\ov\mu}_{C,2}$ and ${\ov\mu}_{C,2}\sim (\langle M_1\rangle_{\rm spec}/\mph)^{1/2}\gg \mu^{\rm pole}_{q,2}$, and the mass scale of gluonia is $\lym^{(\rm spec)}=[\langle M_1\rangle_{\rm spec}\langle M_2\rangle_{\rm spec}/\mph]^{1/3}\sim\la\,[(m_Q/\la)(\la/\mph)^{\nd/(2N_c-N_F)}]^{1/3}\gg\mu(N_{11})\sim\mu(N_{22})$.

\section{Dual theory. Broken flavor symmetry.\\
\hspace*{1.1cm} The region $\mathbf {\mo\ll\mph\ll\la^2/m_Q}$ }

\hspace{5mm} {\bf 10.1\,\,\,  br1 - vacua, $\mathbf{DC_1-DC_2}$ phase}\\

We recall, see (8.1), that the condensates of mions and dual quarks in this vacua are
\bq
\langle M_1\rangle\sim m_Q\mph\,,\quad \langle M_2\rangle\sim \la^2\Bigl (\frac{\la}{\mph}\Bigr )^{\frac{n_1}{N_c-n_1}}\Bigl (\frac{m_Q}{\la}\Bigr )^{\frac{n_2-N_c}{N_c-n_1}}\,,\quad
\frac{\langle M_2\rangle}{\langle M_1\rangle}\sim\Bigl (\frac{\mo}{\mph}\Bigr )^{\frac{N_c}{N_c-n_1}}\ll 1\,,\nonumber
\eq
\bq
\langle N_2\rangle=\langle{\ov q_2} q_2(\mu=\la)\rangle=\frac{\langle M_1\rangle\la}{\mph}\sim m_Q\la\gg\langle N_1\rangle\,,\nonumber
\eq
and so some potentially relevant masses look here as
\bq
{\ov\mu}_{C,2}^{\,2}\sim\langle N_2\rangle=\frac{\langle M_1\rangle\la}{\mph}\sim m_Q\la\gg {\ov\mu}_{C,1}^{\,2}\sim\langle N_1\rangle\,,\quad \mu^{\rm pole}_{q,1}\sim\la\Bigl(\frac{m_Q\mph}{\la^2}\Bigr )^{N_F/3\nd}\gg \mu^{\rm pole}_{q,2}\,,
\eq
\bq
{\ov\mu}_{\rm gl,2}\sim\la\Bigl (\frac{\langle M_1\rangle}{\mph\la}\Bigr )^{N_F/3N_c}\sim\la\Bigl (\frac{m_Q}{\la}\Bigr )^{N_F/3N_c}\gg{\ov\mu}_{\rm gl,1}\,,\quad \frac{{\ov\mu}_{\rm gl,2}}{\mu^{\rm pole}_{q,1}}\sim \Bigl (\frac{\mo}{\mph}\Bigr )^{N_F/3\nd}\ll 1\,.\nonumber
\eq

From (10.1)
\bq
\frac{\mu^{\rm pole}_{q,1}}{{\ov\mu}_{C,1}}>1\quad {\rm at}\quad \mph>{\ov\mu}_{\Phi,1}=\la\Bigl (\frac{\la}{m_Q}\Bigr )^{\ov\sigma}\,,\quad \ov\sigma=\frac{\bo(N_c-n_1)+3\nd(2N_c-N_F)}{2N_F(N_c-n_1)+3\nd N_c}>0\,,
\eq
\bq
\frac{\mu^{\rm pole}_{q,1}}{{\ov\mu}_{C,2}}>1\quad {\rm at}\quad \mph>{\ov\mu}_{\Phi,2}=\la\Bigl (\frac{\la}{m_Q}\Bigr )^{\bo/2N_F}\,,\quad \mo\ll{\ov\mu}_{\Phi,1}\ll{\ov\mu}_{\Phi,2}\ll\frac{\la^2}{m_Q}\,.\nonumber
\eq

Therefore, the mass hierarchies in the region $\mo\ll\mph\ll{\ov\mu}_{\Phi,1}$ look as ${\ov \mu}_{C,2}\gg{\ov \mu}_{C,1}\gg\omp$ and the phase is $DC_1-DC_2$. In the regions ${\ov\mu}_{\Phi,1}\ll\mph\ll{\ov\mu}_{\Phi,2}$
and ${\ov\mu}_{\Phi,2}\ll\mph\ll\la^2/m_Q$ the mass hierarchies look, respectively, as ${\ov \mu}_{C,2}\gg
\omp\gg{\ov \mu}_{C,1}$ and $\omp\gg{\ov \mu}_{C,2}\gg{\ov \mu}_{C,1}$ and the phase is $HQ_1-DC_2$.\\

We start with the $DC_1-DC_2$ phase and recall that the largest constituent mass ${\ov\mu}_{C,2}$ is formed not at the scale $\mu\sim {\ov\mu}_{C,2}$ but at the lower scale $\mu\sim {\ov\mu}_{C,1}$, see \cite{ch2}. Hence, after integrating out simultaneously all dual quarks at $\mu < {\ov\mu}_{C,1}$ and then the dual gluons at $\mu<\lym^{\rm (br1)}$, one obtains the Lagrangian (6.4) with the only difference that the factor $z_M$ is now $z_M=(\la/{\ov\mu}_{C,1})^{2\bd/N_F}$. As one can check, all $N_F^2$ mions with masses $\mu(M)\sim (\la^2/z_M\mph)$ are still much heavier than all $N_F^2$ nions and so the Lagrangian of nions is (6.7). The hybrid nions $N_{12}$ and $N_{21}$ are massless, while the masses of $n_1^2$ nions $N_{11}$ and $n_2^2$ nions $N_{22}$ look now as $\mu(N_{11})\sim \mu(N_{22})\sim m_Q\mph/\la$\,. Because ${\ov\mu}_{C,1}\gg\mu^{\rm conf}_o$ at $\mo<\mph<{\ov\mu}_{\Phi,1}$, the above mion masses are their pole masses and the mions are dynamically relevant in the range of scales $\mu^{\rm pole}(M)\sim (\la^2/z_M\mph)<\mu<\la$.

Besides, there is in the mass spectrum \,:\, a)\, a large number of hadrons made of dual constituent quarks with the masses ${\ov\mu}_{C,2}$ and ${\ov\mu}_{C,1},\,\, {\ov\mu}_{C,2}\gg{\ov\mu}_{C,1}\gg\mu^{\rm pole}(M)$,\, b)\, a large number of gluonia with the mass scale $\sim\lym^{(\rm br1)}=(\langle M_1\rangle_{\rm br1}\langle M_2\rangle_{\rm br1}/\mph)^{1/3}\sim (m_Q\langle M_2\rangle_{\rm br1})^{1/3}\ll{\ov\mu}_{C,1}$ made of dual $SU(\nd)$ gluons.\\

\hspace{5mm} {\bf 10.2\,\,\,  br1 - vacua, $\mathbf{HQ_1-DC_2}$ phase}\\

{\bf 10.2.1}\,\,\, The region ${\ov\mu}_{\Phi,1}\ll\mph\ll{\ov\mu}_{\Phi,2}\,,\,\, {\ov \mu}_{C,2}>\omp >{\ov \mu}_{C,1}$\,. Here, the largest dynamical mass ${\ov\mu}_{C,2}$ is formed not at the scale $\mu\sim{\ov \mu}_{C,2}$ but at the lower scale $\mu\sim \omp\ll {\ov\mu}_{C,2}$\,, see \cite{ch2}. Hence, proceeding as in \cite{ch2} (see section 4), i.e. integrating out the constituent quarks ${\ov q}_2, q_2$ and the quarks ${\ov q}_1, q_1$ as heavy ones at $\mu<\omp$ and then all $SU(\nd)$ dual gluons at $\mu<\lym^{\rm (br1)}$, the Lagrangian of $N_F^2$ mions $M$ and $n_2^2$ nions $N_{22}$ takes now the form
\bq
K=\Biggl [\,\frac{z_M(\la,\omp)}{\la^2}\rm {Tr}\,(M^\dagger M)+\rm {Tr\,\sqrt {N^{\dagger}_{22} N_{22}}}\, \Biggr ]\,,
\eq
\bq
W=(N_c-n_1)S-W_{NM}+W_M\,,\quad S=\Biggl (\frac{\det N_{22}}{\la^{\bd}\det (M_{11}/\la)} \Biggr )^{\frac{1}{N_c-n_1}}\,,
\nonumber
\eq
\bq
\quad W_{NM}={\rm Tr}\,\frac{N_{22}}{\la}\Bigl (M_{22}-M_{21}\frac{1}{M_{11}}M_{12}\Bigr )\,,\quad W_M=m_Q{\rm Tr} M-\frac{1}{2\mph}\Biggl ({\rm Tr} (M^2)-\frac{1}{N_c}({\rm Tr} M)^2\Biggr ),\nonumber
\eq
\bq
z_q(\la,\omp)=\Bigl (\frac{\omp}{\la}\Bigr )^{\bd/N_F}\,,\quad z_M=z_M(\la,\omp)=1/z^2_q(\la,\omp)\,.
\eq
From (10.3),(10.4) the masses of $n_1^2$ mions $M_{11}$\,, $n_2^2$ mions $M_{22}$ and $n_2^2$ nions $N_{22}$ look here as
\bq
\mu^{\rm pole}(M_{11})\sim \mu^{\rm pole}(M_{22})\sim \frac{\la^2}{z_M(\la,\omp)\mph}\gg \mu(N_{22})\sim \frac{m_Q\mph}{\la}\,,
\eq
while $2n_1n_2$ hybrid mions $M_{12}$ and $M_{21}$ are massless.

On the whole, the mass spectrum in this region includes\,: a) a large number of 22-flavored hadrons made of non-relativistic weakly confined constituent quarks ${\ov q}_2,\, q_2$ with the masses ${\ov\mu}_{C,2}\gg\omp$ (the string tension is ${\sqrt\sigma}\sim \lym^{\rm (br1)}\ll\omp\ll {\ov\mu}_{C,2}$)\,, \,\, b) a large number of 11-flavored hadrons made of non-relativistic weakly confined quarks ${\ov q}_1,\, q_1$ with the masses $\omp\gg \lym^{(\rm br1)}$, \,\,c) corresponding heavy hybrid hadrons with the masses $\sim({\ov\mu}_{C,2}+\omp)$,\,\, d) a large number of gluonia with the mass scale $\sim\lym^{(\rm br1)}$, \,\, e) $M_{11},\, M_{22}$ mions and $N_{22}$ nions with the masses (10.5)\,,\,\ f) the hybrid mions $M_{12}\,,\, M_{21}$ are massless.\\

{\bf 10.2.2}\,\,\, The region ${\ov\mu}_{\Phi,2}\ll\mph\ll\la^2/m_Q,\,\, {\ov \mu}_{C,2}\ll\omp$. Here, the largest physical mass is $\omp$. Hence, unlike the constituent quarks ${\ov q}_2,\, q_2$ with the soft non-perturbative dynamical masses ${\ov\mu}_{C,2}$ above, the quarks ${\ov q}_1,\, q_1$ with the hard perturbative masses $\omp$ can now be integrated out at $\mu<\omp$ {\it independently} of other degrees of freedom. After integrating them out, the lower energy theory at $\mu<\omp$ has $N^{\,\prime}_F=(N_F-n_1)=n_2$ flavors, $\nd$ colors, ${\rm\ov b}^{\,\prime}_o=(3\nd-N^{\,\prime}_F)
=(3\nd-n_2)$, and the new scale factor $\Lambda^{\prime}_q$ of its gauge coupling is
\bq
\Bigl [\Lambda^{\prime}_q(M_{11})\Bigr ]^{{\rm\ov b}^{\,\prime}_o}={\it z}^{n_2}_{q}(\la,\omp)\la^{\bd}\det \Bigl (\frac{M_{11}}{\la}\Bigr )\,,\quad \Lambda^{\prime}_q=\langle\Lambda^{\prime}_q (M_{11})\rangle\sim \omp\,.
\eq

Because $3/2<N^{\,\prime}_F=(N_F-n_1)/\,\nd <3$ at $n_1<\bo/2$, this lower energy theory will be in the conformal regime with the anomalous dimensions of ${\ov q}_2,\, q_2$ quarks and $M_{22}$ mions (the mions $M_{11}$ and hybrids $M_{12}, M_{21}$ do not evolve any more at $\mu<\omp$)
\bq
\gamma_q^{\,\prime}=\gamma_q^{\,\prime, \,\rm conf}=\frac{\bd^{\,\prime}}{N_F^{\,\prime}}=\frac{\bd+n_1}{N_F-n_1}\,,\quad \,\gamma_M^{\,\prime}=\gamma_M^{\,\prime, \,\rm conf}= -2\gamma_q^{\,\prime, \,\rm conf}\,,
\eq
while, because $1<N^{\,\prime}_F=(N_F-n_1)/\,\nd <3/2$ at $\bo/2<n_1\leq N_F/2$, it will be in the strong coupling regime
with the anomalous dimensions
\bq
\gamma_q^{\,\prime}=\gamma_q^{\,\prime, \,\rm str}=\frac{2\nd-n_2}{n_2-\nd}=\frac{2\nd-n_2}{N_c-n_1}\,,\quad \gamma_M^\prime=\gamma_M^{\,\prime, \,\rm str}=-(1+\gamma_q^{\,\prime,\,\rm str})=-\frac{\nd}{N_c-n_1}\,.
\eq

Instead of (10.1), the constituent and pole masses of the ${\ov q}_2,\, q_2$ quarks look now as, see (10.4),
\bq
{\ov \mu}^{\,\prime}_{C,2}=\Bigl [ z_q(\la,\omp)\Bigl (\langle N_2\rangle\sim m_Q\la\Bigr )\Bigr ]^{1/2}\,,\quad
{\tilde\mu}_{q,2}^{\rm pole}=\frac{\langle M_2\rangle}{{\it z}_q(\la,\omp)\la}\,\Bigl (\,\frac{\omp}{{\tilde\mu}_
{q,2}^{\rm pole}}\,\Bigr )^{\gamma^\prime_q}\,,\nonumber
\eq
\bq
\Bigl (\frac{{\tilde\mu}_{q,2}^{\rm pole}}{{\ov \mu}^{\,\prime}_{C,2}}\Bigr )_{\rm conf}\sim \Bigl (\frac{\mo}{\mph}\Bigr )^{\frac{N_c(n_1+\bd)}{6\nd(N_c-n_1)}}\ll 1\,,\quad
\Bigl (\frac{{\tilde\mu}_{q,2}^{\rm pole}}{{\ov \mu}^{\,\prime}_{C,2}}\Bigr )_{\rm str}\sim \Bigl (\frac{\mo}{\mph}\Bigr )^{N_c/2\nd}\ll 1\,,
\eq
so that the lower energy theory is, on the whole, in the $HQ_1-DC_2$ phase in any case.

After integrating out the heaviest ${\ov q}_1, q_1$ quarks as heavy ones at $\mu<\omp$ the Lagrangian looks as, see (10.3),(10.4),
\bq
K=\Biggl [\,\frac{z_M(\la,\omp)}{\la^2}\rm {Tr}\,(M^\dagger M)+{\it z_q}(\la,\omp)\Bigl (q_2^\dagger q_2+{\ov q}_2^\dagger{\ov q}_2\Bigr )\,\Biggr ]\,,
\eq
\bq
W=\Bigl [-\frac{2\pi}{{\ov \alpha}(\mu)}{\ov s}\,\Bigr ]-\frac{1}{\la}{\rm Tr}\, {\ov q}_2\Bigl (M_{22}-M_{21}\frac{1}{M_{11}}M_{12}\Bigr )q_2+W_M\,,\nonumber
\eq
with $\Lambda^{\prime}_q(M_{11})$ of the gauge coupling given in (10.6).

Therefore, after integrating then out the constituent ${\ov q}_2,\, q_2$ quarks at $\mu<{\ov \mu}^{\,\prime}_{C,2}$ and, finally, all $SU(\nd)$ gluons at $\mu<\lym^{({\rm br1})}$\,, the Lagrangian of mions $M$ and nions $N_{22}$ has the superpotential as in (10.3) while the Kahler term is
\bq
K=\frac{z_M(\la,\rm\omp)}{\la^2}\,\rm {Tr}\,\Biggl ( M_{11}^\dagger M_{11}+M_{12}^\dagger M_{12}+M_{21}^\dagger M_{21}
+{\it z^\prime_M(\omp,{\ov \mu}^{\,\prime}_{\rm C,2})}\, M_{22}^\dagger M_{22}\Biggr )+
\eq
\bq
+{\it z_q(\la,\rm\omp)}\rm {Tr\,\sqrt {N^{\dagger}_{22} N_{22}}}\,,\quad
{\it z^\prime_M}(\omp,{\ov \mu}^{\,\prime}_{C,2})=\Bigl (\frac{{\ov \mu}^{\,\prime}_{C,2}}{\omp}\Bigr )^{\gamma^\prime_M}\gg 1\,.\nonumber
\eq

From (10.11), the mass of mions $M_{11}$ is
\bq
\mu(M_{11})\sim \frac{\la^2}{z_M(\la,\omp)\mph}\sim\frac{\la^2}{\mph}
\Bigl(\frac{m_Q\mph}{\la^2}\Bigr)^{2\bd/3\nd}\sim\Bigl (\frac{\mo}{\mph}\Bigr )^{N_c/\nd}\omp\,,
\eq
while the masses of mions $M_{22}$ and nions $N_{22}$ look as
\bq
\mu(M_{22})\sim \mu(N_{22})\sim\Biggl (\frac{z_q(\la,\omp) m_Q\la}{z_M^\prime(\omp,{\ov \mu}^{\,\prime}_{\rm C,2})}\Biggr )^{1/2}\,.\nonumber
\eq
At $2n_1<\bo$ this is
\bq
\mu (M_{22})\sim \mu (N_{22})\sim \Bigl (\frac{\mo}{\mph}\Bigr )^{3N_c/n_2}\omp\,,
\eq
and at $\bo<2n_1\leq N_F$ this is
\bq
\mu (M_{22})\sim \mu (N_{22})\sim \Bigl (\frac{\mo}{\mph}\Bigr )^{\frac{N_c(n_2+N_c-n_1)}{4\nd (N_c-n_1)}}\omp\,.
\eq
And finally, the hybrid mions $M_{12}$ and $M_{21}$ are massless.\\

On the whole, the mass spectrum includes in this case. a) A large number of heaviest 11-flavored hadrons made of  weakly confined quarks ${\ov q}_1,\, q_1$  with the masses $\omp$, see (10.1), (the string tension is ${\sqrt\sigma}\sim \lym^{\rm (br1)}\ll {\ov\mu}_{C,2}\ll\omp$)\,. \,\, b) A large number of 22-flavored hadrons made of non-relativistic and weakly confined quarks ${\ov q}_2,\, q_2$ with the constituent masses ${\ov\mu}^{\,\prime}_{C,2}$ (10.9). \,\,c) Corresponding heavy hybrid hadrons with the masses $\sim(\omp+{\ov\mu}^{\,\prime}_{C,2})$.\,\, d) A large number of gluonia with the mass scale $\sim\lym^{(\rm br1)}$ made of dual $SU(\nd)$ gluons. \,\, e) $n_1^2$ mions $M_{11}$ with the masses (10.12).\,\, f) $n_2^2$  mions $M_{22}$ and $n_2^2$  nions $N_{22}$ with the masses (10.13),(10.14).\,\,
g) And finally, the hybrid mions $M_{12}\,,\, M_{21}$ are massless.\\

\hspace{5mm} {\bf 10.3\,\,\, br2 and special vacua}.\\

At $n_2<N_c$ there are also $\rm br2$ - vacua, see section 3. For these, all their properties can be obtained by the replacement $n_1\leftrightarrow n_2$ in formulas of the preceding sections 10.1-10.2\,. The only difference is that, because $n_2\geq N_F/2$, there is no analog of the conformal regime with $n_1<\bo/2$ at $\mph>{\ov\mu}_{\Phi,2}$ and
$\mu<\omp$, i.e. at $\mph>{\ov\mu}_{\Phi,2}$ the lower energy theory at $\mu<\omp$ is always in the strong coupling regime.

As for the special vacua (see section 3), all their properties can also be obtained with $n_1=\nd,\, n_2=N_c$
in formulas of the preceding sections 10.1-10.2\,.\\

\section{Broken $\mathcal{N}=2$ SQCD}

\hspace*{3mm} We consider now ${\cal N}=2$ SQCD with $N_c$ colors, $N_F$ flavors of light quarks, the scale factor $\Lambda_2$ of the gauge coupling, and with ${\cal N}=2$ broken down to ${\cal N}=1$ by the large mass parameter $\mu_X\gg \Lambda_2$ of the adjoint field $X=X^a\lambda^a,\, {\rm Tr}(\lambda^a \lambda^b)=\delta^{ab}/2$. At very high scales $\mu\gg\mu_X$ the Lagrangian looks as (the exponents with gluons are implied in the Kahler term K)
\bq
K=\frac{1}{g^2(\mu,\Lambda_2)}{\rm Tr}\,(X^\dagger X)+{\rm Tr}\,({\textbf{Q}}^\dagger \textbf{Q}+\textbf{Q}\ra \ov{ \textbf{Q}})\,,\quad
\eq
\bq
W=-\frac{2\pi}{\alpha(\mu,\Lambda_2)}S+\mu_X {\rm Tr}\,(X^2)+\sqrt{2}\,{\rm Tr}\,(\ov {\textbf{Q}}X
\textbf{Q})+ m\,{\rm Tr}\,(\ov {\textbf{Q}}\textbf{Q}).\nonumber
\eq

The running mass of $X$ is $\mu_X(\mu)=g^2(\mu)\mu_X$, so that at scales $\mu<\mu^{\rm pole}_X=g^2(\mu^{\rm pole}_X)\mu_X$ the field $X$ decouples from the dynamics and the RG evolution becomes those of ${\cal N}=1$ SQCD. The matching of ${\cal N}=2$ and ${\cal N}=1$ couplings at $\mu=\mu^{\rm pole}_X$ looks as ($\Lambda_2$ and $\la$ are the scale factors of ${\cal N}=2$ and ${\cal N}=1$ gauge couplings, $\la$ is held fixed when $\mu_X\gg\la$ is varied, ${\rm b}_2=2N_c-N_F\,,\,\bo=3N_c-N_F$)
\bq
\frac{2\pi}{\alpha(\mu=\mu^{\rm pole}_X,\Lambda_2)}=\frac{2\pi}{\alpha(\mu=\mu^{\rm pole}_X,\la)}\,,
\quad \mu_X\gg\la\gg\Lambda_2\,,\nonumber
\eq
\bq
b_2\ln\frac{\mu^{\rm pole}_X}{\Lambda_2}
=\bo\ln\frac{\mu^{\rm pole}_X}{\la}-N_F\ln z_Q(\la,\mu^{\rm pole}_X)+N_c\ln\frac{1}{g^2(\mu=\mu^{\rm pole}_X)}\,,
\eq
\bq
\la^{\bo}=\frac{\Lambda^{b_2}_2\mu^{N_c}_X}{z^{N_F}_Q (\la,\mu^{\rm pole}_X)}=
z^{N_F}_Q (\mu^{\rm pole}_X,\la)\Lambda^{b_2}_2\mu^{N_c}_X\,,\quad
z_Q(\la,\mu=\mu^{\rm pole}_X)\sim\Bigl (\ln\frac{\mu^{\rm pole}_X}{\la}\Bigr )^{\frac{N_c}{\bo}}\gg 1.\nonumber
\eq

Although the field $X$ becomes too heavy and does not propagate any more at $\mu<\mu^{\rm pole}_X$, the loops of light quarks and gluons which are still active at $\la<\mu<\mu^{\rm pole}_X$ if the next largest physical mass $\mu_H$ is below $\la$ and at $\mu_H<\mu<\mu^{\rm pole}_{X}$ if $\mu_H>\la$, induce him a non-trivial logarithmic renormalization factor $z_X(\mu^{\rm pole}_X,\mu<\mu^{\rm pole}_X)\ll 1$.

Therefore, finally, at scales $\la\ll\mu\ll\mu^{\rm pole}_X$ if $\mu_H<\la$ and at $\mu_H\ll\mu\ll\mu^{\rm pole}
_X$ if $\mu_H>\la$, the Lagrangian of the broken ${\cal N}=2$ - theory with $0<N_F<2N_c$ can
be written as
\bq
K=\frac{z_X(\mu^{\rm pole}_X,\mu)}{g^2(\mu^{\rm pole}_X)}\,{\rm Tr}\,(X^\dagger X)+z_Q(\mu^{\rm pole}_X,\mu)\,{\rm Tr}\,({\textbf{Q}}
^\dagger \textbf{Q}+\textbf{Q}\ra \ov{\textbf{Q}})\,,
\eq
\bq
W=-\frac{2\pi}{\alpha(\mu,\la)}S+\mu_X {\rm Tr}\,(X^2)+\sqrt{2}\,{\rm Tr}\,(\ov{\textbf{Q}} X \textbf{Q})+ m\,{\rm Tr}\,
(\ov{\textbf{Q}}\textbf{Q})\,.\nonumber
\eq
\bq
z_X(\mu^{\rm pole}_X,\mu)\sim\Biggl (\,\frac{\ln\, (\mu/\la)}{\ln\, (\mu^{\rm pole}_X/\la)}\,\Biggr)^
{{\rm b}_2/\bo}\ll 1\,,
\eq
\bq
z_Q(\mu^{\rm pole}_X,\mu)=z_Q(\mu^{\rm pole}_X,\la)z_Q(\la,\mu),\quad z_Q(\la,\mu)\sim \Bigl (\ln\frac{\mu}
{\la}\Bigr)^{N_c/\bo}\gg 1\,.\nonumber
\eq

In all cases when the field $X$ remains too heavy and dynamically irrelevant, it can be integrated out in (11.3) and one obtains
\bq
K=z_Q(\mu^{\rm pole}_X,\mu)\,{\rm Tr}\,({\textbf{Q}}^\dagger \textbf{Q}+\textbf{Q}\ra \ov{\textbf{Q}})\,,
\eq
\bq
W_Q=-\frac{2\pi}{\alpha(\mu,\la)}S+ m\,{\rm Tr}(\ov{\textbf{Q}} \textbf{Q})-\frac{1}{2\mu_X}\Biggl ({\rm Tr}\,(\ov{\textbf{Q}}\textbf{Q})^2-\frac{1}{N_c}\Bigl({\rm Tr}\,\ov{\textbf{Q}} \textbf{Q} \Bigr)^2 \Biggr ).\nonumber
\eq

Now we redefine the normalization of the quarks fields
\bq
\textbf{Q}=\frac{1}{z^{1/2}_Q(\mu^{\rm pole}_X,\la)}\,Q\,,\quad \ov{\textbf{Q}}=\frac{1}{z^{1/2}_Q(\mu^{\rm pole}_X,
\la)}\,{\ov Q}\,,
\eq
\bq
K=z_Q(\la,\mu){\rm Tr}\Bigl (\,Q^\dagger Q+(Q\ra {\ov Q})\,\Bigr ),\quad W= -\frac{2\pi}{\alpha(\mu,\la)}S+W_Q\,,
\eq
\bq
W_Q=\frac{m}{z_Q(\mu^{\rm pole}_X,\la)}\,{\rm Tr}({\ov Q} Q)-\frac{1}{2 z^2_Q(\mu^{\rm pole}_X,\la)\mu_X}\Biggl ({\rm Tr}
\,({\ov Q}Q)^2-\frac{1}{N_c}\Bigl({\rm Tr}\,{\ov Q} Q \Bigr)^2 \Biggr ).
\eq
Comparing this with (1.3) and choosing
\bq
\frac{m}{z_Q(\mu^{\rm pole}_X,\la)}=m_Q\ll\la\,, \quad z^2_Q(\mu^{\rm pole}_X,\la)\mu_X=\mph\gg\la
\eq
it is seen that with this matching the $\Phi$ - theory and the broken ${\cal N}=2$ SQCD will be equivalent.

Therefore, until both $X$ and $\Phi$ fields remain dynamically irrelevant, all results obtained above for the $\Phi$ - theory will be applicable to the broken ${\cal N}=2$ SQCD as well. Besides, the $\Phi$ and $X$ fields remain dynamically irrelevant in the same region of parameters, i.e. at $N_F<N_c$ and at $\mu_H>\mu_o$ if $N_F>N_c$\,, see (4.1).

Moreover, some general properties of both theories such as {\it the multiplicity of vacua with unbroken or broken flavor symmetry and the values of vacuum condensates of corresponding chiral superfields} (i.e. $\langle{\ov Q}_j Q_i\rangle$ and $\langle S\rangle$, see section 3) {\it are the same in these two theories, independently of whether the fields $\Phi$ and $X$ are irrelevant or relevant}.

Nevertheless, once the fields $\Phi$ and $X$ become relevant, the phase states, the RG evolution, the mass spectra etc., {\it become very different in these two theories}. The properties of the $\Phi$ - theory were described in detail above in the text. In general, once $X$ becomes sufficiently light and dynamically relevant, the dynamics of the broken ${\cal N}=2$ SQCD with $\mu_X\gg\la$ becomes complicated (we expect that the field $X$ will be higgsed, with $\mu_{\rm gl}\sim \mu_o$, see (4.1)\,) and is outside the scope of this paper.\\

Finally, we trace now a transition to the slightly broken ${\cal N}=2$ theory with small $\mu_X\ll\Lambda_2$
and fixed $\Lambda_2$. For this, we write first the appropriate form of the effective superpotential obtained from (11.7),(11.8)
\bq
W^{\rm eff}_Q=-\nd S+\frac{m}{z_Q(\mu^{\rm pole}_X,\la)}\,{\rm Tr}({\ov Q} Q)-\frac{1}{2 z^2_Q(\mu^{\rm pole}_X,\la)\mu_X}\Biggl ({\rm Tr}\,({\ov Q}Q)^2-\frac{1}{N_c}\Bigl({\rm Tr}\,{\ov Q} Q \Bigr)^2 \Biggr ),\nonumber
\eq
\bq
S=\Biggl (\,\,\frac{\det {\ov Q}Q}{\la^{\bo}}\,\,\Biggr )^{1/\nd}\,,\quad \la^{\bo}=z^{N_F}_Q(\mu^{\rm pole}_X,\la)\Lambda^{b_2}_2\mu^{N_c}_X
\eq
and restore now the original normalization of the quark fields $\ov{\textbf{Q}}, \textbf{Q}$ appropriate for the slightly broken ${\cal N}=2$ theory with varying $\mu_X\ll\Lambda_2$ and fixed $\Lambda_2$, see (11.6),
\bq
W^{\rm eff}_Q=-\nd S+m\,{\rm Tr}(\ov{\textbf{Q}} \textbf{Q})-\frac{1}{2 \mu_X}\Biggl ({\rm Tr}\,(\ov{\textbf{Q}}\textbf{Q})^2
-\frac{1}{N_c}\Bigl({\rm Tr}\,\ov{\textbf{Q}} \textbf{Q} \Bigr)^2 \Biggr ),\quad
S=\Bigl (\frac{\det \ov{\textbf{Q}}\textbf{Q}}{\Lambda^{b_2}_2\mu^{N_c}_X} \Bigr )^{1/\nd}
\eq

One can obtain now from (11.11) the values of the quark condensates $\langle\ov{\textbf{Q}}_j \textbf{Q}_i\rangle$ at fixed $\Lambda_2$ and small $\mu_X\ll\Lambda_2$. Clearly, in comparison with $\langle{\ov Q}_j Q_i\rangle$ in section 3, the results for $\langle\ov{\textbf{Q}}_j \textbf{Q}_i\rangle$ are obtained by the replacement\,:\, $m_Q\ra m\,,\, \mph\ra\mu_X\,,\,\la^{\bo}\ra\Lambda^{b_2}_2\mu^{N_c}_X$, while the multiplicities of vacua are the same. From (11.11), the dependence of $\langle\ov{\textbf{Q}}_j \textbf{Q}_i\rangle$ and $\langle S\rangle$ on $\mu_X$ is trivial in all vacua, $\sim \mu_X$.

With the above replacements, the expressions for $\langle{\ov Q}_j Q_i\rangle$ in section 3 in the region $\la\ll\mph\ll\mo$ correspond here to the hierarchy $m\ll\Lambda_2$, while those in the region $\mph\gg\mo$ correspond here to $m\gg\Lambda_2$. In the language of \cite{APS} used in \cite{CKM} (see sections 6-9 therein), the correspondence between the $r$ - vacua \cite{APS,CKM} of the slightly broken ${\cal N}=2$ theory with $0< \mu_X/\Lambda_2\ll 1,\,\, 0< m/\Lambda_2\ll 1$ and our vacua in section 3 looks as
\footnote{\,
This correspondence is based on comparison of multiplicities of our vacua at $\mph\ll\mo$ described in section 3 and those of $r$ - vacua at $m\ll\Lambda_2$ and $\mu_X\ll\Lambda_2$ as these last are given in \cite{CKM}.
}
\,: \,a)\, \,$r=n_1$,\, b)\, our L - vacua with the unbroken or the L - type ones with spontaneously broken flavor symmetry correspond, respectively, to the first group of vacua of the non-baryonic branches with $r=0$ and $r\geq 1,\, r\neq\nd$ in \cite{CKM}\,,\,\, c)\, our S - vacua with the unbroken flavor symmetry and $\rm br2$ - vacua with the spontaneously broken flavor symmetry correspond to the first type from the second group of vacua of the baryonic branches  with, respectively, $r=0$ and $1\leq r<\nd$ in \cite{CKM},\,\, d)\, our special vacua with $n_1=\nd,\, n_2=N_c$ correspond to the second type of vacua from this group, see \cite{CKM}.

\section{Conclusions}

\hspace {4mm} The mass spectra and phase states of the ${\cal N}=1$ SQCD-like $\Phi$ - theory (and its dual variant, the $d\Phi$ - theory) with additional colorless flavored fion fields $\Phi_{ij}$ were described above in the text in some details, within the dynamical scenario $\#1$ (see \cite{ch1,ch2} for more details about this scenario with the coherent colorless diquark condensate).

In comparison with the standard ${\cal N}=1$ SQCD with the superpotential $W=m_Q{\rm Tr}({\ov Q}Q)$ and the only small parameter $m_Q/\la\ll 1$ which serves as the infrared regulator, the $\Phi$ - theory includes two independent competing small parameters which serve as infrared regulators, $m_Q/\la\ll 1$ and $\la/\mph\ll 1$, see (1.3). Due to this the dynamics of this theory is much richer. Two main qualitatively new elements in this $\Phi$ - theory are\,:

a) the appearance of a large number of vacua with the spontaneously broken vectorial flavor symmetry, $U(N_F)\ra U(n_1)\times U(n_2)$,\,\,

b)\, in a number of cases with $N_F>N_c$, due to their interactions with the light quarks, the seemingly heavy and dynamically irrelevant fion fields $\Phi$\, `{\it return back\,}' and there appear two additional generations of light $\Phi$ - particles, see the section 4.\\

This is not a purpose of these conclusions to repeat in a shorter form all results obtained above in the main text for the phase states and mass spectra of the direct and dual theories at different values of $\mph/\la\gg 1$. We only point out here that, similarly to the ordinary SQCD \cite{ch1,ch2,ch3}, the direct $\Phi$ -theory and its dual variant, the $d\Phi$ - theory, are also not equivalent (at least, within the scenario $\#1$ considered in this paper).

We will try only to formulate here in a few words the most general qualitative property of SQCD-like theories which emerged from the studies in \cite{ch1, ch2, ch3} and in this paper. This is {\it the extreme sensitivity of their dynamical behavior in the IR region of momenta, of their mass spectra and even of the phase states, to the values of small parameters in the Lagrangian which serve as infrared regulators}.\\

The $\Phi$-theory with $\mph\gg\la$ considered in this paper is tightly connected with the $X$-theory which is the ${\cal N}=2$ SQCD broken down to ${\cal N}=1$ by the large mass parameter $\mu_X\gg\Lambda_2$ of the adjoint fields $X$. The multiplicity of vacua and the numerical values of the quark and gluino condensates, $\langle{\ov Q}_j Q_i\rangle$ and $\langle S\rangle$, are the same in both theories (under the appropriate matching of parameters, see the section 11). Moreover, in all those cases when the fields $\Phi$ are dynamically irrelevant in the $\Phi$-theory, the fields $X$ are also dynamically irrelevant in the $X$-theory and these two theories are completely equivalent. But even in these cases, this does not mean that these two theories are simply equivalent to the ordinary SQCD with small unimportant corrections. First, the whole physics in a large number of additional vacua with the spontaneously broken flavor symmetry is completely different. And second, even in vacua with the unbroken flavor symmetry, these theories evolve to the standard SQCD with small corrections not at $\mph=(\rm{several})\la$ as one can naively expect, but only {\it at parametrically large values} of $\mph/\la$ (and, besides, these values differ parametrically in the direct and dual theories, see sections 5 and 6).

But when the fields $\Phi$ and $X$ become dynamically relevant the phase states, the mass spectra, etc. become very different in the $\Phi$ and $X$ - theories.\\

We have described also in section 11 the connections between the values of the quark condensates in different vacua in the strongly broken ${\cal N}=2$ SQCD  with large varying $\mu_X\gg\la$ and fixed $\la$ with those in the slightly broken ${\cal N}=2$ SQCD with small varying $\mu_X\ll\Lambda_2$ and fixed $\Lambda_2$.\\  

This work was supported in part by  Ministry of Education and Science of the Russian Federation and RFBR grant 12-02-00106-a.

{\appendix
\section{The RG flow in the $\mathbf \Phi$ - theory at $\mathbf{\mu>\la}$ }

{\bf \quad A.1\quad} We first consider the $\Phi$ - theory at $N_c<N_F<2N_c$ where it is taken as UV-free. We start with the canonically normalized Kahler term $K$ at the very high scale $\mu\sim \mu_{\rm UV}$ and the running couplings and mass parameters
\bq
K={\rm Tr}\,\Bigl ({\widehat\Phi}^\dagger {\widehat\Phi}\Bigr )+{\rm Tr}\Bigl (\,{\widehat Q}^\dagger {\widehat Q}+
( {\widehat Q}\ra {\widehat{\ov Q}})\,\Bigr )\,,\quad W=-\frac{2\pi}{\alpha(\mu)}S+W_{\Phi}+W_Q\,,\quad \nonumber
\eq
\bq
W_{\Phi}=\frac{\mph(\mu)}{2}\Biggl [{\rm Tr}\,({\widehat\Phi}^2)-\frac{1}{\nd}\Bigl ({\rm Tr}\,{\widehat\Phi}\Bigr )^2\Biggr ]\,,\quad W_Q=- f(\mu){\rm Tr}\,\Bigl ( {\widehat{\ov Q}}{\widehat\Phi}{\widehat Q}\Bigr )+{\rm Tr}\,
\Bigl ( {\widehat{\ov Q}}\, m_ Q(\mu) {\widehat Q}\Bigr )\,.
\eq
Now, instead of running parameters, we introduce $\mu$-independent ones, $\la,\,\, \mph$ and $m_Q$ ($\mph\gg\la$ and $m_Q\ll\la$ in the main text),
\bq
\frac{1}{a(\mu)}=\frac{2\pi}{N_c\alpha(\mu)}=\frac{\bo}{N_c}\ln\frac{\mu}{\la}-\frac{N_F}{N_c}\ln z_Q(\la,\mu)+\ln\frac{1}{a(\mu)}+C_a\,,\quad \bo=3N_c-N_F\,,
\eq
\bq
a_f(\mu)=\frac{N_c f^2(\mu)}{2\pi}=\frac{a_f=N_c f^2/2\pi}{z_{\Phi}(\la,\mu)z^2_Q(\la,\mu)}\,,\quad
\mph(\mu)\equiv\frac{f^2\mph}{z_{\Phi}(\la,\mu)}\,,\quad m_Q(\mu)\equiv\frac{m_Q}{z_Q(\la,\mu)}\,,\nonumber
\eq
where $z_Q(\la,\mu\gg\la)\gg 1$ and $z_{\Phi}(\la,\mu)$ are the perturbative renormalization factors (logarithmic in this case) in the theory with {\it all fields massless}, $a_f$ is taken as $a_f\sim 1/(\rm several)$ and $C_a$ is also $O(1)$ (it will be omitted for simplicity). Therefore, after redefinitions of the quark and $\Phi$ fields,
the Lagrangian at the very high scale can be rewritten as
\bq
K=z_{\Phi}(\la,\mu)\frac{1}{f^2}{\rm Tr}\,(\Phi^\dagger\Phi)+z_Q(\la,\mu){\rm Tr}\Bigl (\,Q^\dagger Q+(Q\ra {\ov Q})\,
\Bigr )\,,
\eq
\bq
W_{\Phi}=\frac{\mph}{2}\Biggl [{\rm Tr}\,(\Phi^2)-\frac{1}{\nd}\Bigl ({\rm Tr}\,\Phi\Bigr )^2\Biggr ]\,,\quad
W_Q=-{\rm Tr}\,\Bigl ( {\ov Q}\Phi Q\Bigr )+{\rm Tr}\,\Bigl ( {\ov Q}m_Q Q\Bigr )\,.\nonumber
\eq

From (A.2)
\bq
\frac{d a_f(\mu)}{d\ln\mu}=\beta_f=-a_f(\mu)\Bigl (2\gamma_Q(\mu)+\gamma_{\Phi}(\mu)\Bigr ),\quad \gamma_Q=\frac{d\ln z_Q(\mu)}{d\ln\mu},\quad \gamma_{\Phi}=\frac{d\ln z_{\Phi}(\mu)}{d\ln\mu}\,.
\eq
In the approximation of leading logarithms at large $\mu$
\bq
\gamma_Q(\mu)\simeq  \frac{2 C_F}{N_c}a(\mu)-\frac{N_F}{N_c}a_f(\mu),\quad \gamma_{\Phi}(\mu)\simeq -a_f(\mu)\,,\quad
\frac{2 C_F}{N_c}=\frac{N_c^2-1}{N_c^2}\simeq 1\,.
\eq
From (A.4),(A.5), there is the UV free solution
\bq
a(\mu)\simeq \frac{N_c}{\bo}\frac{1}{\ln (\mu/\la)}\,,\quad
a_f(\mu)\sim a_f\Bigl (\frac{1}{\ln(\mu/\la)}\Bigr )^{\frac{2N_c}{\bo}}\ll a(\mu),\quad 1<\frac{2N_c}{\bo}<2\,,
\eq
\bq
z_Q(\la,\mu)\sim \Bigl (\ln\frac{\mu}{\la}\Bigr )^{N_c/\bo}\gg 1,\quad z_{\Phi}(\la,\mu)\sim 1\,.
\eq
It is seen from (A.6) that the Yukawa coupling $a_f(\mu)$ is parametrically small in comparison with the gauge coupling $a(\mu)$ and, up to small corrections, it has no effect on the RG evolution at large $\mu$.

The first physical mass parameter which influences the RG flow with lowering the scale $\mu$ is $\mu^{\rm pole}_{1}(\Phi)=\mph(\mu=\mu^{\rm pole}_{1}(\Phi)\,)=f^2\mph/z_{\Phi}(\la,\mu^{\rm pole}_{1}(\Phi))\sim f^2\mph\gg\la$, so that $\mu_{\Phi}(\mu)$ becomes $\mu_{\Phi}(\mu)\sim f^2\mph>\mu$ at $\mu<\mu^{\rm pole}_{1}(\Phi)$ and the fields $\Phi$ become too heavy. They do not propagate any more and do not influence the RG evolution until $\mu_{\Phi}(\mu)>\mu$. Nevertheless, the anomalous dimension $\gamma_{\Phi}(\mu)$ remains small but nonzero even at $\mu<\mu^{\rm pole}_{1}(\Phi)$ due to loops of still active light quarks (and gluons interacting with quarks) and, instead of (A.5), the anomalous dimensions look at $\mu<\mu^{\rm pole}_{1}(\Phi)$ as
\bq
\gamma_Q(\mu)\simeq a(\mu),\quad \gamma_{\Phi}(\mu)\simeq -a_f(\mu)\,,
\eq
while (A.6),(A.7) remain the same. Hence, although the heavy fields $\Phi_{ij}$ decouple at $\la<\mu<\mu^{\rm pole}_{1}(\Phi)$, the RG flow remains parametrically the same because their role even at $\mu>\mu^{\rm pole}_{1}(\Phi)$ was small.

Therefore, finally, at scales $\la<\mu<\mu^{\rm pole}_{1}(\Phi)$ if there is no physical masses $\mu_H>\la$ and at
$\mu_H<\mu<\mu^{\rm pole}_{1}(\Phi)$ if $\mu_H>\la$, the Lagrangian of the $\Phi$ - theory with $N_c<N_F<2N_c$ light flavors can be written as
\bq
K=\frac{1}{f^2}{\rm Tr}\,(\Phi^\dagger \Phi)+z_Q(\la,\mu){\rm Tr}\Bigl (\,Q^\dagger Q+(Q\ra {\ov Q})\,\Bigr )\,,\quad
W=-\frac{2\pi}{\alpha(\mu,\la)}S+W_{\Phi}+W_Q\,,\quad \nonumber
\eq
\bq
W_{\Phi}=\frac{\mph}{2}\Biggl [{\rm Tr}\,(\Phi^2)-\frac{1}{\nd}\Bigl ({\rm Tr}\,\Phi\Bigr )^2\Biggr ]\,,\quad
W_Q={\rm Tr}\,\Bigl ( {\ov Q}\,m^{\rm tot}_Q Q\Bigr )\,,\quad m^{\rm tot}_Q=m_Q-\Phi\,,
\eq
with $z_Q(\la,\mu)$ given in (A.7).\\

{\bf \quad A.2\quad} We consider now the case $1\leq N_F<N_c$\,. Although the $\Phi$ - theory is not UV free in this case and requires UV completion at $\mu>\mu_{\rm UV}$, the RG flow at $\mu_H<\mu\ll\mu_{\rm UV}$ is very specific (see below, the quarks are really higgsed in this case at $\mu_H=\mu_{\rm gl},\,\,\la\ll\mu_{\rm gl}\ll\mph\ll\mu_{UV}$, see section 2). We take from the beginning $a_f$ in (A.2) to be sufficiently small, $a_f\ll 1$, and calculate the behavior of $a(\mu)$ and $a_f(\mu)$ at $\la\ll\mu\ll\mu_{UV}$ in the massless theory which follows from their definitions in (A.2). Then, by definition, in the theory with $\la\ll\mu^{\rm pole}_1(\Phi)\ll\mu_{UV}$, the behavior  of $a(\mu)$ and $a_f(\mu)$ at $\mu^{\rm pole}_1(\Phi)\ll\mu\ll\mu_{UV}$ will be the same while, in general, it can be different at $\mu<\mu^{\rm pole}_1(\Phi)$.

There is the same solution (A.6) also at $1\leq N_F<N_c$, with a difference that $2/3<2N_c/\bo<1$ now and $a_f\ll 1$. Hence, starting with $\mu>\la,\,\, a_f(\mu)$ begins first {\it to decrease with increasing} $\mu$, but more slowly now than $a(\mu)\sim\ln^{-1}(\mu/\la)$. Due to this, $\beta_f(\mu)$ in (A.4) changes a sign at $\mu\sim\ov\mu$,
\bq
a_f({\ov\mu})\sim a({\ov\mu})\,\,\ra \,\, \ln\frac{{\ov\mu}}{\la}\sim \Bigl (\frac{1}{a_f}\Bigr )^{\frac{\bo}{N_c-N_F}}
\gg 1,\,\, a_f({\ov\mu})\sim\frac{1}{\ln ({\ov\mu}/\la)}\sim \Bigl (a_f\Bigr )^{\frac{\bo}{N_c-N_F}}\ll a_f\ll 1
\eq
and then $a_f(\mu)$ begins to grow
\bq
a_f(\mu>\ov\mu)\sim \frac{1}{\ln(\mu_{UV}/\mu)}\,,\quad \ln\Bigl (\frac{\mu_{UV}}{\ov\mu}\Bigr )\sim \Bigl (\frac{1}{a_f}\Bigr )^{\frac{\bo}{N_c-N_F}}\gg 1
\eq
with further increasing $\mu>{\ov\mu}$. Therefore, $z_{\Phi}(\la,\mu<{\ov\mu})\sim 1$ in the massless theory.

For our purposes in section 2 it will be sufficient to have $\mu_{\rm gl}\ll\mu^{\rm pole}_1(\Phi)\sim a_f\mph\ll{\ov\mu}\ll\mu_{\rm UV}$. This leads to a sufficiently weak logarithmic restriction
\bq
\frac{1}{a_f}\gg \Bigl (\ln\frac{\mph}{\la}\Bigr )^{\frac{N_c-N_F}{\bo}},\quad  0<\frac{N_c-N_F}{\bo}<\frac{1}{3}\,,
\eq
and then $z_{\Phi}(\la,\mu<\mu^{\rm pole}_1(\Phi))$ remains $\sim 1$ also in the $\Phi$ - theory with massive fields
$\Phi$.

\section{There is no vacua with $\mathbf{\langle S\rangle=0}$ at $\mathbf{m_Q\neq 0}$}

The purpose of this appendix is to show that the gluino condensate $\langle S\rangle\neq 0$ at $m_Q\neq 0$ in all vacua with the broken flavor symmetry, $U(N_F)\ra U(n_1)\times U(n_2)$, in both the direct and dual theories. \\

\hspace*{1cm} {\bf 1\,.\,\, Direct theory}\\

We assume that there is at $N_c<N_F<2N_c$ a large number of {\it additional} vacua with either $1\leq n_1\leq N_c-1$ components $\langle{\ov Q}_1Q_1=\Pi_1\rangle=0$, or $n_2\geq n_1$ components $\langle{\ov Q}_2Q_2=\Pi_2\rangle=0$. Even in this case the relations at $\mu=\la$
\bq
\langle \Pi_1+\Pi_2\rangle-\frac{1}{N_c}{\rm Tr}\,\langle\Pi\rangle=m_Q\mph,\quad
\langle S\rangle=\frac{1}{\mph}\langle\Pi_1\rangle\langle\Pi_2\rangle,\quad
\langle\Pi_1\rangle\neq\langle\Pi_2\rangle\,,
\eq
\bq
\langle m^{\rm tot}_{Q,1}\rangle=\langle m_Q-\Phi_1\rangle=\frac{\langle\Pi_2\rangle}{\mph}\,,\quad
\langle m^{\rm tot}_{Q,2}\rangle=\langle m_Q-\Phi_2\rangle=\frac{\langle\Pi_1\rangle}{\mph}\,,\nonumber
\eq
following from the Konishi anomalies (1.2),(1.4) remain valid. Therefore, one obtains from (B.1) that either
\bq
\langle \Pi_2\rangle=0\,, \quad \langle\Pi_1\rangle=\frac{N_c}{N_c-n_1}\, m_Q\mph\,, \quad \langle S\rangle=0\,,\quad 1\leq n_1\leq N_c-1\,,
\eq
\bq
\langle m^{\rm tot}_{Q,1}\rangle=\frac{\langle \Pi_2\rangle}{\mph}=0\,,\quad \langle m^{\rm tot}_{Q,2}\rangle
=\frac{\langle \Pi_1\rangle}{\mph}=\frac{N_c}{N_c-n_1}\, m_Q\,,\nonumber
\eq
or
\bq
\langle \Pi_1\rangle=0\,, \quad \langle\Pi_2\rangle=\frac{N_c}{N_c-n_2}\, m_Q\mph\,,
\quad \langle S\rangle=0\,, \quad n_2\neq N_c\,,
\eq
\bq
\quad \langle m^{\rm tot}_{Q,2}\rangle=\frac{\langle \Pi_1\rangle}{\mph}=0\,,\quad \langle m^{\rm tot}_{Q,1}\rangle=
\frac{\langle \Pi_2\rangle}{\mph}=\frac{N_c}{N_c-n_2}\, m_Q\, \nonumber
\eq
in these vacua. We will show below that this assumption is not self-consistent. I.e., we will start with (B.2) or (B.3) and calculate then explicitly $\langle S\rangle\neq 0$ in these vacua. For this, using a holomorphic dependence of $\langle S\rangle$ on $\mph$, it will be sufficient to calculate $\langle S\rangle\neq 0$ in some range of most convenient values of $\mph$. Hence, we take $m_Q\mph\sim \la^2$.\\

In vacua (B.2) with $\langle\Pi_2\rangle=0,\, \langle\Pi_1\rangle\sim m_Q\mph\sim \la^2$ the quarks ${\ov Q}_1,\, Q_1$ are higgsed with $\langle{\ov Q}_1\rangle=\langle Q_1\rangle\sim \la$. At $n_1<N_c-1$ the lower energy theory at $\mu<\la$ contains $SU(N_c-n_1)$ unbroken gauge symmetry with the scale factor of the gauge coupling $(\Lambda^\prime)^{\bo^\prime}\sim\la^{\bo}/\det \Pi_{11},\, \langle\Lambda^\prime\rangle\sim\la$, $n^2_1$ pions $\Pi_{11}$ and ${\ov Q}_2,\, Q_2$ quarks with zero condensate and the running mass $\langle m^{\rm tot}_{Q,2}\rangle=\langle m_Q-\Phi_2\rangle=\langle\Pi_1\rangle/\mph\sim m_Q$ at $\mu=\la$. For this reason, the variants with the $DC_2$ or $Higgs_2$ phases of these quarks are excluded, they will be always in the heavy quark $HQ_2$ - phase. At all $n_1<N_c-1$, proceeding as in \cite{ch1, ch2, ch3}, i.e. lowering the scale down to $\mu<m^{\rm pole}_{Q,2}\sim m_Q/z_Q(\la,m^{\rm pole}_{Q,2})$ and integrating out ${\ov Q}_2, Q_2$ quarks as heavy particles, there remains the pure $SU(N_c-n_1)$ Yang-Mills theory (and $n^2_1$ pions $\Pi_{11}$) with the scale factor of its gauge coupling
\bq
\lym^3=\Biggl (\frac{\la^{\bo}\det (m_Q-\Phi_{22})}{\det\Pi_{11}}\Biggr )^{1/(N_c-n_1)},
\eq
and, finally, with the Lagrangian of the form (2.22) at $\mu<\langle\lym\rangle$. From (B.4)
\bq
\langle S\rangle=\langle\lym^3\rangle\sim \la^3\Bigl (\frac{\la}{\mph}\Bigr )^{\frac{n_1}{N_c-n_1}}\Bigl (\frac{m_Q}
{\la}\Bigr )^{\frac{n_2-n_1}{N_c-n_1}}\neq 0\,.
\eq

At $n_1=N_c-1$ the gauge group will be broken completely and (B.4) originates from the instanton contribution.

The vacua (B.3) with $\langle\Pi_1\rangle=0$ are considered the same way and one obtains (B.4),(B.5) with the replacement
$n_1\leftrightarrow n_2$. (In vacua (B.3) the cases with $n_2>N_c$ are excluded from the beginning as the rank of $\langle Q_2\rangle$ is $\leq N_c$ and the unbroken $U(n_2)$ flavor symmetry cannot be maintained; the case $n_2=N_c$ is also excluded as $\langle\Pi_1\rangle\neq 0$ in this case, see (B.1)\,). Hence, only the cases with $n_2\leq N_c-1$ remain).

On the whole, the assumption about the existence of additional vacua (B.2) or (B.3) with $\langle\Pi_1\rangle=0$ or $\langle\Pi_2\rangle=0$ at $N_c<N_F<2N_c$ is not self-consistent.\\

\hspace*{1cm} {\bf 2\,.\,\, Dual theory}\\

The dual analog of (B.1)-(B.3) looks as, see (1.8),
\bq
\langle M_1+M_2\rangle-\frac{1}{N_c}{\rm Tr}\,\langle M\rangle=m_Q\mph,\quad
\langle S\rangle=\frac{1}{\mph}\langle M_1\rangle\langle M_2\rangle,\quad
\langle M_1\rangle\neq\langle M_2\rangle\,,
\eq
By assumption, there is a large number of {\it additional} vacua with either
\bq
\langle M_2\rangle=0\,, \quad \langle M_1\rangle=\frac{N_c}{N_c-n_1}\, m_Q\mph\,, \quad \langle S\rangle=0\,,
\quad 1\leq n_1\leq N_c-1\,,
\eq
\bq
\langle N_1\rangle=\langle m^{\rm tot}_{Q,1}\rangle\la=\frac{\langle M_2\rangle\la}{\mph}=0\,,\quad \langle N_2\rangle=\langle m^{\rm tot}_{Q,2}\rangle\la
=\frac{\langle M_1\rangle\la}{\mph}=\frac{N_c}{N_c-n_1}\, m_Q\la\,,\nonumber
\eq
or
\bq
\langle M_1\rangle=0\,, \quad \langle M_2\rangle=\frac{N_c}{N_c-n_2}\, m_Q\mph\,, \quad \langle S\rangle=0\,,
\quad n_2\neq N_c\,,
\eq
\bq
\langle N_2\rangle=\langle m^{\rm tot}_{Q,2}\rangle\la=\frac{\langle M_1\rangle\la}{\mph}=0\,,\quad \langle N_1\rangle=\langle m^{\rm tot}_{Q,1}\rangle\la
=\frac{\langle M_2\rangle\la}{\mph}=\frac{N_c}{N_c-n_2}\, m_Q\la\,.\nonumber
\eq
In this case, it is more convenient for our purposes to choose the regions $\la\ll\mph\ll\mo$ at $3N_c/2<N_F<2N_c$ and $\la\ll\mph\ll\la(\la/m_Q)^{1/2}$ at $N_c<N_F<3N_c/2$.

We start with (B.7). It is not difficult to check that in these ranges of $\mph$ and at all $N_c<N_F<2N_c$ the largest mass is ${\ov\mu}_{\rm gl,2}\gg \mu^{\rm pole}_{q,1}$ due to higgsing of ${\ov q}_2, q_2$ quarks. Hence, in these (B.7) vacua, the cases with $n_2>\nd$ are excluded from the beginning as the rank of $\langle q_2\rangle$ is $\leq \nd$ and the unbroken $U(n_2)$ flavor symmetry cannot be maintained. But this excludes all such vacua as $n_1\leq N_c-1$ and $n_2=N_F-n_1\geq
\nd+1$.

Therefore, there remain only (B.8) vacua. In these, in the above ranges of $\mph$, the largest mass is ${\ov\mu}_{\rm gl,1}\gg \mu^{\rm pole}_{q,2}$ due to higgsing of ${\ov q}_1, q_1$ quarks. Hence, one obtains from similar considerations that $n_1\leq\nd-1$ (the case $n_1=\nd$ is also excluded from (B.6),(B.8)\,). And similarly, because their condensate $\langle N_2\rangle=0$, the quarks ${\ov q}_2, q_2$ will be always in the heavy quark $HQ_2$ - phase only. Hence, at all $n_1<N_c-1$, proceeding as in \cite{ch1, ch2, ch3}, i.e. integrating out first higgsed gluons and ${\ov q}_1, q_1$ quarks at $\mu<{\ov\mu}_{\rm gl,1}$, then ${\ov q}_2, q_2$ quarks with unhiggsed colors at $\mu<\mu^{\rm pole}_{q,2}$ and, finally,
unhiggsed gluons at $\mu<\langle\lym\rangle$, one obtains the low energy Lagrangian of the form (9.11) with
\bq
\lym^3=\Biggl (\frac{\la^{\bd}\det\Bigl (M_{22}/\la\Bigr )}{\det N_{11}}\Biggr )^{1/(\nd-n_1)}\,,\quad
\langle S\rangle=\langle\lym^3\rangle\sim \la^3\Bigl (\frac{\mph}{\la}\Bigr )^{\frac{n_2}{n_2-N_c}}\Bigl (\frac{m_Q}
{\la}\Bigr )^{\frac{n_2-n_1}{n_2-N_c}}\neq 0\,.
\eq

At $n_1=\nd-1$ the dual gauge group will be broken completely and (B.9) originates from the instanton contribution.

On the whole, the assumption about the existence of additional vacua (B.7) or (B.8) with $\langle M_1\rangle=0$ or $\langle M_2\rangle=0$ at $N_c<N_F<2N_c$ is also not self-consistent.\\

\end{document}